\documentclass[11pt,titlepage,a4paper]{article}
\usepackage{bozhomac}  
\usepackage{bozhlogo}  
\usepackage{cite}	
\usepackage{varioref}	

\title{\bfseries	\vspace*{-1.7in}
{\huge Lagrangian quantum field theory\\[1ex] in momentum picture}
 \\[1.3ex]
{\LARGE IV.\ Commutation relations for free fields}
}

\vspace{1.7ex}

\author{
Bozhidar Z.\ Iliev
\thanks{Laboratory of Mathematical Modeling in Physics,
Institute for Nuclear Research and \mbox{Nuclear} Energy,
Bulgarian Academy of Sciences,
Boul.\ Tzarigradsko chauss\'ee~72, 1784 Sofia, Bulgaria}
\thanks{E-mail address: bozho@inrne.bas.bg}
\thanks{URL: http://theo.inrne.bas.bg/$\sim$bozho/}
}

\date{	
 \vspace{2.27ex}\ShortTitle{QFT in momentum picture: IV}\\[0.27ex]
 \vspace{3.27ex}
\small
	\begin{tabular}{r@{$\colon\to~$}l}
 \vspace{0.27ex} Produced	& \fbox{\today}	\\[0.27ex]
	\end{tabular} \\[1.27ex]
\normalsize
\small
	\begin{tabular}{r@{$\colon~$}l}
\normalsize\sffamily\bfseries
\vspace{0.27ex} http://www.arXiv.org e-Print archive No. &
\normalsize\sffamily\bfseries
arXiv:0704.0066[hep-th]				\\[0.27ex]
	\end{tabular} \\[-0.27ex]
 \vspace{4.27ex}{\Huge\BOZHO}	\\[4.27ex]
 \vspace{0.27ex}\Subject{Quantum field theory}
								\\[2.27ex]
	\begin{tabular}{r@{\hspace{0.512em}}|@{\hspace{0.512em}}l}
 \vspace{0.27ex}\MSC[2000]{81Q99, 81S05, 81T99\\\hspace{0pt}}
&
 \vspace{0.27ex}\PACS[2003]{03.70.+k, 11.10.Ef, 11.10.-z,\\
				11.90.+t, 12.90.+b}
	\end{tabular} \\[1.27ex]
 \vspace{0.27ex}\KeyWords{%
Commutation relations, Anticommutation relations, Free quantum fields\\
Paracommutation relations, Parafermi and parabose commutation relations\\
Heisenberg relations (equations), Euler-Lagrange equations, Equations of
motion\\
Lagrangians for free fields, Momentum operator, Angular momentum operator\\
Spin and orbital angular momentum operators,
Normal ordering%
}\\[0.27ex]
}

\newcommand{\tU}{{\tope{U}}}		
\newcommand{\ta}{{\Tilde{a}}}		
\newcommand{\opsi}{\overline{\psi}}	
\newcommand{\bpsi}{\Breve{\psi}}	
\newcommand{\bk}{\boldsymbol{k}}  	

 \newcommand{\Hil}{\mathcal{F}}		
	\newcommand{\base}{\mathit{M}}	

\newcommand{\ope}[2][{}]{\lindex[\mathcal{#2}]{}{#1}} 
\newcommand{\tope}[2][{}]{\ope[#1]{\Tilde{#2}}} 
\begin{document}		

\renewcommand{\thepage}{\roman{page}}

\renewcommand{\thefootnote}{\fnsymbol{footnote}} 
\maketitle				
\renewcommand{\thefootnote}{\arabic{footnote}}   

\tableofcontents		


\begin{abstract}

	Possible (algebraic) commutation relations in the Lagrangian quantum
theory of free (scalar, spinor and vector) fields are considered from
mathematical view-point. As sources of these relations are employed the
Heisenberg equations/relations for the dynamical variables and a specific
condition for uniqueness of the operators of the dynamical variables (with
respect to some class of Lagrangians). The paracommutation relations or some
their generalizations are pointed as the most general ones that entail the
validity of all Heisenberg equations. The simultaneous fulfillment of the
Heisenberg equations and the uniqueness requirement turn to be impossible.
This problem is solved via a redefinition of the dynamical variables,
similar to the normal ordering procedure and containing it as a special
case. That implies corresponding changes in the admissible commutation
relations. The introduction of the concept of the vacuum makes narrow the
class of the possible commutation relations; in particular, the mentioned
redefinition of the dynamical variables is reduced to normal ordering. As a
last restriction on that class is imposed the requirement for existing of an
effective procedure for calculating vacuum mean values. The standard bilinear
commutation relations are pointed as the only known ones that satisfy all of
the mentioned conditions and do not contradict to the existing data.

\end{abstract}

\renewcommand{\thepage}{\arabic{page}}


\section {Introduction}
\label{Introduction}

	 The main subject of this paper is an analysis of possible (algebraic)
commutation relations in the Lagrangian quantum theory%
\footnote{~%
In this paper we considered only the Lagrangian (canonical) quantum field
theory in which the quantum fields are represented as operators, called field
operators, acting on some Hilbert space, which in general is unknown if
interacting fields are studied. These operators are supposed to satisfy some
equations of motion, from them are constructed conserved quantities
satisfying conservation laws, etc. From the view\ndash point of present\ndash
day quantum field theory, this approach is only a preliminary stage for more
or less rigorous formulation of the theory in which the fields are
represented via operator\ndash valued distributions, a fact required even for
description of free fields. Moreover, in non\ndash perturbative directions,
like constructive and conformal field theories, the main objects are the
vacuum mean (expectation) values of the fields and from these are
reconstructed the Hilbert space of states and the acting on it fields.
Regardless of these facts, the Lagrangian (canonical) quantum field theory is
an inherent component of the most of the ways of presentation of quantum
field theory adopted explicitly or implicitly in books
like~\cite{Bogolyubov&Shirkov,Bjorken&Drell,Roman-QFT,Ryder-QFT,
Akhiezer&Berestetskii,Ramond-FT,Bogolyubov&et_al.-AxQFT,Bogolyubov&et_al.-QFT}.
Besides, the Lagrangian approach is a source of many ideas for other
directions of research, like the axiomatic quantum field
theory~\cite{Roman-QFT,Bogolyubov&et_al.-AxQFT,Bogolyubov&et_al.-QFT}.%
}
of free fields.
These relations are considered only from mathematical view\ndash point and
physical consequence of them, like the statistics of many\ndash particle
systems, are not investigated.

	The canonical quantization method finds its origin in the classical
Hamiltonian mechanics~\cite{Dirac-PQM,Dirac-LQM} and naturally leads to the
canonical (anti)commutation
relations~\cite{Bjorken&Drell-2,Roman-QFT,Itzykson&Zuber}. These relations can
be obtained from different assumptions (see,
e.g.,~\cite{Bogolyubov&Shirkov,
bp-QFTinMP-scalars,bp-QFTinMP-spinors,bp-QFTinMP-vectors}) and are one of the
basic corner stones of the present-day quantum field theory.

	Theoretically there are possible also non-canonical commutation
relations. The best known example of them being the so\ndash called
paracommutation
relations~\cite{Green-1953,Greenberg&Messiah-1965,Ohnuki&Kamefuchi}. But,
however, it seems no one of the presently known particles/fields obeys them.

	In the present work is shown how different classes of commutation
relations, understood in a broad sense as algebraic connections between
creation and/or annihilation operators, arise from the Lagrangian formalism,
when applied to three types of Lagrangians describing free scalar, spinor and
vector fields. Their origin is twofold. One one hand, a requirement for
uniqueness of the dynamical variables (that can be calculated from
Lagrangians leading to identical Euler\ndash Lagrange equation) entails a
number of specific commutation relations. On another hand, any one of the
so\ndash called Heisenberg
relations/equations~\cite{Bjorken&Drell-2,Roman-QFT}, implies corresponding
commutation relations; for example, the paracommutation relations arise from
the Heisenberg equations regarding the momentum operator, when `charge
symmetric' Lagrangian is employed.%
\footnote{~%
Ordinary~\cite{Bjorken&Drell-2,Roman-QFT}, the commutation relations are
postulated and the validity of the Heisenberg relations is then verified.
We follow the opposite method by postulating the Heisenberg equations and,
then, looking for commutation relations that are compatible with them.%
}
The combination of the both methods leads to strong, generally incompatible,
restrictions on the admissible types of commutation relations.

	The introduction of the concept of vacuum, combined with the
mentioned uniqueness of the operators of the dynamical variables, changes the
situation and requires a redefinition of these operators in a way similar to
the one known as the normal
ordering~\cite{Bjorken&Drell-2,Bogolyubov&Shirkov,Roman-QFT,Itzykson&Zuber},
which is its special case. Some natural assumptions reduce the former to the
letter one; in particular, in that way are excluded the paracommutation
relations. However, this does not reduce the possible commutation relations
to the canonical ones. Further, the requirement to be available an
effective procedure for calculating vacuum mean (expectation) values, to
which reduce all predictable results in the theory, puts new restriction,
whose only realistic solution at the time being seems to be the standard
canonical (anti)commutation relations.

	The layout of the work is as follows.

	Sect.~\ref{Sect2} gives an idea of the momentum picture of motion and
discusses the relations between the creation and annihilation operators in
it and in Heisenberg picture.
	In Sect.~\ref{Sect3} are reviewed some basic results
from~\cite{bp-QFTinMP-scalars,bp-QFTinMP-spinors,bp-QFTinMP-vectors}, part of
which can be found also in papers
like~\cite{Bjorken&Drell-2,Bogolyubov&Shirkov,Roman-QFT,Itzykson&Zuber}. In
particular, the explicit expression of the dynamical variables via the
creation and annihilation operators are presented (without assuming some
commutation relations or normal ordering) and it is pointed to the existence
of a family of such variables for a given system of Euler\ndash Lagrange
equations for free fields.
	The last fact is analyzed in Sect.~\ref{Sect4},
where a number of its consequences, having a sense of commutation relations,
are drawn. The Heisenberg relations and the commutation relations between the
dynamical variables are reviewed and analyzed in Sect.~\ref{Sect5}. It is
pointed that the letter should be consequences from the former ones. Arguments
are presented that the Heisenberg equation concerning the angular momentum
operator should be split into two independent ones, representing its
`orbital' and `spin' parts, respectively.

	Sect.~\ref{Sect6} contains a method for assigning commutation
relations to the Heisenberg equations. It is shown that the Heisenberg
equation involving the `orbital' part of the angular momentum gives rise to a
differential, not algebraic, commutation relation and the one concerning the
`spin' part of the angular momentum implies a complicated integro\ndash
differential connections between the creation and annihilation operators.
Special attention is paid to the paracommutation relations, whose particular
kind are the ordinary ones, which ensure the validity of the Heisenberg
equations concerning the momentum operator. Partially is analyzed the problem
for compatibility of the different types of commutation relations derived. It
is proved that some generalization of the paracommutation relations ensures the
fulfillment of all of the Heisenberg relations.

	Sect.~\ref{Sect7} is devoted to consequences from the commutation
relations derived in Sect.~\ref{Sect6} under the conditions for uniqueness of
the dynamical variables presented in Sect.~\ref{Sect4}. Generally, these
requirements are incompatible with the commutation relations. To overcome the
problem, it is proposed a redefinition of the dynamical variables via a method
similar to (and generalizing) the normal ordering. This, of course, entails
changes in the commutation relations, the new versions of which happen to be
compatible with the uniqueness conditions and ensure the validity of the
Heisenberg relations.

	The concept of the vacuum is introduced in Sect.~\ref{Sect8}. It
reduces (practically) the redefinition of the operators of the dynamical
variables to the one obtained via the normal ordering procedure in the
ordinary quantum field theory, but, without additional suppositions, does
not reduce the commutation relations to the standard bilinear ones. As a last
step in specifying the commutation relations as much as possible, we introduce
the requirement the theory to supply an effective way for calculating vacuum
mean values of (anti\ndash normally ordered) products of creation and
annihilation operators to which are reduced all predictable results, in
particular the mean values of the dynamical variables. The standard bilinear
commutation relation seems to be the only ones know at present that survive
that last condition, however their uniqueness in this respect is not
investigated.

	Sect.~\ref{Sect9} deals with the same problems as described above but
for systems containing at least two different quantum fields. The main
obstacle is the establishment of commutation relations between
creation/annihilation operators concerning different fields. Argument is
presented that they should contain commutators or anticommutators of these
operators. The major of corresponding commutation relations are explicitly
written and the results obtained turn to be similar to the ones just
described, only in `multifield' version.

	Section~\ref{Conclusion} closes the paper by summarizing its main
results.

\vspace{1.2ex}

	The books~\cite{Bogolyubov&Shirkov,Roman-QFT,Bjorken&Drell} will be
used as standard reference works on quantum field theory. Of course, this is
more or less a random selection between the great number of (text)books and
papers on the theme to which the reader is referred for more details or other
points of view. For this end, e.g.,~\cite{Itzykson&Zuber,Ryder-QFT,Schweber} or
the literature cited
in~\cite{Bogolyubov&Shirkov,Roman-QFT,Bjorken&Drell,Itzykson&Zuber,
Ryder-QFT,Schweber} may be helpful.

	Throughout this paper $\hbar$ denotes the Planck's constant (divided
by $2\pi$), $c$ is the velocity of light in vacuum, and $\iu$ stands for the
imaginary unit. The superscripts $\dag$ and $\top$ mean respectively
Hermitian conjugation and transposition (of operators or matrices), the
superscript $\ast$ denotes complex conjugation, and the symbol $\circ$
denotes compositions of mappings/operators.

	By $\delta_{fg}$, or $\delta_f^g$ or $\delta^{fg}$  ($:=1$ for $f=g$,
$:=0$ for $f=g$) is denoted the Kronecker $\delta$\ndash symbol, depending on
arguments $f$ and $g$, and $\delta^n(y)$, $y\in\field[R]^n$, stands for the
$n$\ndash dimensional Dirac $\delta$\ndash function; $\delta(y):=\delta^1(y)$
for $y\in\field[R]$.

	The Minkowski spacetime is denoted by $\base$. The Greek indices run
from 0 to $\dim\base-1=3$. All Greek indices will be raised and lowered by
means of the standard 4\ndash dimensional Lorentz metric tensor
$\eta^{\mu\nu}$ and its inverse $\eta_{\mu\nu}$ with signature
$(+\,-\,-\,-)$. The Latin indices $a,b,\dots$ run from 1 to $\dim\base-1=3$
and, usually, label the spacial components of some object. The Einstein's
summation convention over indices repeated on different levels is assumed
over the whole range of their values.

	At last, we ought to give an explanation why this work appears under
the general title ``Lagrangian quantum field theory in momentum picture''
when in it all considerations are done, in fact, in Heisenberg picture with
possible, but not necessary, usage of the creation and annihilation operators
in momentum picture. First of all, we essentially employ the obtained
in~\cite{bp-QFTinMP-scalars,bp-QFTinMP-spinors,bp-QFTinMP-vectors}
expressions for the dynamical variables in momentum picture for three types
of Lagrangians. The corresponding operators in Heisenberg picture, which in
fact is used in this paper, can be obtained via a direct calculation, as it
is partially done in, e.g.,~\cite{Bogolyubov&Shirkov} for one of the
mentioned types of Lagrangians. The important point here is that in
Heisenberg picture it suffice to be used only the standard Lagrangian
formalism, while in momentum picture one has to suppose the commutativity
between the components of the momentum operator and the validity of the
Heisenberg relations for it (see below equations~\eref{2.1} and~\eref{2.28}).
Since for the analysis of the commutation relations we intend to do the
fulfillment of these relations is not necessary (they are subsidiary
restrictions on the Lagrangian formalism), the Heisenberg picture of motion
is the natural one that has to be used. For this reason, the expression for
the dynamical variables obtained
in~\cite{bp-QFTinMP-scalars,bp-QFTinMP-spinors,bp-QFTinMP-vectors} will be
used simply as their Heisenberg counterparts, but expressed via the creation
and annihilation operators in momentum picture. The only real advantage one
gets in this way is the more natural structure of the orbital angular
momentum operator. As the commutation relations considered below are
algebraic ones, it is inessential in what picture of motion they are written
or investigated.


\section{The momentum picture}
	\label{Sect2}

	Since the momentum picture of motion will be used only partially in
this work, below is presented only its definition and the connection between
the creation/annihilation operators in it and in Heisenberg picture. Details
concerning the momentum picture can be found
in~\cite{bp-QFT-pictures,bp-QFT-MP} and in the corresponding
sections devoted to it
in~\cite{bp-QFTinMP-scalars,bp-QFTinMP-spinors,bp-QFTinMP-vectors}.

	Let us consider a system of quantum fields, represented in Heisenberg
picture of motion by field operators $\tope{\varphi}_i(x)\colon\Hil\to\Hil$,
$i=1,\dots,n\in\field[N]$, acting on the system's Hilbert space $\Hil$ of
states and depending on a point $x$ in Minkowski spacetime $\base$. Here and
henceforth, all quantities in Heisenberg picture will be marked by a tilde
(wave) ``$\tope{\mspace{6mu}}\mspace{3mu}$'' over their kernel symbols. Let
$\tope{P}_\mu$ denotes the system's (canonical) momentum vectorial operator,
defined via the energy\ndash momentum tensorial operator $\tope{T}^{\mu\nu}$
of the system, viz.
	\begin{equation}
			\label{2.0}
\tope{P}_\mu
:=
\frac{1}{c}\int\limits_{x^0=\const} \tope{T}_{0\mu}(x) \Id^3\bs x .
	\end{equation}
Since this operator is Hermitian, $\tope{P}_\mu^\dag=\tope{P}_\mu$, the
operator
	\begin{equation}	\label{12.112}
\ope{U}(x,x_0)
 =
\exp\Bigl( \iih \sum_\mu (x^\mu-x_0^\mu)\tope{P}_{\mu}  \Bigr) ,
	\end{equation}
where $x_0\in\base$ is arbitrarily fixed and $x\in\base$,%
\footnote{~%
The notation $x_0$, for a fixed point in $\base$, should not be confused with
the zeroth covariant coordinate $\eta_{0\mu}x^\mu$ of $x$ which, following
the convention $x_\nu:=\eta_{\nu\mu}x^\mu$, is denoted by the same symbol
$x_0$. From the context, it will always be clear whether $x_0$ refers to a
point in $\base$ or to the zeroth covariant coordinate of a point
$x\in\base$.%
}
is unitary, \ie
\(
\ope{U}^\dag(x_0,x)
:= (\ope{U}(x,x_0))^\dag
 = \ope{U}^{-1}(x,x_0)
:= (\ope{U}(x,x_0))^{-1}
\)
and, via the formulae
	\begin{align}	\label{12.113}
\tope{X}\mapsto \ope{X}(x)
	&= \ope{U}(x,x_0) (\tope{X})
\\			\label{12.114}
\tope{A}(x)\mapsto \ope{A}(x)
	&= \ope{U}(x,x_0)\circ (\tope{A}(x)) \circ \ope{U}^{-1}(x,x_0) ,
	\end{align}
realizes the transition to the \emph{momentum picture}. Here $\tope{X}$ is a
state vector in system's Hilbert space of states $\Hil$ and
$\tope{A}(x)\colon\Hil\to\Hil$ is (observable or not) operator\ndash valued
function of $x\in\base$ which, in particular, can be polynomial or convergent
power series in the field operators $\tope{\varphi}_i(x)$; respectively
$\ope{X}(x)$ and $\ope{A}(x)$ are the corresponding quantities in momentum
picture.
	In particular, the field operators transform as
	\begin{align}	\label{12.115}
\tope{\varphi}_i(x)\mapsto \ope{\varphi}_i(x)
     = \ope{U}(x,x_0)\circ \tope{\varphi}_i(x) \circ \ope{U}^{-1}(x,x_0) .
	\end{align}
	Notice, in~\eref{12.112} the multiplier $(x^\mu-x_0^\mu)$ is regarded
as a real parameter (in which $\tope{P}_\mu$ is linear). Generally,
$\ope{X}(x)$ and $\ope{A}(x)$ depend also on the point $x_0$ and, to be
quite correct, one should write $\ope{X}(x,x_0)$ and $\ope{A}(x,x_0)$ for
$\ope{X}(x)$ and $\ope{A}(x)$, respectively. However, in the most situations
in the present work, this dependence is not essential or, in fact, is not
presented at all. For that reason, we shall \emph{not} indicate it explicitly.

	The momentum picture is most suitable in quantum field theories in
which the components $\tope{P}_\mu$ of the momentum operator commute between
themselves and satisfy the Heisenberg relations/equations with the field
operators, \ie when $\tope{P}_\mu$ and $\tope{\varphi}_i(x)$ satisfy the
relations:
	\begin{align}	\label{2.1}
& [\tope{P}_\mu, \tope{P}_\nu ]_{\_} = 0
\\			\label{2.28}
& [\tope{\varphi}_i(x), \tope{P}_\mu]_{\_} = \ih\pd_\mu \tope{\varphi}_i(x).
	\end{align}
Here $[A,B]_{\pm}:=A\circ B \pm B\circ A$, $\circ$ being the composition of
mappings sign, is the commutator/anticommutator of operators (or matrices)
$A$ and $B$.

	However, the fulfillment of the relations~\eref{2.1} and~\eref{2.28}
will not be supposed in this paper until Sect.~\ref{Sect6} (see also
Sect.~\ref{Sect5}).

	Let $a_s^{\pm}(\bk)$ and $a_s^{\dag\,\pm}(\bk)$ be the
creation/annihilation operators of some free particular field (see
Sect.~\ref{Sect3} below for a detailed explanation of the notation). We have
the connections
	\begin{gather}	\label{2.28-1}
	\begin{split}
\ta_s^\pm(\bk)
& = \e^{ \pm\frac{1}{\ih} x^\mu k_\mu }
\ope{U}^{-1}(x,x_0)\circ a_s^\pm(\bk) \circ \ope{U}(x,x_0)
\\
\ta_s^{\dag\,\pm}(\bk)
& = \e^{ \pm\frac{1}{\ih} x^\mu k_\mu }
\ope{U}^{-1}(x,x_0)\circ a_s^{\dag\,\pm}(\bk) \circ \ope{U}(x,x_0)
	\end{split}
\Bigg\} \quad  k_0=\sqrt{m^2c^2+\bk^2}
\\\intertext{whose explicit form is}	\label{2.28-2}
	\begin{split}
\ta_s^\pm(\bk)
& = \e^{ \pm\frac{1}{\ih} x_0^\mu k_\mu } a_s^\pm(\bk)
\\
\ta_s^{\dag\,\pm}(\bk)
& = e^{ \pm\frac{1}{\ih} x_0^\mu k_\mu } a_s^{\dag\,\pm}(\bk)
	\end{split}
\Bigg\} \quad  k_0=\sqrt{m^2c^2+\bk^2} .
	\end{gather}

	Further it will be assumed
$\ta_s^{\pm}(\bk)$ and $\ta_s^{\dag\,\pm}(\bk)$
to be defined in Heisenberg picture, independently of
$a_s^{\pm}(\bk)$ and $a_s^{\dag\,\pm}(\bk)$,
by means of the standard Lagrangian formalism. What concerns the operators
$a_s^{\pm}(\bk)$ and $a_s^{\dag\,\pm}(\bk)$,
we shall regard them as \emph{defined} via~\eref{2.28-2}; this makes them
independent from the momentum picture of motion. The fact that the so\ndash
defined operators
$a_s^{\pm}(\bk)$ and $a_s^{\dag\,\pm}(\bk)$
coincide with the creation/annihilation operators in momentum picture (under
the conditions~\eref{2.1} and~\eref{2.28}) will be inessential in the almost
whole text.


\section
[Lagrangians, Euler-Lagrange equations and dynamical variables]
{Lagrangians, Euler-Lagrange equations\\ and dynamical variables}
\label{Sect3}

	In~\cite{bp-QFTinMP-scalars,bp-QFTinMP-spinors,bp-QFTinMP-vectors} we
have investigated the Lagrangian quantum field theory of respectively scalar,
spin~$\frac{1}{2}$ and vector free fields. The main Lagrangians from which it
was derived are respectively (see \emph{loc.\ cit.}\ or,
e.g.~\cite{Bjorken&Drell-2,Bogolyubov&Shirkov,Roman-QFT,Itzykson&Zuber}):
	\begin{subequations}	\label{3.1}
	\begin{align}
			\label{3.1a}
  \tope{L}'_{\mathrm{sc}}
= \tope{L}'_{\mathrm{sc}} (\tope{\varphi},\tope{\varphi}^\dag)
= &
- \frac{1}{1+\tau(\tope{\varphi})} m^2c^4
	\tope{\varphi}(x)\circ \tope{\varphi}^\dag(x)
+ \frac{1}{1+\tau(\tope{\varphi})} c^2\hbar^2
	(\pd_\mu\tope{\varphi}(x)) \circ (\pd^\mu\tope{\varphi}^\dag(x))
\\			\label{3.1b}
	\begin{split}
  \tope{L}'_{\mathrm{sp}}
= \tope{L}'_{\mathrm{sp}} (\tope{\psi},\tope{\bpsi})
= &
- \frac{1}{2}\ih c\{
  \tope{\bpsi}^\top(x) C^{-1}\gamma^\mu\circ(\pd_\mu\tope{\psi}(x))
\\  &
- (\pd_\mu\tope{\bpsi}^\top(x)) C^{-1}\gamma^\mu\circ \tope{\psi}(x)
\}
+ mc^2 \tope{\bpsi}^\top(x) C^{-1}\circ \tope{\psi}(x)
	\end{split}
\\			\label{3.1c}
	\begin{split}
  \tope{L}'_{\mathrm{v}}
= \tope{L}'_{\mathrm{v}} (\tope{U},\tope{U}^\dag)
= &
\frac{m^2c^4}{1+\tau(\tU)} \tU_\mu^\dag \circ \tU^\mu
\\ &
+
\frac{c^2\hbar^2}{1+\tau(\tU)}
\bigl\{ - (\pd_\mu\tU_\nu^\dag) \circ (\pd^\mu\tU^\nu)
        + (\pd_\mu\tU^{\mu\dag}) \circ (\pd_\nu\tU^\nu) \bigr\}
	\end{split}
	\end{align}
	\end{subequations}
Here it is used the following notation: $\tope{\varphi}(x)$ is a scalar field,
a tilde (wave) over a symbol means that it is in Heisenberg picture,
the dagger $\dag$ denotes Hermitian conjugation,
$\tope{\psi}:=(\tope{\psi}_0,\tope{\psi}_1,\tope{\psi}_2,\tope{\psi}_3)$ is a
4\ndash spinor field,
$\tope{\bpsi} := C\tope{\opsi}^\top := C(\tope{\psi}^\dag \gamma^0)$
is its charge conjugate with $\gamma^\mu$ being the Dirac gamma matrices and
the matrix $C$ satisfies the equations
$C^{-1}\gamma^\mu C = - \gamma^\mu$ and $C^\top=-C$,  $U_\mu$ is a vector
field, $m$ is the field's mass (parameter) and the function
	\begin{equation}	\label{3.2}
\tau(A) :=
	\begin{cases}
1  &\text{for $A^\dag=A$ (Hermitian operator)}
\\
0  &\text{for $A^\dag\not=A$ (non-Hermitian operator)}
	\end{cases}
\ ,
	\end{equation}
with $A\colon\Hil\to\Hil$ being an operator on the systems Hilbert space
$\Hil$ of states, takes care of is the field charged (non\ndash Hermitian) or
neutral (Hermitian, uncharged). Since a spinor field is a charged one, we
have $\tau(\tope{\psi})=0$; sometimes below the number $0=\tau(\tope{\psi})$
will be written explicitly for unification of the notation.

	We have explored also the consequences from the `charge conjugate'
Lagrangians
	\begin{subequations}	\label{3.3}
	\begin{align}
			\label{3.3a}
  \tope{L}^{\prime\prime}_{\mathrm{sc}}
&= \tope{L}^{\prime\prime}_{\mathrm{sc}} (\tope{\varphi},\tope{\varphi}^\dag)
:= \tope{L}^{\prime}_{\mathrm{sc}} (\tope{\varphi}^\dag,\tope{\varphi})
\\			\label{3.3b}
  \tope{L}^{\prime\prime}_{\mathrm{sp}}
&= \tope{L}^{\prime\prime}_{\mathrm{sp}} (\tope{\psi},\tope{\bpsi})
:= \tope{L}^{\prime}_{\mathrm{sp}} (\tope{\bpsi},\tope{\psi})
\\			\label{3.3c}
  \tope{L}^{\prime\prime}_{\mathrm{v}}
&= \tope{L}^{\prime\prime}_{\mathrm{v}} (\tope{U},\tope{U}^\dag)
:= \tope{L}^{\prime}_{\mathrm{v}} (\tope{U}^\dag,\tope{U}) ,
	\end{align}
	\end{subequations}
as well as from the `charge symmetric' Lagrangians
	\begin{subequations}	\label{3.4}
	\begin{align}
			\label{3.4a}
  \tope{L}^{\prime\prime\prime}_{\mathrm{sc}}
&= \tope{L}^{\prime\prime\prime}_{\mathrm{sc}}
					(\tope{\varphi},\tope{\varphi}^\dag)
:=
\frac{1}{2}\bigl(
\tope{L}^{\prime}_{\mathrm{sc}} + \tope{L}^{\prime\prime}_{\mathrm{sc}}
\bigr)
=
\frac{1}{2}\bigl\{
\tope{L}'_{\mathrm{sc}} (\tope{\varphi},\tope{\varphi}^\dag) +
\tope{L}^{\prime}_{\mathrm{sc}} (\tope{\varphi}^\dag,\tope{\varphi})
\bigr\}
\\			\label{3.4b}
  \tope{L}^{\prime\prime\prime}_{\mathrm{sp}}
&= \tope{L}^{\prime\prime\prime}_{\mathrm{sp}} (\tope{\psi},\tope{\bpsi})
:=
\frac{1}{2}\bigl(
\tope{L}^{\prime}_{\mathrm{sp}} + \tope{L}^{\prime\prime}_{\mathrm{sp}}
\bigr)
=
\frac{1}{2}\bigl\{
\tope{L}'_{\mathrm{sp}} (\tope{\psi},\tope{\bpsi}) +
\tope{L}^{\prime}_{\mathrm{sp}} (\tope{\bpsi},\tope{\psi})
\bigr\}
\\			\label{3.4c}
  \tope{L}^{\prime\prime\prime}_{\mathrm{v}}
&= \tope{L}^{\prime\prime\prime}_{\mathrm{v}} (\tope{U},\tope{U}^\dag)
:=
\frac{1}{2}\bigl(
\tope{L}^{\prime}_{\mathrm{v}} + \tope{L}^{\prime\prime}_{\mathrm{v}}
\bigr)
=
\frac{1}{2}\bigl\{
\tope{L}'_{\mathrm{v}} (\tope{U},\tope{U}^\dag) +
\tope{L}^{\prime}_{\mathrm{v}} (\tope{U}^\dag,\tope{U})
\bigr\} .
	\end{align}
	\end{subequations}

	It is essential to be noted, for a massless, $m=0$, vector field to
the Lagrangian formalism are added as subsidiary conditions the Lorenz
conditions
	\begin{equation}	\label{3.5}
\pd^\mu \tU_\mu = 0 \quad \pd^\mu \tU_\mu^\dag = 0
	\end{equation}
on the solutions of the corresponding Euler-Lagrange equations. Besides, if
the opposite is not stated explicitly, no other restrictions, like the
(anti)commutation relations, are supposed to be imposed on the above
Lagrangians. And a technical remark, for convenience, the fields
$\tope{\varphi}$, $\tope{\psi}$ and $\tU$ and their charge conjugate
$\tope{\varphi}^\dag$, $\tope{\bpsi}$ and $\tU^\dag$, respectively, are
considered as independent field variables.

	Let $\tope{L}^{\prime}$ denotes any one of the Lagrangians~\eref{3.1}
and $\tope{L}^{\prime\prime}$ (resp.\ $\tope{L}^{\prime\prime\prime}$) the
corresponding to it Lagrangian given via~\eref{3.3} (resp.~\eref{3.4}).
Physically the difference between
$\tope{L}^{\prime}$ and $\tope{L}^{\prime\prime}$ is that the particles for
$\tope{L}^{\prime}$ are antiparticles for $\tope{L}^{\prime\prime}$ and
\emph{vice versa}. Both of the Lagrangians
$\tope{L}^{\prime}$ and $\tope{L}^{\prime\prime}$ are \emph{not} charge
symmetric, \ie the arising from them theories are not invariant under the
change particle$\leftrightarrow$antiparticle (or, in mathematical terms,
under some of the changes
$\tope{\varphi} \leftrightarrow \tope{\varphi}^\dag$,
$\tope{\psi} \leftrightarrow \tope{\bpsi}$,
$\tU \leftrightarrow \tU^\dag$)
unless some additional hypotheses are made. Contrary to this, the Lagrangian
$\tope{L}^{\prime\prime\prime}$ \emph{is} charge symmetric and, consequently,
the formalism on its base is invariant under the change
particle$\leftrightarrow$antiparticle.%
\footnote{~%
Besides, under the same assumptions, the Lagrangian
$\tope{L}^{\prime\prime\prime}$ does not admit quantization via
anticommutators (commutators) for integer (half\ndash integer) spin
field, while
$\tope{L}^{\prime}$ and $\tope{L}^{\prime\prime}$
do not make difference between integer and half\ndash integer spin fields. %
}

	The Euler-Lagrange equations for the Lagrangians
$\tope{L}^{\prime}$, $\tope{L}^{\prime\prime}$ and
$\tope{L}^{\prime\prime\prime}$
happen to
coincide~\cite{bp-QFTinMP-scalars,bp-QFTinMP-spinors,bp-QFTinMP-vectors}:%
\footnote{~%
Rigorously speaking, the Euler-Lagrange equations for the
Lagrangian~\eref{3.4b} are identities like $0=0$ ---
see~\cite{bp-QFT-action-principle}. However, bellow we shall handle this
exceptional case as pointed in~\cite{bp-QFTinMP-spinors}.%
}
	\begin{equation}	\label{3.6}
\frac{\pd \tope{L}^{\prime}}{\pd\chi}
- \frac{\pd}{\pd x^\mu}
\Bigl( \frac{\pd\tope{L}^{\prime}}{\pd(\pd_\mu\chi)} \Bigr)
\equiv
\frac{\pd \tope{L}^{\prime\prime}}{\pd\chi}
- \frac{\pd}{\pd x^\mu}
\Bigl( \frac{\pd\tope{L}^{\prime\prime}}{\pd(\pd_\mu\chi)} \Bigr)
\equiv
\frac{\pd \tope{L}^{\prime\prime\prime}}{\pd\chi}
- \frac{\pd}{\pd x^\mu}
  \Bigl( \frac{\pd\tope{L}^{\prime\prime\prime}}{\pd(\pd_\mu\chi)} \Bigr)
=
0 ,
	\end{equation}
where
\(
\chi =
\tope{\varphi}, \tope{\varphi}^\dag,
\tope{\psi}, \tope{\bpsi},
\tU, \tU^\dag
\)
for respectively scalar, spinor and vector field.

	Since the creation and annihilation operators are defined only on the
base of Euler\ndash Lagrange
equations~\cite{Bjorken&Drell-2,Bogolyubov&Shirkov,Roman-QFT,Itzykson&Zuber,
bp-QFTinMP-scalars,bp-QFTinMP-spinors,bp-QFTinMP-vectors}, we can assert that
these operators are identical for the Lagrangians
$\tope{L}^{\prime}$, $\tope{L}^{\prime\prime}$ and
$\tope{L}^{\prime\prime\prime}$.
We shall denote these operators by
 $a_s^{\pm}(\bk)$ and $a_s^{\dag\,\pm}(\bk)$
with the convention that $a_s^{+}(\bk)$ (resp.\ $a_s^{\dag\,+}(\bk)$) creates
a particle (resp.\ antiparticle) with 4\ndash momentum
$(\sqrt{m^2c^2+\bk^2},\bk)$, polarization $s$ (see below) and charge $(-q)$
(resp.\ $(+q)$)%
\footnote{~%
For a neutral field, we put $q=0$.%
}
and $a_s^{\dag\,-}(\bk)$ (resp.\ $a_s^{-}(\bk)$) annihilates/destroys such
a particle (resp.\ antiparticle).
	Here and henceforth $\bk\in\field[R]^3$ is interpreted as
(anti)particle's 3\ndash momentum and the values of the polarization index
$s$ depend on the field considered: $s=1$ for a scalar field, $s=1$ or
$s=1,2$ for respectively massless ($m=0$) or massive ($m\not=0$) spinor
field, and $s=1,2,3$ for a vector field.%
\footnote{~%
For convenience, in~\cite{bp-QFTinMP-spinors}, we have set $s=0$ if $m=0$ and
$s=1,2$ if $m\not=0$ for a spinor field. For a massless vector field, one may
set $s=1,2$, thus eliminating the `unphysical' value $s=3$ for $m=0$ ---
see~\cite{Bjorken&Drell-2,Bogolyubov&Shirkov,bp-QFTinMP-vectors}.
In~\cite{bp-QFTinMP-scalars}, for a scalar field, the notation
 $\varphi_0^{\pm}(\bk)$ and $\varphi_0^{\dag\,\pm}(\bk)$
is used for
 $a_1^{\pm}(\bk)$ and $a_1^{\dag\,\pm}(\bk)$, respectively.%
}
Since massless vector field's modes with $s=3$ may enter only in the spin and
orbital angular momenta operators~\cite{bp-QFTinMP-vectors}, we, for
convenience, shall assume that the polarization indices $s,t,\dots$ take the
values from~1 to $2j+1-\delta_{0m}(1-\delta_{0j})$, where $j=0,\frac{1}{2},1$
is the spin for scalar, spinor and vector field, respectively, and
$\delta_{0m}:=1$ for $m=0$ and $\delta_{0m}:=0$ for $m\not=0$;%
\footnote{~%
In this way the case $(j,s,m)=(1,3,0)$ is excluded from further
considerations; if  $(j,m)=(1,0)$ and $q=0$, the case considered further in
this work corresponds to an electromagnetic field in Coulomb gauge, as the
modes with $s=3$ are excluded~\cite{bp-QFTinMP-vectors}. However, if the
case $(j,s,m)=(1,3,0)$ is important for some reasons, the reader can easily
obtain the corresponding results by applying the ones
from~\cite{bp-QFTinMP-vectors}.%
}
if the value $s=3$ is important when $j=1$ and $m=0$, it will be
commented/considered separately. Of course, the creation and annihilation
operators are different for different fields; one should write, e.g.,
${\lindex[a]{j}{}}_{s}^{\pm}(\bk)$ for $a_{s}^{\pm}(\bk)$, but we shall not
use such a complicated notation and will assume the dependence on $j$ to be
an implicit one.

	The following settings will be frequently used throughout this chapter:
	\begin{equation}
			\label{3.12}
	\begin{split}
& j :=
	\begin{cases}
0  	     &\text{for scalar field} \\
\frac{1}{2}  &\text{for spinor field} \\
1	     &\text{for vector field}
	\end{cases}
\qquad
\tau :=
	\begin{cases}
1  &\text{for $q=0$ (neutral (Hermitian) field)} \\
0  &\text{for $q\not=0$ (charged (non-Hermitian) field)}
	\end{cases}
\\
& \varepsilon
:= (-1)^{2j}
=
	\begin{cases}
+ 1  &\text{for integer $j$ (bose fields)} \\
- 1  &\text{for half-integer $j$ (fermi fields)}
	\end{cases}
	\end{split}
	\end{equation} 
\vspace{-3ex}
	\begin{equation}	\label{3.12-1}
[A,B]_\pm := [A,B]_{\pm1} := A\circ B \pm B\circ A ,
	\end{equation}
where $A$ and $B$ are operators on the system's Hilbert space $\Hil$ of
states.

	The dynamical variables corresponding to
$\tope{L}^{\prime}$, $\tope{L}^{\prime\prime}$ and
$\tope{L}^{\prime\prime\prime}$
are, however, completely different, unless some additional conditions are
imposed on the Lagrangian
formalism~\cite{bp-QFTinMP-scalars,bp-QFTinMP-spinors,bp-QFTinMP-vectors}. In
particular, the momentum operators $\tope{P}_{\mu}^\omega$, charge operators
$\tope{Q}^\omega$, spin operators $\tope{S}_{\mu\nu}^\omega$ and
orbital operators $\tope{L}_{\mu\nu}^\omega$, where
$\omega=\prime,\prime\prime,\prime\prime\prime$, for these Lagrangians
are~\cite{bp-QFTinMP-scalars,bp-QFTinMP-spinors,bp-QFTinMP-vectors}:
	\begin{subequations}	\label{3.7}
	\begin{align}	\label{3.7a}
\tope{P}_\mu^{\prime}
& =
\frac{1}{1+\tau}
\sum_{s=1}^{2j+1-\delta_{0m}(1-\delta_{0j})} \int \Id^3\bk
k_\mu |_{ k_0=\sqrt{m^2c^2+{\bs k}^2} }
\{
a_s^{\dag\,+}(\bk)\circ a_s^-(\bk) + \varepsilon
a_s^{\dag\,-}(\bk)\circ a_s^+(\bk)
\}
\displaybreak[1]\\	\label{3.7b}
\tope{P}_\mu^{\prime\prime}
& =
\frac{1}{1+\tau}
\sum_{s=1}^{2j+1-\delta_{0m}(1-\delta_{0j})} \int \Id^3\bk
k_\mu |_{ k_0=\sqrt{m^2c^2+{\bs k}^2} }
\{
a_s^{+}(\bk)\circ a_s^{\dag\,-}(\bk) + \varepsilon
a_s^{-}(\bk)\circ a_s^{\dag\,+}(\bk)
\}
\displaybreak[1]\\	\label{3.7c}
\tope{P}_\mu^{\prime\prime\prime}
& =
\frac{1}{2(1+\tau)}
\sum_{s=1}^{2j+1-\delta_{0m}(1-\delta_{0j})} \int \Id^3\bk
k_\mu |_{ k_0=\sqrt{m^2c^2+{\bs k}^2} }
\{
[ a_s^{\dag\,+}(\bk) , a_s^-(\bk)]_\varepsilon +
[a_s^+(\bk) , a_s^{\dag\,-}(\bk) ]_\varepsilon
\}
	\end{align}
	\end{subequations}
\vspace{-4ex} 
	\begin{subequations}	\label{3.8}
	\begin{align}	\label{3.8a}
\tope{Q}^{\prime}
& =
+ q
\sum_{s=1}^{2j+1-\delta_{0m}(1-\delta_{0j})} \int \Id^3\bk
\{
a_s^{\dag\,+}(\bk)\circ a_s^-(\bk) - \varepsilon
a_s^{\dag\,-}(\bk)\circ a_s^+(\bk)
\}
\displaybreak[1]\\	\label{3.8b}
\tope{Q}^{\prime\prime}
& =
- q
\sum_{s=1}^{2j+1-\delta_{0m}(1-\delta_{0j})} \int \Id^3\bk
\{
a_s^{+}(\bk)\circ a_s^{\dag\,-}(\bk) - \varepsilon
a_s^{-}(\bk)\circ a_s^{\dag\,+}(\bk)
\}
\displaybreak[1]\\	\label{3.8c}
\tope{Q}^{\prime\prime\prime}
& =
\frac{1}{2} q
\sum_{s=1}^{2j+1-\delta_{0m}(1-\delta_{0j})} \int \Id^3\bk
\{
[ a_s^{\dag\,+}(\bk) , a_s^-(\bk)]_\varepsilon -
[a_s^+(\bk) , a_s^{\dag\,-}(\bk) ]_\varepsilon
\}
	\end{align}
	\end{subequations}
\vspace{-4ex} 
	\begin{subequations}	\label{3.9}
	\begin{align}	\label{3.9a}
	\begin{split}
\tope{S}_{\mu\nu}^{\prime}
& =
\frac{(-1)^{j-1/2} j \hbar }{1+\tau}
\sum_{s,s'=1}^{2j+1-\delta_{0m}(1-\delta_{1j})} \int \Id^3\bk
\bigl\{
\sigma_{\mu\nu}^{s s',-}(\bk) a_{s}^{\dag\,+}(\bk)\circ a_{s'}^{-}(\bk)
\\ &
\hphantom{=
\frac{(-1)^{j-1/2} j \hbar }{1+\tau}
\sum_{s,s'=1}^{2j+1-\delta_{0m}(1-\delta_{1j})} \int \Id^3\bk
}
+
\sigma_{\mu\nu}^{s s',+}(\bk) a_{s}^{\dag\,-}(\bk)\circ a_{s'}^{+}(\bk)
\bigr\}
	\end{split}
\displaybreak[1]\\	\label{3.9b}
	\begin{split}
\tope{S}_{\mu\nu}^{\prime\prime}
& = \varepsilon
\frac{(-1)^{j-1/2} j \hbar }{1+\tau}
\sum_{s,s'=1}^{2j+1-\delta_{0m}(1-\delta_{1j})} \int \Id^3\bk
\bigl\{
\sigma_{\mu\nu}^{s s',+}(\bk) a_{s'}^{+}(\bk)\circ a_{s}^{\dag\,-}(\bk)
\\ &
\hphantom{=
\frac{(-1)^{j-1/2} j \hbar }{1+\tau}
\sum_{s,s'=1}^{2j+1-\delta_{0m}(1-\delta_{1j})} \int \Id^3\bk
}
+
\sigma_{\mu\nu}^{s s',-}(\bk) a_{s'}^{-}(\bk)\circ a_{s}^{\dag\,+}(\bk)
\bigr\}
	\end{split}
\displaybreak[1]\\	\label{3.9c}
	\begin{split}
\tope{S}_{\mu\nu}^{\prime\prime\prime}
& =
\frac{(-1)^{j-1/2} j \hbar }{2(1+\tau)}
\sum_{s,s'=1}^{2j+1-\delta_{0m}(1-\delta_{1j})} \int \Id^3\bk
\bigl\{
\sigma_{\mu\nu}^{s s',-}(\bk)
	[ a_{s}^{\dag\,+}(\bk) , a_{s'}^{-}(\bk)]_\varepsilon
\\ &
\hphantom{=
\frac{(-1)^{j-1/2} j \hbar }{1+\tau}
\sum_{s,s'=1}^{2j+1-\delta_{0m}(1-\delta_{1j})} \int \Id^3\bk
}
+
\sigma_{\mu\nu}^{s s',+}(\bk)
	[ a_{s}^{\dag\,-}(\bk) , a_{s'}^{+}(\bk) ]_\varepsilon
\bigr\}
	\end{split}
	\end{align}
	\end{subequations}
\vspace{-4ex} 
	\begin{subequations}	\label{3.10}
	\begin{align}	\label{3.10a}
	\begin{split}
\tope{L}_{\mu\nu}^{\prime}
= &
x_{0\,\mu} \tope{P}_\nu^{\prime} -
x_{0\,\nu} \tope{P}_\mu^{\prime}
\\
&
+
\frac{(-1)^{j-1/2} j \hbar }{1+\tau}
\sum_{s,s'=1}^{2j+1-\delta_{0m}(1-\delta_{1j})} \int \Id^3\bk
\bigl\{
l_{\mu\nu}^{s s',-}(\bk) a_s^{\dag\,+}(\bk) \circ a_{s'}^-(\bk)
\\ &
\hphantom{+
\frac{(-1)^{j-1/2} j \hbar }{1+\tau}
\sum_{s,s'=1}^{2j+1-\delta_{0m}(1-\delta_{1j})} \int \Id^3\bk
}
+
l_{\mu\nu}^{s s',+}(\bk) a_s^{\dag\,-}(\bk) \circ a_{s'}^+(\bk)
\bigr\}
\\
&
+
\frac{\ih}{2(1+\tau)}
\sum_{s=1}^{2j+1-\delta_{0m}(1-\delta_{0j})} \int \Id^3\bk
\Bigl\{
a_s^{\dag\,+}(\bk)
\Bigl( \xlrarrow{ k_\mu \frac{\pd}{\pd k^\nu} }
     - \xlrarrow{ k_\nu \frac{\pd}{\pd k^\mu} } \Bigr)
\circ a_s^-(\bk)
\\ &
- \varepsilon
a_s^{\dag\,-}(\bk)
\Bigl( \xlrarrow{ k_\mu \frac{\pd}{\pd k^\nu} }
     - \xlrarrow{ k_\nu \frac{\pd}{\pd k^\mu} } \Bigr)
\circ a_s^+(\bk)
\Bigr\} \Big|_{ k_0=\sqrt{m^2c^2+{\bs k}^2} }
	\end{split}
\\ \displaybreak[1] 	\label{3.10b} 
	\begin{split}
\tope{L}_{\mu\nu}^{\prime\prime}
= &
x_{0\,\mu} \tope{P}_\nu^{\prime\prime} -
x_{0\,\nu} \tope{P}_\mu^{\prime\prime}
\\
&
+ \varepsilon
\frac{(-1)^{j-1/2} j \hbar }{1+\tau}
\sum_{s,s'=1}^{2j+1-\delta_{0m}(1-\delta_{1j})} \int \Id^3\bk
\bigl\{
l_{\mu\nu}^{s s',+}(\bk) a_{s'}^{+}(\bk) \circ a_{s}^{\dag\,-}(\bk)
\\ &
\hphantom{+
\frac{(-1)^{j-1/2} j \hbar }{1+\tau}
\sum_{s,s'=1}^{2j+1-\delta_{0m}(1-\delta_{1j})} \int \Id^3\bk
}
+
l_{\mu\nu}^{s s',-}(\bk) a_{s'}^{-}(\bk) \circ a_{s}^{\dag\,+}(\bk)
\bigr\}
\\
&
+
\frac{\ih}{2(1+\tau)}
\sum_{s=1}^{2j+1-\delta_{0m}(1-\delta_{0j})} \int \Id^3\bk
\Bigl\{
a_s^{+}(\bk)
\Bigl( \xlrarrow{ k_\mu \frac{\pd}{\pd k^\nu} }
     - \xlrarrow{ k_\nu \frac{\pd}{\pd k^\mu} } \Bigr)
\circ a_s^{\dag\,-}(\bk)
\\ &
- \varepsilon
a_s^{-}(\bk)
\Bigl( \xlrarrow{ k_\mu \frac{\pd}{\pd k^\nu} }
     - \xlrarrow{ k_\nu \frac{\pd}{\pd k^\mu} } \Bigr)
\circ a_s^{\dag\,+}(\bk)
\Bigr\} \Big|_{ k_0=\sqrt{m^2c^2+{\bs k}^2} }
	\end{split}
\\ \displaybreak[1]	\label{3.10c} 
	\begin{split}
\tope{L}_{\mu\nu}^{\prime\prime\prime}
= &
x_{0\,\mu} \tope{P}_\nu^{\prime\prime\prime} -
x_{0\,\nu} \tope{P}_\mu^{\prime\prime\prime}
\\
&
+ \frac{(-1)^{j-1/2} j \hbar }{2(1+\tau)}
\sum_{s,s'=1}^{2j+1-\delta_{0m}(1-\delta_{1j})} \int \Id^3\bk
\bigl\{
l_{\mu\nu}^{s s',-}(\bk)
	[ a_{s}^{\dag\,+}(\bk) , a_{s'}^{-}(\bk)]_\varepsilon
\\ &
\hphantom{=
\frac{(-1)^{j-1/2} j \hbar }{1+\tau}
\sum_{s,s'=1}^{2j+1-\delta_{0m}(1-\delta_{1j})} \int \Id^3\bk
}
+
l_{\mu\nu}^{s s',+}(\bk)
	[ a_{s}^{\dag\,-}(\bk) , a_{s'}^{+}(\bk) ]_\varepsilon
\bigr\}
\\
&
+
\frac{\ih}{4(1+\tau)}
\sum_{s=1}^{2j+1-\delta_{0m}(1-\delta_{0j})} \int \Id^3\bk
\Bigl\{
a_s^{\dag\,+}(\bk)
\Bigl( \xlrarrow{ k_\mu \frac{\pd}{\pd k^\nu} }
     - \xlrarrow{ k_\nu \frac{\pd}{\pd k^\mu} } \Bigr)
\circ a_s^-(\bk)
\\ & - \varepsilon
a_s^{-}(\bk)
\Bigl( \xlrarrow{ k_\mu \frac{\pd}{\pd k^\nu} }
     - \xlrarrow{ k_\nu \frac{\pd}{\pd k^\mu} } \Bigr)
\circ a_s^{\dag\,+}(\bk)
+
a_s^{+}(\bk)
\Bigl( \xlrarrow{ k_\mu \frac{\pd}{\pd k^\nu} }
     - \xlrarrow{ k_\nu \frac{\pd}{\pd k^\mu} } \Bigr)
\circ a_s^{\dag\,-}(\bk)
\\ &
- \varepsilon
a_s^{\dag\,-}(\bk)
\Bigl( \xlrarrow{ k_\mu \frac{\pd}{\pd k^\nu} }
     - \xlrarrow{ k_\nu \frac{\pd}{\pd k^\mu} } \Bigr)
\circ a_s^+(\bk)
\Bigr\} \Big|_{ k_0=\sqrt{m^2c^2+{\bs k}^2} } \ .
	\end{split}
	\end{align}
	\end{subequations}
	Here we have used the following notation: $(-1)^{n+1/2}:=(-1)^{n}\iu$
for all $n\in\field[N]$ and $\iu:=+\sqrt{-1}$,
	\begin{multline}	\label{3.11}
A(\bk) \xlrarrow{ k_\mu\frac{\pd}{\pd k^\nu} } \circ B(\bk)
:=
-
\Bigl( k_\mu\frac{\pd A(\bk)}{\pd k^\nu} \Bigr) \circ B(\bk)
+
\Bigl( A(\bk) \circ k_\mu\frac{\pd B(\bk)}{\pd k^\nu} \Bigr)
\\ =
k_\mu \Bigl(  A(\bk) \xlrarrow{ \frac{\pd}{\pd k^\nu} } \circ B(\bk) \Bigr)
	\end{multline}
for operators $A(\bk)$ and $B(\bk)$ having $C^1$ dependence on $\bk$,%
\footnote{~%
More generally, if $\omega\colon\{\Hil\to\Hil\}\to\{\Hil\to\Hil\}$ is a
mapping on the operator space over the system's Hilbert space, we put
 $A\xlrarrow{\omega}\circ B := -\omega(A)\circ B + A\circ \omega(B)$
for any $A,B\colon\Hil\to\Hil$. Usually~\cite{Bjorken&Drell,Itzykson&Zuber},
this notation is used for $\omega=\pd_\mu$.%
}
and
$\sigma_{\mu\nu}^{ss',\pm}(\bk)$ and $l_{\mu\nu}^{ss',\pm}(\bk)$ are some
functions of $\bk$ such that%
\footnote{~%
For the explicit form of these functions,
see~\cite{bp-QFTinMP-scalars,bp-QFTinMP-spinors,bp-QFTinMP-vectors}; see also
equation~\eref{6.55} below.%
}
	\begin{equation}	\label{3.13}
	\begin{split}
&
\sigma_{\mu\nu}^{ss',\pm}(\bk) = - \sigma_{\nu\mu}^{ss',\pm}(\bk)
\quad
l_{\mu\nu}^{ss',\pm}(\bk) = - l_{\nu\mu}^{ss',\pm}(\bk)
\\
&
\sigma_{\mu\nu}^{ss',\pm}(\bk) =  l_{\nu\mu}^{ss',\pm}(\bk) = 0
\qquad\text{for $j=0$ (scalar field)}
\\
&   \sigma_{\mu\nu}^{ss',-}(\bk)
= - \sigma_{\mu\nu}^{ss',+}(\bk)
=:  \sigma_{\mu\nu}^{ss'}(\bk)
= - \sigma_{\mu\nu}^{s's}(\bk)
= - \sigma_{\nu\mu}^{ss'}(\bk)
\qquad\text{for $j=1$ (vector field)}
\\
&   l_{\mu\nu}^{ss',-}(\bk)
= - l_{\mu\nu}^{ss',+}(\bk)
=:  l_{\mu\nu}^{ss'}(\bk)
= - l_{\mu\nu}^{s's}(\bk)
= - l_{\nu\mu}^{ss'}(\bk)
\qquad\text{for $j=1$ (vector field)}.
	\end{split}
	\end{equation}

	A technical remark must be make at this point. The
equations~\eref{3.7}--\eref{3.10} were derived
in~\cite{bp-QFTinMP-scalars,bp-QFTinMP-spinors,bp-QFTinMP-vectors} under some
additional conditions, represented by equations~\eref{2.1} and~\eref{2.28},
which are considered bellow in Sect.~\ref{Sect5} and ensure the effectiveness
of the momentum picture of motion~\cite{bp-QFT-MP} used
in~\cite{bp-QFTinMP-scalars,bp-QFTinMP-spinors,bp-QFTinMP-vectors}. However,
as it is partially proved, e.g., in~\cite{Bogolyubov&Shirkov}, when the
quantities~\eref{3.7}--\eref{3.10} are expressed via the Heisenberg creation
and annihilation operators (see~\eref{2.28-2}), they remain valid, up to a
phase factor, and without making the mentioned assumptions, \ie these
assumptions are needless when one works entirely in Heisenberg picture. For
this reason, we shall consider~\eref{3.7}--\eref{3.10} as pure consequence of
the Lagrangian formalism.

	We should emphasize, in~\eref{3.9} and~\eref{3.10} with
$\tope{S}_{\mu\nu}^{\omega}$ and $\tope{L}_{\mu\nu}^{\omega}$,
$\omega=\prime,\prime\prime,\prime\prime\prime$,
are denoted the spin and orbital, respectively, operators for
$\tope{L}^\omega$, which are the spacetime\ndash independent parts of the
spin and orbital, respectively, angular momentum
operators~\cite{bp-QFT-angular-momentum-operator,bp-QFTinMP-spinors}; if the
last operators are denoted by
$\tope{\underline{S}}_{\mu\nu}^{\omega}$ and
$\tope{\underline{L}}_{\mu\nu}^{\omega}$, the total angular momentum
operator of a system with Lagrangian $\tope{L}^\omega$
is~\cite{bp-QFT-angular-momentum-operator}
	\begin{equation}	\label{3.14}
\tope{M}_{\mu\nu}^{\omega}
=
\tope{\underline{L}}_{\mu\nu}^{\omega} +
\tope{\underline{S}}_{\mu\nu}^{\omega}
=
\tope{L}_{\mu\nu}^{\omega} + \tope{S}_{\mu\nu}^{\omega} ,
\qquad
\omega=\prime,\prime\prime,\prime\prime\prime
	\end{equation}
and
 $\tope{\underline{S}}_{\mu\nu}^{\omega} = \tope{S}_{\mu\nu}^{\omega}$
(and hence
$\tope{\underline{L}}_{\mu\nu}^{\omega} = \tope{L}_{\mu\nu}^{\omega}$)
iff $\tope{\underline{S}}_{\mu\nu}^{\omega}$ is a conserved operator or,
equivalently, iff the system's canonical energy\ndash momentum tensor is
symmetric.%
\footnote{~%
In~\cite{bp-QFT-angular-momentum-operator,bp-QFTinMP-spinors} the spin and
orbital operators are labeled with an additional left superscript $\circ$,
which, for brevity, is omitted in the present work as in it only these
operators, not
$\tope{\underline{S}}_{\mu\nu}^{\omega}$ and
$\tope{\underline{L}}_{\mu\nu}^{\omega}$,
will be considered. Notice, the operators
$\tope{\underline{S}}_{\mu\nu}^{\omega}$ and
$\tope{\underline{L}}_{\mu\nu}^{\omega}$
are, generally, time\ndash dependent while the orbital and spin ones are
conserved, as a result of which the total angular momentum is a conserved
operator too~\cite{bp-QFT-angular-momentum-operator,bp-QFTinMP-spinors}.%
}

	Going ahead (see Sect.~\ref{Sect6}), we would like to note that the
expressions~\eref{3.7c} and, consequently, the Lagrangian
$\tope{L}^{\prime\prime\prime}$ are the base from which the paracommutation
relations were first derived~\cite{Green-1953}.

	And a last remark. Above we have expressed the dynamical variables in
\emph{Heisenberg picture via the creation and annihilation operators in
momentum picture}.
If one works entirely in Heisenberg picture, the operators~\eref{2.28-2},
representing the creation and annihilation operators in Heisenberg picture,
should be used. Besides, by virtue of the equations
	\begin{gather}
			\label{3.15}
(a_s^{\pm}(\bk))^\dag = a_s^{\dag\,\mp}(\bk) \quad
(a_s^{\dag\,\pm}(\bk))^\dag = a_s^{\mp}(\bk)
\\			\label{3.16}
\bigl( \ta_s^\pm(\bk) \bigr)^\dag = \ta_s^{\dag\,\mp}(\bk) \quad
\bigl( \ta_s^{\dag\,\pm}(\bk) \bigr)^\dag = \ta_s^\mp(\bk) ,
	\end{gather}
some of the relations concerning $a_{s}^{\dag\,\pm}(\bk)$, \eg the
Euler-Lagrange and Heisenberg equations, are consequences of the similar ones
regarding $a_{s}^{\pm}(\bk)$. In view of~\eref{2.28-2}, we shall
consider~\eref{3.7}--\eref{3.10} as obtained form the corresponding
expressions in Heisenberg picture by making the replacements
$\ta_{s}^{\pm}(\bk)\mapsto a_{s}^{\pm}(\bk)$ and
$\ta_{s}^{\dag\,\pm}(\bk)\mapsto a_{s}^{\dag\,\pm}(\bk)$.
So,~\eref{3.7}--\eref{3.10} will have, up to a phase factor, a sense of
dynamical variables in Heisenberg picture expressed via the
creation/annihilation operators in momentum picture.


\section {On the uniqueness of the dynamical variables}
\label{Sect4}

	Let $\ope{D}=\ope{P}_\mu,\ope{Q},\ope{S}_{\mu\nu},\ope{L}_{\mu\nu}$
denotes some dynamical variable, \viz the momentum, charge, spin, or orbital
operator, of a system with Lagrangian $\ope{L}$. Since the Euler\ndash
Lagrange equations for the Lagrangians
$\ope{L}^{\prime}$, $\ope{L}^{\prime\prime}$ and
$\ope{L}^{\prime\prime\prime}$
coincide (see~\eref{3.6}), we can assert that any field satisfying these
equations admits at least three classes of conserved operators, \viz
$\ope{D}^{\prime}$, $\ope{D}^{\prime\prime}$ and
\(
\ope{D}^{\prime\prime\prime}
=\frac{1}{2}\bigl( \ope{D}^{\prime} + \ope{D}^{\prime\prime} \bigr) .
\)
Moreover, it can be proved that the Euler\ndash Lagrange equations for the
Lagrangian
	\begin{equation}	\label{4.1}
\ope{L}_{\alpha,\beta}
:= \alpha \ope{L}^{\prime} + \beta \ope{L}^{\prime\prime}
\qquad \alpha+\beta\not=0
	\end{equation}
do not depend on $\alpha,\beta\in\field[C]$ and coincide with~\eref{3.6}.
Therefore there exists a two parameter family of conserved dynamical
variables for these equations given via
	\begin{equation}	\label{4.2}
\ope{D}_{\alpha,\beta}
:= \alpha \ope{D}^{\prime} + \beta
\ope{D}^{\prime\prime} \qquad \alpha+\beta\not=0 .
	 \end{equation}
Evidently
 $\ope{L}^{\prime\prime\prime} = \ope{L}_{\frac{1}{2},\frac{1}{2}}$ and
 $\ope{D}^{\prime\prime\prime} = \ope{D}_{\frac{1}{2},\frac{1}{2}}$.
Since the Euler\ndash Lagrange equations~\eref{3.6} are linear and
homogeneous (in the cases considered), we can, without a lost of generality,
restrict the parameters $\alpha,\beta\in\field[C]$  to such that
	\begin{equation}	\label{4.3}
\alpha + \beta =1 ,
	\end{equation}
which can be achieved by an appropriate renormalization (by a factor
$(\alpha+\beta)^{-1/2}$) of the field operators. Thus any field satisfying
the Euler\ndash Lagrange equations~\eref{3.6} admits the family
$\ope{D}_{\alpha,\beta}$, $\alpha + \beta =1$, of conserved operators.
Obviously, this conclusion is valid if in~\eref{4.1} we replace the particular
Lagrangians $\ope{L}^{\prime}$ and $\ope{L}^{\prime\prime}$ (see~\eref{3.1}
and~\eref{3.3}) with any two Lagrangians (of one and the same field
variables) which lead to identical Euler\ndash Lagrange equations.  However,
the essential point in our case is that
$\ope{L}^{\prime}$ and $\ope{L}^{\prime\prime}$
do not differ only by a full divergence, as a result of which the operators
$\ope{D}_{\alpha,\beta}$ are different for different pairs $(\alpha,\beta)$,
$\alpha+\beta=1$.%
\footnote{~%
Note, no commutativity or some commutation relations between the field
operators and their charge (or Hermitian) conjugate are presupposed, i.e., at
the moment, we work in a theory without such relations and normal ordering.%
}

	Since one expects a physical system to possess uniquely defined
dynamical characteristics, \eg energy and total angular momentum, and the
Euler\ndash Lagrange equations are considered (in the framework of Lagrangian
formalism) as the ones governing the spacetime evolution of the system
considered, the problem arises when the dynamical operators
$\ope{D}_{\alpha,\beta}$, $\alpha + \beta =1$, are independent of the
particular choice of $\alpha$ and $\beta$, \ie of the initial Lagrangian one
starts off. Simple calculation show that the operators~\eref{4.2}, under the
condition~\eref{4.3}, are independent of the particular values of the
parameters $\alpha$ and $\beta$ if and only if
	\begin{equation}	\label{4.4}
\ope{D}^{\prime} =\ope{D}^{\prime\prime} .
	\end{equation}
Some consequences of the condition(s)~\eref{4.4} will be considered below,
as well as possible ways for satisfying these restrictions on the Lagrangian
formalism.

	Combining~\eref{3.7}--\eref{3.10} with~\eref{4.4}, for respectively
$\ope{D}=\ope{P}_\mu,\ope{Q},\ope{S}_{\mu\nu},\ope{L}_{\mu\nu}$,
we see that a free scalar, spinor or vector field has a uniquely defined
dynamical variables if and only if the following equations are fulfilled:
	\begin{multline}	\label{4.5}
\sum_{s=1}^{ 2j+1 - \delta_{0m} (1-\delta_{0j}) } \int \Id^3\bk \,
k_\mu\big|_{ k_0=\sqrt{m^2c^2+\bk^2} }
\bigl\{
a_{s}^{\dag\,+}(\bk) \circ a_{s}^{-}(\bk) - \varepsilon
a_{s}^{-}(\bk) \circ a_{s}^{\dag\,+}(\bk)
\\ -
a_{s}^{+}(\bk) \circ a_{s}^{\dag\,-}(\bk) + \varepsilon
a_{s}^{\dag\,-}(\bk) \circ a_{s}^{+}(\bk)
\bigr\}
= 0
	\end{multline}
\vspace{-4ex}
	\begin{multline}	\label{4.6}
q\times
\sum_{s=1}^{ 2j+1 - \delta_{0m} (1-\delta_{0j}) } \int \Id^3\bk \,
\bigl\{
a_{s}^{\dag\,+}(\bk) \circ a_{s}^{-}(\bk) - \varepsilon
a_{s}^{-}(\bk) \circ a_{s}^{\dag\,+}(\bk)
\\ +
a_{s}^{+}(\bk) \circ a_{s}^{\dag\,-}(\bk) - \varepsilon
a_{s}^{\dag\,-}(\bk) \circ a_{s}^{+}(\bk)
\bigr\}
= 0
	\end{multline}
\vspace{-4ex}
	\begin{multline}	\label{4.7}
\sum_{s,s'=1}^{ 2j+1 - \delta_{0m} (1-\delta_{1j}) } \int \Id^3\bk \,
\bigl\{
\sigma_{\mu\nu}^{ss',-}(\bk) a_{s}^{\dag\,+}(\bk) \circ a_{s'}^{-}(\bk)
- \varepsilon
\sigma_{\mu\nu}^{ss',-}(\bk) a_{s'}^{-}(\bk) \circ a_{s}^{\dag\,+}(\bk)
\\ - \varepsilon
\sigma_{\mu\nu}^{ss',+}(\bk) a_{s'}^{+}(\bk) \circ a_{s}^{\dag\,-}(\bk) +
\sigma_{\mu\nu}^{ss',+}(\bk) a_{s}^{\dag\,-}(\bk) \circ a_{s'}^{+}(\bk)
\bigr\}
= 0
	\end{multline}
\vspace{-4ex}
	\begin{multline}	\label{4.8}
\sum_{s,s'=1}^{ 2j+1 - \delta_{0m} (1-\delta_{1j}) } \int \Id^3\bk \,
\bigl\{
l_{\mu\nu}^{ss',-}(\bk) a_{s}^{\dag\,+}(\bk) \circ a_{s'}^{-}(\bk)
- \varepsilon
l_{\mu\nu}^{ss',-}(\bk)	a_{s'}^{-}(\bk) \circ a_{s}^{\dag\,+}(\bk)
\\ - \varepsilon
l_{\mu\nu}^{ss',+}(\bk) a_{s'}^{+}(\bk) \circ a_{s}^{\dag\,-}(\bk) +
l_{\mu\nu}^{ss',+}(\bk) a_{s}^{\dag\,-}(\bk) \circ a_{s'}^{+}(\bk)
\bigr\}
\displaybreak[2] \\ 
+ \frac{1}{2}
\! \sum_{s=1}^{ 2j+1 - \delta_{0m} (1-\delta_{0j}) }
\!\!\!\! \! \int \!\!\!\! \Id^3\bk \,
\Bigl\{
a_{s}^{\dag\,+}(\bk)
\Bigl( \xlrarrow{ k_\mu \frac{\pd}{\pd k^\nu} }
		- \xlrarrow{ k_\nu \frac{\pd}{\pd k^\mu} } \Bigr)
\circ a_{s}^{-}(\bk) + \varepsilon
a_{s}^{-}(\bk)
\Bigl( \xlrarrow{ k_\mu \frac{\pd}{\pd k^\nu} }
		- \xlrarrow{ k_\nu \frac{\pd}{\pd k^\mu} } \Bigr)
\circ a_{s}^{\dag\,+}(\bk)
\\ -
a_{s}^{+}(\bk)
\Bigl( \xlrarrow{ k_\mu \frac{\pd}{\pd k^\nu} }
		- \xlrarrow{ k_\nu \frac{\pd}{\pd k^\mu} } \Bigr)
\circ a_{s}^{\dag\,-}(\bk) - \varepsilon
a_{s}^{\dag\,-}(\bk)
\Bigl( \xlrarrow{ k_\mu \frac{\pd}{\pd k^\nu} }
		- \xlrarrow{ k_\nu \frac{\pd}{\pd k^\mu} } \Bigr)
\circ a_{s}^{+}(\bk)
\Bigr\} \Big|_{ k_0=\sqrt{m^2c^2+{\bs k}^2} }
= 0 .
      \end{multline}

	In~\eref{4.6} is retained the constant factor $q$ as in the
neutral case it is equal to zero and, consequently, the equation~\eref{4.6}
reduces to identity.

	Since the Euler-Lagrange equations do not impose some restrictions on
the creation and annihilation operators, the equations~\eref{4.5}--\eref{4.8}
can be regarded as subsidiary conditions on the Lagrangian formalism and can
serve as equations for (partial) determination of the creation and
annihilation operators. The system of integral
equations~\eref{4.5}--\eref{4.8} is quite complicated and we are not going to
investigate it in the general case. Below we shall restrict ourselves to
analysis of only those solutions of~\eref{4.5}--\eref{4.8}, if any, for which
the integrands in~\eref{4.5}--\eref{4.8} vanish. This means that we shall
replace the system of \emph{integral} equations~\eref{4.5}--\eref{4.8} with
respect to creation and annihilation operators with the following system of
\emph{algebraic} equations (do not sum over $s$ and $s'$ in~\eref{4.12}
and~\eref{4.13}!):
	\begin{multline}	\label{4.9}
a_{s}^{\dag\,+}(\bk) \circ a_{s}^{-}(\bk) - \varepsilon
a_{s}^{-}(\bk) \circ a_{s}^{\dag\,+}(\bk)
-
a_{s}^{+}(\bk) \circ a_{s}^{\dag\,-}(\bk) + \varepsilon
a_{s}^{\dag\,-}(\bk) \circ a_{s}^{+}(\bk)
= 0
	\end{multline}
\vspace{-4ex}
	\begin{multline}	\label{4.10}
a_{s}^{\dag\,+}(\bk) \circ a_{s}^{-}(\bk) - \varepsilon
a_{s}^{-}(\bk) \circ a_{s}^{\dag\,+}(\bk)
+
a_{s}^{+}(\bk) \circ a_{s}^{\dag\,-}(\bk) - \varepsilon
a_{s}^{\dag\,-}(\bk) \circ a_{s}^{+}(\bk)
= 0
\quad\text{ if } q\not=0
	\end{multline}
\vspace{-4ex}
	\begin{multline}	\label{4.11}
\Bigl\{
a_{s}^{\dag\,+}(\bk)
\Bigl( \xlrarrow{ k_\mu \frac{\pd}{\pd k^\nu} }
		- \xlrarrow{ k_\nu \frac{\pd}{\pd k^\mu} } \Bigr)
\circ a_{s}^{-}(\bk) + \varepsilon
a_{s}^{-}(\bk)
\Bigl( \xlrarrow{ k_\mu \frac{\pd}{\pd k^\nu} }
		- \xlrarrow{ k_\nu \frac{\pd}{\pd k^\mu} } \Bigr)
\circ a_{s}^{\dag\,+}(\bk)
\\ -
a_{s}^{+}(\bk)
\Bigl( \xlrarrow{ k_\mu \frac{\pd}{\pd k^\nu} }
		- \xlrarrow{ k_\nu \frac{\pd}{\pd k^\mu} } \Bigr)
\circ a_{s}^{\dag\,-}(\bk) - \varepsilon
a_{s}^{\dag\,-}(\bk)
\Bigl( \xlrarrow{ k_\mu \frac{\pd}{\pd k^\nu} }
		- \xlrarrow{ k_\nu \frac{\pd}{\pd k^\mu} } \Bigr)
\circ a_{s}^{+}(\bk)
\Bigr\} \Big|_{ k_0=\sqrt{m^2c^2+{\bs k}^2} }
= 0
      \end{multline}
\vspace{-4ex}
	\begin{multline}	\label{4.12}
\sum_{s,s'}
 \bigl\{
\sigma_{\mu\nu}^{ss',-}(\bk) a_{s}^{\dag\,+}(\bk)\circ a_{s'}^{-}(\bk)
- \varepsilon
\sigma_{\mu\nu}^{ss',-}(\bk) a_{s'}^{-}(\bk) \circ a_{s}^{\dag\,+}(\bk)
\\ - \varepsilon
\sigma_{\mu\nu}^{ss',+}(\bk) a_{s'}^{+}(\bk) \circ a_{s}^{\dag\,-}(\bk) +
\sigma_{\mu\nu}^{ss',+}(\bk) a_{s}^{\dag\,-}(\bk) \circ a_{s'}^{+}(\bk)
 \bigr\}
= 0
	\end{multline}
\vspace{-4ex}
	\begin{multline}	\label{4.13}
\sum_{s,s'}
 \bigl\{
l_{\mu\nu}^{ss',-}(\bk)  a_{s}^{\dag\,+}(\bk) \circ a_{s'}^{-}(\bk)
- \varepsilon
l_{\mu\nu}^{ss',-}(\bk)	a_{s'}^{-}(\bk) \circ a_{s}^{\dag\,+}(\bk)
\\ - \varepsilon
l_{\mu\nu}^{ss',+}(\bk) a_{s'}^{+}(\bk) \circ a_{s}^{\dag\,-}(\bk) +
l_{\mu\nu}^{ss',+}(\bk) a_{s}^{\dag\,-}(\bk) \circ a_{s'}^{+}(\bk)
 \bigr\}
= 0
      \end{multline}
	Here:
 $s=1,\dots,2j+1-\delta_{0m}(1-\delta_{0j})$
in~\eref{4.9}--\eref{4.11} and
 $s,s'=1,\dots,2j+1-\delta_{0m}(1-\delta_{1j})$
in~\eref{4.12} and~\eref{4.13}. (Notice, by virtue of~\eref{3.13},
the equations~\eref{4.12} and~\eref{4.13} are identically valid for $j=0$,
\ie for scalar fields.) Since all polarization indices enter in~\eref{4.5}
and~\eref{4.6} on equal footing, we do not sum over $s$
in~\eref{4.9}--\eref{4.11}. But in~\eref{4.12} and~\eref{4.13} we have
retain the summation sign as the modes with definite polarization cannot be
singled out in the general case.
	One may obtain weaker versions of~\eref{4.9}--\eref{4.13} by summing
in them over the polarization indices, but we shall not consider these
conditions below regardless of the fact that they also ensure uniqueness of
the dynamical variables.

	At first, consider the equations~\eref{4.9}--\eref{4.11}. Since for a
neutral field, $q=0$, we have $a_s^{\dag\,\pm}(\bk)=a_s^{\pm}(\bk)$, which
physically means coincidence of field's particles and antiparticles, the
equations~\eref{4.9}--\eref{4.11} hold identically in this case.

	Let consider now the case $q\not=0$, \ie the investigated field to be
charged one. Using the standard notation (cf.~\eref{3.12-1})
	\begin{equation}	\label{4.14}
[A,B]_\eta := A\circ B + \eta B\circ A,
	\end{equation}
for operators $A$ and $B$ and $\eta\in\field[C]$, we rewrite~\eref{4.9}
and~\eref{4.10} as
	\begin{align}	\label{4.9'}
			\tag{\ref{4.9}$^\prime$}
[ a_s^{\dag\,+}(\bk) , a_s^{-}(\bk) ]_{-\varepsilon} -
[ a_s^{+}(\bk) , a_s^{\dag\,-}(\bk) ]_{-\varepsilon} & = 0
\\			\label{4.10'}
			\tag{\ref{4.10}$^\prime$}
[ a_s^{\dag\,+}(\bk) , a_s^{-}(\bk) ]_{-\varepsilon} +
[ a_s^{+}(\bk) , a_s^{\dag\,-}(\bk) ]_{-\varepsilon} & = 0
\quad\text{ if } q\not=0 ,
	\end{align}
which are equivalent to
	\begin{equation}	\label{4.15}
[ a_s^{\dag\,\pm}(\bk) , a_s^{\mp}(\bk) ]_{-\varepsilon} = 0
\quad\text{ if } q\not=0 .
	\end{equation}

	Differentiating~\eref{4.15} and inserting the result
into~\eref{4.11}, one can verify that~\eref{4.11} is tantamount to
	\begin{multline}	\label{4.15new}
\Bigl\{
\Bigl[
a_{s}^{\dag\,+}(\bk)
,
\Bigl( k_\mu \frac{\pd}{\pd k^\nu} - k_\nu \frac{\pd}{\pd k^\mu} \Bigr)
\circ a_{s}^{-}(\bk)
\Bigr]_{-\varepsilon}
\\ -
\Bigl[
a_{s}^{+}(\bk)
,
\Bigl( k_\mu \frac{\pd}{\pd k^\nu} - k_\nu \frac{\pd}{\pd k^\mu} \Bigr)
\circ a_{s}^{\dag\,-}(\bk)
\Bigr]_{-\varepsilon}
\Bigr\} \Big|_{ k_0=\sqrt{m^2c^2+{\bs k}^2} }
= 0
\quad\text{ if } q\not=0 ,
      \end{multline}

	Consider now~\eref{4.12} and~\eref{4.13}. By means of the
shorthand~\eref{4.14}, they read
	\begin{align}	\label{4.16}
\sum_{s,s'} \bigl\{
\sigma_{\mu\nu}^{ss',-}(\bk)
[ a_s^{\dag\,+}(\bk) , a_{s'}^{-}(\bk) ]_{-\varepsilon} +
\sigma_{\mu\nu}^{ss',+}(\bk)
[ a_{s}^{\dag\,-}(\bk) , a_{s'}^{+}(\bk) ]_{-\varepsilon}
\bigr\}
& = 0
\\			\label{4.17}
\sum_{s,s'} \bigl\{
l_{\mu\nu}^{ss',-}(\bk)
[ a_s^{\dag\,+}(\bk) , a_{s'}^{-}(\bk) ]_{-\varepsilon} +
l_{\mu\nu}^{ss',+}(\bk)
[ a_{s}^{\dag\,-}(\bk) , a_{s'}^{+}(\bk) ]_{-\varepsilon}
\bigr\}
& = 0  .
	\end{align}
For a scalar field, $j=0$, these conditions hold identically, due
to~\eref{3.13}. But for $j\not=0$ they impose new restrictions on the
formalism. In particular, for vector fields, $j=1$ and $\varepsilon=+1$ they
are satisfied iff (see~\eref{3.13})
	\begin{equation}	\label{4.18}
[ a_s^{\dag\,+}(\bk) , a_{s'}^{-}(\bk) ]_{-\varepsilon} -
[ a_{s}^{\dag\,-}(\bk) , a_{s'}^{+}(\bk) ]_{-\varepsilon}-
[ a_{s'}^{\dag\,+}(\bk) , a_{s}^{-}(\bk) ]_{-\varepsilon} +
[ a_{s'}^{\dag\,-}(\bk) , a_{s}^{+}(\bk) ]_{-\varepsilon}
= 0 .
	\end{equation}
One can satisfy~\eref{4.16} and~\eref{4.17} if the following generalization
of~\eref{4.15} holds
	\begin{equation}	\label{4.19}
[ a_s^{\dag\,\pm}(\bk) , a_{s'}^{\mp}(\bk) ]_{-\varepsilon} = 0.
	\end{equation}

	For spin $j=\frac{1}{2}$ (and hence $\varepsilon=-1$ --
see~\eref{3.12}), the conditions~\eref{4.12} and~\eref{4.13} cannot be
simplified much, but, if one requires the vanishment of the operator
coefficients after $\sigma_{\mu\nu}^{ss',\pm}(\bk)$ and
$l_{\mu\nu}^{ss',\pm}(\bk)$, one gets
	\begin{equation}	\label{4.19new}
a_s^{\dag\,\pm}(\bk) \circ a_{s'}^{\mp}(\bk)  = 0
\qquad j=\frac{1}{2} \quad \varepsilon=-1 .
	\end{equation}

	Excluding some special cases, \eg neutral scalar field ($q=0$ and
$j=0$), the equations~\eref{4.15} and~\eref{4.19new} are unacceptable from
many viewpoints. The main of them is that they are incompatible with the
ordinary (anti)commutation relations (see, e.g.,
e.g.~\cite{Bjorken&Drell-2,Bogolyubov&Shirkov,Itzykson&Zuber,Ohnuki&Kamefuchi}
or Sect.~\ref{Sect6}, in particular, equations~\eref{6.12} bellow); for
example,~\eref{4.19new} means that the acts of creation and annihilation of
(anti)particles with identical characteristics should be mutually
independent, which contradicts to the existing theory and experimental data.

	Now we shall try another way for achieving uniqueness of the
dynamical variables for free fields. Since in~\eref{4.9}--\eref{4.13}
naturally appear (anti)commutators between creation and annihilation
operators and these (anti)commutators vanish under the standard normal
ordering~\cite{Bjorken&Drell-2,Bogolyubov&Shirkov,Itzykson&Zuber,
Ohnuki&Kamefuchi}, one may suppose that the normally ordered expressions of
the dynamical variables may coincide. Let us analyze this method.

    Recall~\cite{Bjorken&Drell-2,Bogolyubov&Shirkov,Itzykson&Zuber,Roman-QFT},
the normal ordering operator $\ope{N}$ (for free field theory) is a linear
operator on the operator space of the system considered such that to a
product (composition)  $c_1\circ\dots\circ c_n$ of $n\in\field[N]$ creation
and/or annihilation operators $c_1,\dots\,c_n$ it assigns the operator
$(-1)^f c_{\alpha_1}\circ\cdots c_{\alpha_n}$. Here
$(\alpha_1,\dots,\alpha_n)$ is a permutation of $(1,\dots,n)$, all creation
operators stand to the left of all annihilation ones, the relative order
between the creation/annihilation operators is preserved, and $f$ is equal to
the number of transpositions among the fermion operators ($j=\frac{1}{2}$)
needed to be achieved the just\ndash described order (``normal order'') of the
operators $c_1\circ\dots\circ c_n$ in $c_{\alpha_1}\circ\cdots
c_{\alpha_n}$.%
\footnote{~%
We have slightly modified the definition given
in~\cite{Bjorken&Drell-2,Bogolyubov&Shirkov,Itzykson&Zuber,Roman-QFT} because
no (anti)commutation relations are presented in our exposition till the
moment. In this paper we do not concern the problem for elimination of the
`unphysical' operators $a_{3}^{\pm}(\bk)$ and $a_{3}^{\dag\,\pm}(\bk)$ from the
spin and orbital momentum operators when $j=1$; for details,
see~\cite{bp-QFTinMP-vectors}, where it is proved that, for an electromagnetic
field, $j=1$ and $q=0$, one way to achieve this is by adding to the number
$f$ above the number of transpositions between $a_{s}^{\pm}(\bk)$, $s=1,2$,
and $a_{3}^{\pm}(\bk)$ needed for getting normal order.%
}
In particular this means that
	\begin{alignat}{2}
\notag
\ope{N}\bigl( a_{s}^{+}(\bk)\circ a_{t}^{\dag\,-}(\bs p) \bigr)
& = a_{s}^{+}(\bk)\circ a_{t}^{\dag\,-}(\bs p)
& \quad
\ope{N}\bigl( a_{s}^{\dag\,+}(\bk)\circ a_{t}^{-}(\bs p) \bigr)
& = a_{s}^{\dag\,+}(\bk)\circ a_{t}^{-}(\bs p)
\\    \label{4.20}
\ope{N}\bigl( a_{s}^{-}(\bk)\circ a_{t}^{\dag\,+}(\bs p) \bigr)
& = \varepsilon a_{t}^{\dag\,+}(\bs p)\circ a_{s}^{-}(\bk)
& \quad
\ope{N}\bigl( a_{s}^{\dag\,-}(\bk)\circ a_{t}^{+}(\bs p) \bigr)
& = \varepsilon a_{t}^{+}(\bs p)\circ a_{s}^{\dag\,-}(\bk)
	\end{alignat}
and, consequently, we have
	\begin{equation}	\label{4.21}
\ope{N}\bigl(
[ a_{s}^{\dag\,\pm}(\bk) , a_{t}^{\mp}(\bs p) ]_{-\varepsilon}
\bigr)
= 0
\quad
\ope{N}\bigl(
[ a_{s}^{\pm}(\bk) , a_{t}^{\dag\,\mp}(\bs p) ]_{-\varepsilon}
\bigr)
= 0,
	\end{equation}
due to $\varepsilon:=(-1)^{2j}=\pm1$ (see~\eref{3.12}). (In fact, below only
the equalities~\eref{4.20} and~\eref{4.21}, not the general definition of a
normal product, will be applied.)

	Applying the normal ordering operator to~\eref{4.9'}, \eref{4.10'},
\eref{4.16} and~\eref{4.17}, we, in view of~\eref{4.21}, get the identity
$0=0$, which means that the conditions~\eref{4.9}, \eref{4.10}, \eref{4.12}
and~\eref{4.13} are identically satisfied after normal ordering. This is
confirmed by the application of $\ope{N}$ to~\eref{3.7} and~\eref{3.8}, which
results respectively in (see~\eref{4.20})
	\begin{multline}	\label{4.22}
  \ope{N}(\tope{P}_\mu^{\prime})
= \ope{N}(\tope{P}_\mu^{\prime\prime})
\\ =
\frac{1}{1+\tau}
\sum_{s=1}^{2j+1-\delta_{0m}(1-\delta_{0j})} \int \Id^3\bk
k_\mu |_{ k_0=\sqrt{m^2c^2+{\bs k}^2} }
\{
a_s^{\dag\,+}(\bk)\circ a_s^-(\bk) +
a_s^+(\bk) \circ a_s^{\dag\,-}(\bk)
\}
	\end{multline}
\vspace{-4ex}
	\begin{gather}	\label{4.23}
  \ope{N}(\tope{Q}^{\prime})
= \ope{N}(\tope{Q}^{\prime\prime})
=
\frac{1}{1+\tau}
\sum_{s=1}^{2j+1-\delta_{0m}(1-\delta_{0j})} \int \Id^3\bk
\{
a_s^{\dag\,+}(\bk)\circ a_s^-(\bk) -
a_s^+(\bk) \circ a_s^{\dag\,-}(\bk)
\} .
	\end{gather}
Therefore the normal ordering ensures the uniqueness of the momentum and
charge operators, if we redefine them respectively as
	\begin{equation}	\label{4.24}
\tope{P}_\mu:= \ope{N}(\tope{P}_\mu^{\prime})
\quad
\tope{Q}:=\ope{N}(\tope{Q}^{\prime}).
	\end{equation}

	Putting
\(
\omega_{\mu\nu}
:= k_\mu \frac{\pd}{\pd k^\nu} -  k_\nu \frac{\pd}{\pd k^\mu}
\)
and using~\eref{4.20}, one can verify that
	\begin{equation}	\label{4.25}
	\begin{split}
\ope{N}\bigl(
a_{s}^{+}(\bk) \xlrarrow{\omega_{\mu\nu}} \circ a_{s}^{\dag\,-}(\bk)
\bigr)
& = a_{s}^{+}(\bk) \xlrarrow{\omega_{\mu\nu}}\circ a_{s}^{\dag\,-}(\bk)
\\
\ope{N}\bigl(
a_{s}^{\dag\,+}(\bk) \xlrarrow{\omega_{\mu\nu}} \circ a_{s}^{-}(\bk)
\bigr)
& = a_{s}^{\dag\,+}(\bk) \xlrarrow{\omega_{\mu\nu}} \circ a_{s}^{-}(\bk)
\\
\ope{N}\bigl(
a_{s}^{-}(\bk)  \xlrarrow{\omega_{\mu\nu}}  \circ a_{s}^{\dag\,+}(\bk)
\bigr)
&  =
-\varepsilon
a_{s}^{\dag\,+}(\bk)  \xlrarrow{\omega_{\mu\nu}}  \circ a_{s}^{-}(\bk)
\\
\ope{N}\bigl(
a_{s}^{\dag\,-}(\bk)  \xlrarrow{\omega_{\mu\nu}}  \circ a_{s}^{+}(\bk)
\bigr)
&  =
- \varepsilon
a_{s}^{+}(\bk)  \xlrarrow{\omega_{\mu\nu}}  \circ a_{s}^{\dag\,-}(\bk).
	\end{split}
	\end{equation}
As a consequence of these equalities, the action of $\ope{N}$ on the l.h.s.\
of~\eref{4.11} vanishes. Combining this result with the mentioned fact that
the normal ordering converts~\eref{4.12} and~\eref{4.13} into identities, we
see that the normal ordering procedure ensures also uniqueness of the spin and
orbital operators if we redefine them respectively as:
	\begin{align}	\label{4.29}
	\begin{split}
\tope{S}_{\mu\nu}
& := \ope{N}( \tope{S}_{\mu\nu}^{\prime} )
:= \ope{N}( \tope{S}_{\mu\nu}^{\prime\prime} )
=
\frac{(-1)^{j-1/2} j \hbar }{1+\tau}
\\ & \!\!\! \times
\sum_{s,s'=1}^{2j+1-\delta_{0m}(1-\delta_{1j})} \int \Id^3\bk
\bigl\{
\sigma_{\mu\nu}^{s s',-}(\bk) a_{s}^{\dag\,+}(\bk)\circ a_{s'}^{-}(\bk)
+ \varepsilon
\sigma_{\mu\nu}^{s s',+}(\bk) a_{s'}^{+}(\bk) \circ  a_{s}^{\dag\,-}(\bk)
\bigr\}
	\end{split}
\displaybreak[1] \\	\label{4.30}
	\begin{split}
\tope{L}_{\mu\nu}
& := \ope{N}( \tope{L}_{\mu\nu}^{\prime} )
:= \ope{N}( \tope{L}_{\mu\nu}^{\prime\prime} )
=
x_{0\,\mu} \tope{P}_\nu -
x_{0\,\nu} \tope{P}_\mu
+
\frac{(-1)^{j-1/2} j \hbar }{1+\tau}
\\ & \times
\sum_{s,s'=1}^{2j+1-\delta_{0m}(1-\delta_{1j})} \int \Id^3\bk
\bigl\{
l_{\mu\nu}^{s s',-}(\bk) a_s^{\dag\,+}(\bk) \circ a_{s'}^-(\bk)
+ \varepsilon
l_{\mu\nu}^{s s',+}(\bk) a_{s'}^+(\bk) \circ a_s^{\dag\,-}(\bk)
\bigr\}
\\
&
+
\frac{\ih}{2(1+\tau)}
\sum_{s=1}^{2j+1-\delta_{0m}(1-\delta_{0j})} \int \Id^3\bk
\Bigl\{
a_s^{\dag\,+}(\bk)
\Bigl( \xlrarrow{ k_\mu \frac{\pd}{\pd k^\nu} }
     - \xlrarrow{ k_\nu \frac{\pd}{\pd k^\mu} } \Bigr)
\circ a_s^-(\bk)
\\ &
+
a_s^+(\bk)
\Bigl( \xlrarrow{ k_\mu \frac{\pd}{\pd k^\nu} }
     - \xlrarrow{ k_\nu \frac{\pd}{\pd k^\mu} } \Bigr)
\circ a_s^{\dag\,-}(\bk)
\Bigr\} \Big|_{ k_0=\sqrt{m^2c^2+{\bs k}^2} } \ ,
	\end{split}
	\end{align}
where~\eref{3.13} was applied.


\section{Heisenberg relations}
\label{Sect5}

	The conserved operators, like momentum and charge operators, are
often identified with the generators of the corresponding transformations
under which the action operator is
invariant~\cite{Bjorken&Drell-2,Bogolyubov&Shirkov,Roman-QFT,Itzykson&Zuber}.
This leads to a number of commutation relations between the components of these
operators and between them and the field operators. The relations of the letter
set are known/referred as the \emph{Heisenberg relations} or \emph{equations}.
Both kinds of commutation relations are from pure geometric origin and,
consequently, are completely external to the Lagrangian formalism; one of the
reasons being that the mentioned identification is, in general, unacceptable
and may be carried out only on some subset of the system's Hilbert space of
states~\cite{bp-QFT-momentum-operator,bp-QFT-angular-momentum-operator}.
Therefore their validity in a pure Lagrangian theory is questionable and
should be verified~\cite{Bjorken&Drell-2}. However, the considered relations
are weaker conditions than the identification of the corresponding operators
and there are strong evidences that these relations should be valid in a
realistic quantum field theory~\cite{Bjorken&Drell-2,Bogolyubov&Shirkov};
e.g., the commutativity between the momentum and charge operators (see
below~\eref{5.12}) expresses the experimental fact that the 4\ndash momentum
and charge of any system are simultaneously measurable quantities.

	It is known~\cite{Bjorken&Drell-2,Bogolyubov&Shirkov}, in a pure
Lagrangian approach, the field equations, which are usually identified with the
Euler\ndash Lagrange,~%
\footnote{~%
Recall, there are Lagrangians whose classical Euler-Lagrange equations are
identities. However, their correct and rigorous
treatment~\cite{bp-QFT-action-principle} reveals that they entail field
equations which are mathematically correct and physically sensible.%
}
are the only
restrictions on the field operators. Besides, these equations do not determine
uniquely the field operators and the letter can be expressed through the
creation and annihilation operators. Since the last operators are left
completely arbitrary by a pure Lagrangian formalism, one is free to impose on
them any system of \emph{compatible} restrictions. The best known examples of
this kind are the famous canonical (anti)commutation relations and their
generalization, the so\ndash called paracommutation
relations~\cite{Green-1953,Ohnuki&Kamefuchi}. In general, the problem for
compatibility of such subsidiary to the Lagrangian formalism system of
restrictions  with, for instance, the Heisenberg relations is open and
requires particular investigation~\cite{Bjorken&Drell-2}. For example, even
the canonical (anti)commutation relations for electromagnetic field in
Coulomb gauge are incompatible with the Heisenberg equation involving the
(total) angular momentum operator unless the gauge symmetry of this field is
taken into account~\cite[\S~84]{Bjorken&Drell-2}. However, the
(para)commutation relations are, by construction, compatible with the
Heisenberg relations regarding momentum operator (see~\cite{Green-1953} or
below Subsect.~\ref{Subsect6.1}). The ordinary approach is to be imposed a
system of equations on the creation and annihilation operators and, then, to
be checked its compatibility with, e.g., the Heisenberg relations. In the
next sections we shall investigate the opposite situation: assuming the
validity of (some of) the Heisenberg equations, the possible restrictions on
the creation and annihilation operators will be explored. For this purpose,
below we briefly review the Heisenberg relations and other ones related to
them.

	Consider a system of quantum fields $\tope{\varphi}_i(x)$,
$i=1,\dots,N\in\field[N]$, where $\tope{\varphi}_i(x)$ denote the components
of all fields (and their Hermitian conjugates), and
$\tope{P}_\mu$, $\tope{Q}$ and $\tope{M}_{\mu\nu}$ be its momentum, charge
and (total) angular momentum operators, respectively. The Heisenberg
relations/equations for these operators
are~\cite{Bjorken&Drell-2,Bogolyubov&Shirkov,Roman-QFT,Itzykson&Zuber}
	\begin{align}
			\label{5.1}
 [\tope{\varphi}_i(x), \tope{P}_{\mu}]_{\_}
	&= \ih \frac{\pd \tope{\varphi}_i(x)}{\pd x^\mu}
\\
			\label{5.2}
 [\tope{\varphi}_i(x), \tope{Q}]_{\_}
	&= {e}(\tope{\varphi}_i) q \tope{\varphi}_i(x)
\\			\label{5.3}
 [\tope{\varphi}_i(x), \tope{M}_{\mu\nu}]_{\_}
&=
\ih\{
x_\mu\pd_\nu\tope{\varphi}_i(x) - x_\nu\pd_\mu\tope{\varphi}_i(x) \}
+ \ih \sum_{i'} I_{i'\mu\nu}^{j} \tope{\varphi}_{i'}(x) .
	\end{align}
Here: $q=\const$ is the fields' charge,
${e}(\tope{\varphi}_i) = 0$ if
		$\tope{\varphi}_i^\dag = \tope{\varphi}_i$,
${e}(\tope{\varphi}_i) = \pm 1$ if
		$\tope{\varphi}_i^\dag \not= \tope{\varphi}_i$
with
${e}(\tope{\varphi}_i) + {e}(\tope{\varphi}_i^\dag) = 0$,
and the constants $I_{i\mu\nu}^{i'} = -I_{i\nu\mu}^{i'}$ characterize the
transformation properties of the field operators under 4\ndash rotations.
(If $\varepsilon(\tope{\varphi}_i)\not=0$, it is a convention whether to put
$\varepsilon(\tope{\varphi}_i)=+1$ or $\varepsilon(\tope{\varphi}_i)=-1$ for a
fixed $i$.)

	We would like to make some comments on~\eref{5.3}. Since its r.h.s.\
is a sum of two operators, the first (second) characterizing the pure orbital
(spin) angular momentum properties of the system considered, the idea arises
to split~\eref{5.3} into two independent equations, one involving the orbital
angular momentum operator and another concerning the spin angular momentum
operator. This is supported by the observation that, it seems, no process
is known for transforming orbital angular momentum into spin one and \vv
(without destroying the system). So one may suppose the existence of operators
$\tope{M}_{\mu\nu}^{\mathrm{or}}$ and $\tope{M}_{\mu\nu}^{\mathrm{sp}}$ such
that
	\begin{align}	\label{5.4}
& [\tope{\varphi}_i(x), \tope{M}_{\mu\nu}^{\mathrm{or}}]_{\_}
=
\ih\{
x_\mu\pd_\nu\tope{\varphi}_i(x) - x_\nu\pd_\mu\tope{\varphi}_i(x) \}
\\			\label{5.5}
& [\tope{\varphi}_i(x), \tope{M}_{\mu\nu}^{\mathrm{sp}}]_{\_}
=
\ih \sum_{i'} I_{i\mu\nu}^{i'} \tope{\varphi}_{i'}(x)
\\			\label{5.5-1}
& \tope{M}_{\mu\nu}
= \tope{M}_{\mu\nu}^{\mathrm{or}} + \tope{M}_{\mu\nu}^{\mathrm{sp}} .
	\end{align}

	However, as particular calculations
demonstrate~\cite{bp-QFTinMP-spinors,bp-QFTinMP-vectors,
Akhiezer&Berestetskii}, neither the spin (resp.\ orbital) nor the spin
(resp.\ orbital) angular momentum operator is a suitable candidate for
$\tope{M}_{\mu\nu}^{\mathrm{sp}}$ (resp.\ $\tope{M}_{\mu\nu}^{\mathrm{or}}$).
If we assume the validity of~\eref{5.1}, then equations~\eref{5.4}
and~\eref{5.5}  can be satisfied if we choose
	\begin{align}	\label{5.6}
\tope{M}_{\mu\nu}^{\mathrm{or}} (x)
&
= \tope{L}_{\mu\nu}^{\mathrm{ext}}
:= x_\mu \tope{P}_\nu -  x_\nu \tope{P}_\mu
\\			\label{5.7}
\tope{M}_{\mu\nu}^{\mathrm{sp}} (x)
&
= \tope{M}_{\mu\nu}^{(0)} (x)
:= \tope{M}_{\mu\nu} - \tope{L}_{\mu\nu}^{\mathrm{ext}}
= \tope{S}_{\mu\nu} + \tope{L}_{\mu\nu} -
  \{ x_\mu \tope{P}_\nu -  x_\nu \tope{P}_\mu \}
	\end{align}
with $\tope{M}_{\mu\nu}$ satisfying~\eref{5.3}. These operators are
\emph{not} conserved ones. Such a representation is in agreement with the
equations~\eref{3.10}, according to which the operator~\eref{5.6} enters
additively in the expressions for the orbital operator.%
\footnote{~%
This is evident in the momentum picture of motion, in which $x_\mu$ stands
for $x_{0\,\mu}$ in~\eref{3.10} ---
see~\cite{bp-QFTinMP-scalars,bp-QFTinMP-spinors,bp-QFTinMP-vectors}.%
}
The physical sense of the operator~\eref{5.6} is that it represents the
orbital angular momentum of the system due to its movement as a whole.
Respectively, the operator~\eref{5.7} describes the system's angular momentum
as a result of its internal movement and/or structure.

	Since the spin (orbital) angular momentum is associated with the
structure (movement) of a system, in the operator~\eref{5.7} are mixed the
spin and orbital angular momenta. These quantities can be separated
completely via the following representations of the operators
$\ope{M}_{\mu\nu}^{\mathrm{or}}$ and $\ope{M}_{\mu\nu}^{\mathrm{sp}}$ in
momentum picture (when~\eref{5.1} holds)
	\begin{align}	\label{5.7-1}
\ope{M}_{\mu\nu}^{\mathrm{or}}
& =
x_\mu\ope{P}_\nu -  x_\mu\ope{P}_\mu + \ope{L}_{\mu\nu}^{\mathrm{int}}
\\			\label{5.7-2}
\ope{M}_{\mu\nu}^{\mathrm{sp}}
& 	=
\ope{M}_{\mu\nu} -
(x_\mu\ope{P}_\nu -  x_\mu\ope{P}_\mu ) - \ope{L}_{\mu\nu}^{\mathrm{int}} ,
	\end{align}
where $\ope{L}_{\mu\nu}^{\mathrm{int}}$ describes the `internal' orbital
angular momentum of the system considered and depends on the Lagrangian we
have started off. Generally said, $\ope{L}_{\mu\nu}^{\mathrm{int}}$ is the
part of the orbital angular momentum operator containing derivatives of the
creation and annihilation operators. In particular, for the Lagrangians
$\ope{L}^{\prime}$, $\ope{L}^{\prime\prime}$ and
$\ope{L}^{\prime\prime\prime}$ (see Sect.~\ref{Sect3}), the explicit forms of
the operators~\eref{5.7-1} and~\eref{5.7-2} respectively are:
	\begin{subequations}	\label{5.7-3}
	\begin{align}	\label{5.7-3a}
	\begin{split}
\ope{M}_{\mu\nu}^{\prime\,\mathrm{or}}
= &
x_{ \mu} \ope{P}_\nu^{\prime} -
x_{ \nu} \ope{P}_\mu^{\prime}
\\
&
+
\frac{\ih}{2(1+\tau)}
\sum_{s=1}^{2j+1-\delta_{0m}(1-\delta_{0j})} \int \Id^3\bk
\Bigl\{
a_s^{\dag\,+}(\bk)
\Bigl( \xlrarrow{ k_\mu \frac{\pd}{\pd k^\nu} }
     - \xlrarrow{ k_\nu \frac{\pd}{\pd k^\mu} } \Bigr)
\circ a_s^-(\bk)
\\ &
- \varepsilon
a_s^{\dag\,-}(\bk)
\Bigl( \xlrarrow{ k_\mu \frac{\pd}{\pd k^\nu} }
     - \xlrarrow{ k_\nu \frac{\pd}{\pd k^\mu} } \Bigr)
\circ a_s^+(\bk)
\Bigr\} \Big|_{ k_0=\sqrt{m^2c^2+{\bs k}^2} }
	\end{split}
\displaybreak[1] \\	\label{5.7-3b} 
	\begin{split}
\ope{M}_{\mu\nu}^{\prime\prime\,\mathrm{or}}
= &
x_{ \mu} \ope{P}_\nu^{\prime\prime} -
x_{ \nu} \ope{P}_\mu^{\prime\prime}
\\
&
+
\frac{\ih}{2(1+\tau)}
\sum_{s=1}^{2j+1-\delta_{0m}(1-\delta_{0j})} \int \Id^3\bk
\Bigl\{
a_s^{+}(\bk)
\Bigl( \xlrarrow{ k_\mu \frac{\pd}{\pd k^\nu} }
     - \xlrarrow{ k_\nu \frac{\pd}{\pd k^\mu} } \Bigr)
\circ a_s^{\dag\,-}(\bk)
\\ &
- \varepsilon
a_s^{-}(\bk)
\Bigl( \xlrarrow{ k_\mu \frac{\pd}{\pd k^\nu} }
     - \xlrarrow{ k_\nu \frac{\pd}{\pd k^\mu} } \Bigr)
\circ a_s^{\dag\,+}(\bk)
\Bigr\} \Big|_{ k_0=\sqrt{m^2c^2+{\bs k}^2} }
	\end{split}
\displaybreak[1]\\	\label{5.7-3c} 
	\begin{split}
\ope{M}_{\mu\nu}^{\prime\prime\prime\,\mathrm{or}}
= &
x_{ \mu} \ope{P}_\nu^{\prime\prime\prime} -
x_{ \nu} \ope{P}_\mu^{\prime\prime\prime}
\\
&
+
\frac{\ih}{4(1+\tau)}
\sum_{s=1}^{2j+1-\delta_{0m}(1-\delta_{0j})} \int \Id^3\bk
\Bigl\{
a_s^{\dag\,+}(\bk)
\Bigl( \xlrarrow{ k_\mu \frac{\pd}{\pd k^\nu} }
     - \xlrarrow{ k_\nu \frac{\pd}{\pd k^\mu} } \Bigr)
\circ a_s^-(\bk)
\\ & - \varepsilon
a_s^{-}(\bk)
\Bigl( \xlrarrow{ k_\mu \frac{\pd}{\pd k^\nu} }
     - \xlrarrow{ k_\nu \frac{\pd}{\pd k^\mu} } \Bigr)
\circ a_s^{\dag\,+}(\bk)
+
a_s^{+}(\bk)
\Bigl( \xlrarrow{ k_\mu \frac{\pd}{\pd k^\nu} }
     - \xlrarrow{ k_\nu \frac{\pd}{\pd k^\mu} } \Bigr)
\circ a_s^{\dag\,-}(\bk)
\\ &
- \varepsilon
a_s^{\dag\,-}(\bk)
\Bigl( \xlrarrow{ k_\mu \frac{\pd}{\pd k^\nu} }
     - \xlrarrow{ k_\nu \frac{\pd}{\pd k^\mu} } \Bigr)
\circ a_s^+(\bk)
\Bigr\} \Big|_{ k_0=\sqrt{m^2c^2+{\bs k}^2} } .
	\end{split}
	\end{align}
	\end{subequations}
\vspace{-4ex} 
	\begin{subequations}	\label{5.7-4}
	\begin{align}	\label{5.7-4a}
	\begin{split}
\ope{M}_{\mu\nu}^{\prime\,\mathrm{sp}}
& =
\frac{(-1)^{j-1/2} j \hbar }{1+\tau}
\sum_{s,s'=1}^{2j+1-\delta_{0m}(1-\delta_{1j})} \int \Id^3\bk
\bigl\{
( \sigma_{\mu\nu}^{s s',-}(\bk) + l_{\mu\nu}^{s s',-}(\bk) )
a_{s}^{\dag\,+}(\bk)\circ a_{s'}^{-}(\bk)
\\ &
\hphantom{=
\frac{(-1)^{j-1/2} j \hbar }{1+\tau}
\sum_{s,s'=1}^{2j+1-\delta_{0m}(1-\delta_{1j})} \int \Id^3\bk
}
+
( \sigma_{\mu\nu}^{s s',+}(\bk) + l_{\mu\nu}^{s s',+}(\bk) )
a_{s}^{\dag\,-}(\bk)\circ a_{s'}^{+}(\bk)
\bigr\}
	\end{split}
\displaybreak[1]\\	\label{5.7-4b}
	\begin{split}
\ope{M}_{\mu\nu}^{\prime\prime\,\mathrm{sp}}
& = \varepsilon
\frac{(-1)^{j-1/2} j \hbar }{1+\tau}
\sum_{s,s'=1}^{2j+1-\delta_{0m}(1-\delta_{1j})} \int \Id^3\bk
\bigl\{
( \sigma_{\mu\nu}^{s s',+}(\bk) + l_{\mu\nu}^{s s',+}(\bk) )
a_{s'}^{+}(\bk)\circ a_{s}^{\dag\,-}(\bk)
\\ &
\hphantom{=
\frac{(-1)^{j-1/2} j \hbar }{1+\tau}
\sum_{s,s'=1}^{2j+1-\delta_{0m}(1-\delta_{1j})} \int \Id^3\bk
}
+
( \sigma_{\mu\nu}^{s s',-}(\bk) + \sigma_{\mu\nu}^{s s',-}(\bk) )
a_{s'}^{-}(\bk)\circ a_{s}^{\dag\,+}(\bk)
\bigr\}
	\end{split}
\displaybreak[1]\\	\label{5.7-4c}
	\begin{split}
\ope{M}_{\mu\nu}^{\prime\prime\prime\,\mathrm{sp}}
& =
\frac{(-1)^{j-1/2} j \hbar }{2(1+\tau)}
\sum_{s,s'=1}^{2j+1-\delta_{0m}(1-\delta_{1j})} \int \Id^3\bk
\bigl\{
( \sigma_{\mu\nu}^{s s',-}(\bk) + l_{\mu\nu}^{s s',-}(\bk) )
	[ a_{s}^{\dag\,+}(\bk) , a_{s'}^{-}(\bk)]_\varepsilon
\\ &
\hphantom{=
\frac{(-1)^{j-1/2} j \hbar }{1+\tau}
\sum_{s,s'=1}^{2j+1-\delta_{0m}(1-\delta_{1j})} \int \Id^3\bk
}
+
( \sigma_{\mu\nu}^{s s',+}(\bk) + l_{\mu\nu}^{s s',+}(\bk) )
	[ a_{s}^{\dag\,-}(\bk) , a_{s'}^{+}(\bk) ]_\varepsilon
\bigr\} .
	\end{split}
	\end{align}
	\end{subequations}
Obviously (see Sect.~\ref{Sect2}), the equations~\eref{5.7-4}
have the same form in Heisenberg picture in terms of the
operators~\eref{2.28-2} (only tildes over $\ope{M}$ and $a$ must be added), but
the equations~\eref{5.7-3} change substantially due to the existence of
derivatives of the creation and annihilation operators in
them~\cite{bp-QFTinMP-scalars,bp-QFTinMP-spinors,bp-QFTinMP-vectors}:
	\begin{subequations}	\label{5.7-5}
	\begin{align}	\label{5.7-5a}
	\begin{split}
\tope{M}_{\mu\nu}^{\prime\,\mathrm{or}}
= &
\frac{\ih}{2(1+\tau)}
\sum_{s=1}^{2j+1-\delta_{0m}(1-\delta_{0j})} \int \Id^3\bk
\Bigl\{
\ta_s^{\dag\,+}(\bk)
\Bigl( \xlrarrow{ k_\mu \frac{\pd}{\pd k^\nu} }
     - \xlrarrow{ k_\nu \frac{\pd}{\pd k^\mu} } \Bigr)
\circ \ta_s^-(\bk)
\\ &
- \varepsilon
\ta_s^{\dag\,-}(\bk)
\Bigl( \xlrarrow{ k_\mu \frac{\pd}{\pd k^\nu} }
     - \xlrarrow{ k_\nu \frac{\pd}{\pd k^\mu} } \Bigr)
\circ \ta_s^+(\bk)
\Bigr\} \Big|_{ k_0=\sqrt{m^2c^2+{\bs k}^2} }
	\end{split}
\displaybreak[1] \\	\label{5.7-5b} 
	\begin{split}
\tope{M}_{\mu\nu}^{\prime\prime\,\mathrm{or}}
= &
\frac{\ih}{2(1+\tau)}
\sum_{s=1}^{2j+1-\delta_{0m}(1-\delta_{0j})} \int \Id^3\bk
\Bigl\{
\ta_s^{+}(\bk)
\Bigl( \xlrarrow{ k_\mu \frac{\pd}{\pd k^\nu} }
     - \xlrarrow{ k_\nu \frac{\pd}{\pd k^\mu} } \Bigr)
\circ \ta_s^{\dag\,-}(\bk)
\\ &
- \varepsilon
\ta_s^{-}(\bk)
\Bigl( \xlrarrow{ k_\mu \frac{\pd}{\pd k^\nu} }
     - \xlrarrow{ k_\nu \frac{\pd}{\pd k^\mu} } \Bigr)
\circ \ta_s^{\dag\,+}(\bk)
\Bigr\} \Big|_{ k_0=\sqrt{m^2c^2+{\bs k}^2} }
	\end{split}
\displaybreak[1]\\	\label{5.7-5c} 
	\begin{split}
\tope{M}_{\mu\nu}^{\prime\prime\prime\,\mathrm{or}}
= &
\frac{\ih}{4(1+\tau)}
\sum_{s=1}^{2j+1-\delta_{0m}(1-\delta_{0j})} \int \Id^3\bk
\Bigl\{
\ta_s^{\dag\,+}(\bk)
\Bigl( \xlrarrow{ k_\mu \frac{\pd}{\pd k^\nu} }
     - \xlrarrow{ k_\nu \frac{\pd}{\pd k^\mu} } \Bigr)
\circ \ta_s^-(\bk)
\\ & - \varepsilon
\ta_s^{-}(\bk)
\Bigl( \xlrarrow{ k_\mu \frac{\pd}{\pd k^\nu} }
     - \xlrarrow{ k_\nu \frac{\pd}{\pd k^\mu} } \Bigr)
\circ \ta_s^{\dag\,+}(\bk)
+
\ta_s^{+}(\bk)
\Bigl( \xlrarrow{ k_\mu \frac{\pd}{\pd k^\nu} }
     - \xlrarrow{ k_\nu \frac{\pd}{\pd k^\mu} } \Bigr)
\circ \ta_s^{\dag\,-}(\bk)
\\ &
- \varepsilon
\ta_s^{\dag\,-}(\bk)
\Bigl( \xlrarrow{ k_\mu \frac{\pd}{\pd k^\nu} }
     - \xlrarrow{ k_\nu \frac{\pd}{\pd k^\mu} } \Bigr)
\circ \ta_s^+(\bk)
\Bigr\} \Big|_{ k_0=\sqrt{m^2c^2+{\bs k}^2} } \ .
	\end{split}
	\end{align}
	\end{subequations}
From~\eref{5.7-5} and~\eref{5.7-4} is clear that the operators
$\tope{M}_{\mu\nu}^{\mathrm{or}}$ and $\tope{M}_{\mu\nu}^{\mathrm{sp}}$ so
defined are \emph{conserved} (contrary to~\eref{5.6} and~\eref{5.7}) and do
not depend on the validity of the Heisenberg relations~\eref{5.1}
(contrary to expressions~\eref{5.7-3} in momentum picture).

	The problem for whether the operators~\eref{5.7-4} and~\eref{5.7-5}
satisfy the equations~\eref{5.4} and~\eref{5.5}, respectively, will be
considered in Sect.~\ref{Sect6}.

	There is an essential difference between~\eref{5.4} and~\eref{5.5}:
the equation~\eref{5.5} depends on the particular properties of the operators
$\tope{\varphi}_i(x)$ under 4\ndash rotations via the coefficients
$I_{i\mu\nu}^{i'}$ (see~\eref{5.19} below), while~\eref{5.4} does not depend
on them. This is explicitly reflected in~\eref{5.7-3} and~\eref{5.7-4}: the
former set of equations is valid independently of the geometrical nature of
the fields considered, while the latter one depends on it via the `spin'
(`polarization') functions $\sigma_{\mu\nu}^{ss',\pm}(\bk)$ and
$l_{\mu\nu}^{ss',\pm}(\bk)$.
	Similar remark concerns~\eref{5.3}, on one hand, and~\eref{5.1}
and~\eref{5.2}, on another hand: the particular form of~\eref{5.3}
essentially depends on the geometric properties of $\tope{\varphi}_i(x)$
under 4\ndash rotations, the other equations being independent of them.

	It should also be noted, the relation~\eref{5.3} does not hold for a
canonically quantized electromagnetic field in Coulomb gauge unless some
additional terms it its r.h.s., reflecting the gauge symmetry of the field,
are taken into account~\cite[\S~84]{Bjorken&Drell-2}.

	As it was said above, the relations~\eref{5.1}--\eref{5.3} are from
pure geometrical origin. However, the last discussion,
concerning~\eref{5.4}--\eref{5.7}, reveals that the terms in braces
in~\eref{5.3} should be connected with the momentum operator in the (pure)
Lagrangian approach. More precisely, on the background of
equations~\eref{3.9a}--\eref{3.10c}, the Heisenberg relation~\eref{5.3}
should be replaced with
	\begin{equation}	\label{5.8}
 [\tope{\varphi}_i(x), \tope{M}_{\mu\nu}]_{\_}
=
x_\mu [ \tope{\varphi}_i(x), \tope{P}_\nu ]_{\_} -
x_\nu [ \tope{\varphi}_i(x), \tope{P}_\mu ]_{\_}
+ \ih \sum_{j}
I_{i\mu\nu}^{i'} \tope{\varphi}_{i'}(x) ,
	\end{equation}
which is equivalent to~\eref{5.3} if~\eref{5.1} is true. An advantage of the
last equation is that it is valid in any picture of motion (in the same
form) while~\eref{5.3} holds only in Heisenberg picture.%
\footnote{~%
In other pictures of motion, generally, additional terms in the r.h.s.\
of~\eref{5.3} will appear, \ie the functional form of the r.h.s.\
of~\eref{5.3} is not invariant under changes of the picture of motion,
contrary to~\eref{5.8}.%
}
Obviously,~\eref{5.8} is equivalent to~\eref{5.5} with
$\tope{M}_{\mu\nu}^{\mathrm{sp}}$ defined by~\eref{5.7}.

	The other kind of geometric relations mentioned at the beginning
of this section are connected with the basic relations defining the Lie
algebra of the Poincar\'e group~\cite[pp.~143--147]{Bogolyubov&et_al.-AxQFT},
\cite[sect.~7.1]{Bogolyubov&et_al.-QFT}. They require the fulfillment of the
following equations between the components $\tope{P}_\mu$ of the momentum and
$\tope{M}_{\mu\nu}$ of the angular momentum
operators~\cite{Bogolyubov&et_al.-AxQFT,Bogolyubov&et_al.-QFT,Roman-QFT,
Akhiezer&Berestetskii}:
	\begin{align}
			\label{5.9}
& [ \tope{P}_\mu ,\tope{P}_\nu ]_{\_} = 0
\\			\label{5.10}
& [ \tope{M}_{\mu\nu} , \tope{P}_\lambda ]_{\_}
 = - \ih ( \eta_{\lambda\mu}\tope{P}_\nu - \eta_{\lambda\nu}\tope{P}_\mu ) .
\\			\label{5.11}
& [ \tope{M}_{\varkappa\lambda} , \tope{M}_{\mu\nu} ]_{\_}
=
- \ih \bigl\{
\eta_{\varkappa\mu}	\tope{M}_{\lambda\nu} -
\eta_{\lambda\mu}	\tope{M}_{\varkappa\nu} -
\eta_{\varkappa\nu}	\tope{M}_{\lambda\mu} +
\eta_{\lambda\nu}	\tope{M}_{\varkappa\mu}
\bigr\} .
	\end{align}
We would like to pay attention to the minus sign in the multiplier $(-\ih)$
in~\eref{5.10} and~\eref{5.11} with respect to the above references, where
$\ih$ stands instead of $-\ih$ in these equations. When (a representation of)
the Lie algebra of the Poincar\'e group is considered, this difference in the
sign is insignificant as it can be absorbed into the definition of
$\tope{M}_{\mu\nu}$. However, the change of the sign of the angular momentum
operator, $\tope{M}_{\mu\nu}\mapsto -\tope{M}_{\mu\nu}$, will result in the
change $\ih\mapsto -\ih$ in the r.h.s.\ of~\eref{5.3}. This means that
equations~\eref{5.9}, \eref{5.10} and~\eref{5.3}, when considered together,
require a suitable choice of the signs of the multiplier $\ih$ in their
right hand sides as these signs change simultaneously when
$\tope{M}_{\mu\nu}$  is replaced with $-\tope{M}_{\mu\nu}$. Since
equations~\eref{5.3}, \eref{5.10} and~\eref{5.11} hold, when
$\tope{M}_{\mu\nu}$ is defined according to the Noether's theorem and the
ordinary (anti)commutation relations are
valid~\cite{bp-QFTinMP-scalars,bp-QFTinMP-spinors,bp-QFTinMP-vectors}, we
accept these equations in the way they are written above.

	To the relations~\eref{5.9}--\eref{5.11} should be added the
equations~\cite[p.~78]{Roman-QFT}
	\begin{align}
			\label{5.12}
& [ \tope{Q},\tope{P}_\mu ]_{\_} = 0
\\			\label{5.13}
& [ \tope{Q},\tope{M}_{\mu\nu} ]_{\_} = 0 ,
	\end{align}
which complete the algebra of observables and express, respectively, the
translational and rotational invariance of the charge operator $\tope{Q}$;
physically they mean that the charge and momentum  or the charge and angular
momentum  are simultaneously measurable quantities.

	Since the spin properties of a system are generally independent of
its charge or momentum, one may also expect the validity of the relations%
\footnote{~%
Recall, $\tope{S}_{\mu\nu}$ (resp.\ $\tope{L}_{\mu\nu}$) is the conserved
spin (resp.\ orbital) operator, not the generally non\ndash conserved
spin (resp.\ orbital) angular momentum
operator~\cite{bp-QFT-angular-momentum-operator}.%
}
	\begin{align}
			\label{5.14}
& [ \tope{S}_{\mu\nu},\tope{P}_\mu ]_{\_} = 0
\\			\label{5.15}
& [ \tope{S}_{\mu\nu},\tope{Q} ]_{\_} = 0 .
	\end{align}
But, as the spin describes, in a sense, some of the rotational properties of
the system, equality like
$ [ \tope{S}_{\mu\nu}, \tope{L}_{\varkappa\lambda} ]_{\_} = 0$
is not likely to hold. Indeed, the considerations
in~\cite{bp-QFTinMP-scalars,bp-QFTinMP-spinors,bp-QFTinMP-vectors} reveal
that~\eref{5.14} and~\eref{5.15}, but not the last equation, are true in the
framework of the Lagrangian formalism with added to it standard
(anti)commutation relations. Notice, if~\eref{5.14} and~\eref{5.15} hold,
then, respectively,~\eref{5.10} and~\eref{5.13} are equivalent to
	\begin{align}
			\label{5.16}
& [ \tope{L}_{\mu\nu} , \tope{P}_\lambda ]_{\_}
 = - \ih ( \eta_{\lambda\mu}\tope{P}_\nu - \eta_{\lambda\nu}\tope{P}_\mu ) .
\\			\label{5.17}
& [ \tope{Q},\tope{L}_{\mu\nu} ]_{\_} = 0 .
	\end{align}

	It is intuitively clear, not all of the commutation
relations~\eref{5.1}--\eref{5.3} and~\eref{5.9}--\eref{5.15} are independent:
if $\tope{D}$ denotes some of the operators
$\tope{P}_\mu$, $\tope{Q}$, $\tope{M}_{\mu\nu}$, $\tope{S}_{\mu\nu}$ or
$\tope{L}_{\mu\nu}$ and the commutators
 $[\tope{\varphi}_{i}(x),\tope{D}]_{\_}$, $i=1,\dots,N$, are known, then, in
principle, one can calculate the commutators
$[\Gamma(\tope{\varphi}_{1}(x),\dots,\tope{\varphi}_{N}(x)),\tope{D}]_{\_}$,
where $\Gamma(\tope{\varphi}_{1}(x),\dots,\tope{\varphi}_{N}(x))$ is, for
example, any function/functional bilinear in
$\tope{\varphi}_{1}(x),\dots,\tope{\varphi}_{N}(x)$;
to prove this fact, one should apply the identity
 $[A,B\circ C]_{\_}=[A,B]_{\_}\circ C + B\circ [A,C]_{\_}$
a suitable number of times. In particular, if $\tope{D}_1$ and $\tope{D}_2$
denote any two (distinct) operators of the dynamical variables, and
$[\tope{\varphi}_{i}(x),\tope{D}_1]_{\_}$
is known, then the commutator $[\tope{D}_{1},\tope{D}_2]_{\_}$ can be
calculated explicitly. For this reason, we can expect that:\\
	\indent\hphantom{i}
	(i) Equation~\eref{5.1} implies~\eref{5.9}, \eref{5.10},
\eref{5.12}, \eref{5.14} and~\eref{5.16}.\\
	\indent\hphantom{}
	(ii) Equation~\eref{5.2} implies~\eref{5.12},
\eref{5.13}, \eref{5.15}, and~\eref{5.17}.\\
	\indent
	(iii) Equation~\eref{5.3} implies~\eref{5.10},
\eref{5.11}, and~\eref{5.13}.\\
Besides,~\eref{5.3} may, possibly, entail equations like~\eref{5.11} with $S$
or $L$ for $M$, with an exception of $\tope{M}_{\mu\nu}$ in the l.h.s., \ie
	\begin{equation}	\label{5.18}
	\begin{split}
[ \tope{S}_{\varkappa\lambda} , \tope{M}_{\mu\nu} ]_{\_}
& =
- \ih \bigl\{
\eta_{\varkappa\mu}	\tope{S}_{\lambda\nu} -
\eta_{\lambda\mu}	\tope{S}_{\varkappa\nu} -
\eta_{\varkappa\nu}	\tope{S}_{\lambda\mu} +
\eta_{\lambda\nu}	\tope{S}_{\varkappa\mu}
\bigr\}
\\
[ \tope{L}_{\varkappa\lambda} , \tope{M}_{\mu\nu} ]_{\_}
& =
- \ih \bigl\{
\eta_{\varkappa\mu}	\tope{L}_{\lambda\nu} -
\eta_{\lambda\mu}	\tope{L}_{\varkappa\nu} -
\eta_{\varkappa\nu}	\tope{L}_{\lambda\mu} +
\eta_{\lambda\nu}	\tope{L}_{\varkappa\mu}
\bigr\}
	\end{split}
	\end{equation}
The validity of assertions (i)--(iii) above for free scalar, spinor and
vector fields, when respectively
	\begin{subequations}	\label{5.19}
	\begin{align}
				\label{5.19a}
&
\tope{\varphi}_i(x) \mapsto \tope{\varphi}(x), \tope{\varphi}^\dag(x)
\quad
I_{i\mu\nu}^{i'} \mapsto I_{\mu\nu} = 0
\quad
{e}(\tope{\varphi}) = - {e}(\tope{\varphi}^\dag) = +1
\\				\label{5.19b}
&
\tope{\varphi}_i(x) \mapsto \tope{\psi}(x), \tope{\bpsi}(x)
\quad
I_{i\mu\nu}^{i'} \mapsto I_{\psi\mu\nu} = I_{\bpsi\mu\nu}
= - \frac{\iu}{2} \sigma_{\mu\nu}
\quad
{e}(\tope{\psi}) = - {e}(\tope{\bpsi}) = +1
\\				\label{5.19c}
&
\tope{\varphi}_i(x) \mapsto \tope{U}_\mu(x), \tope{U}_\mu^\dag(x)
\quad
I_{i\mu\nu}^{i'} \mapsto I_{\rho\mu\nu}^\sigma = I_{\rho\mu\nu}^{\dag\,\sigma}
= \delta_{\mu}^{\sigma} \eta_{\nu\rho} -
  \delta_{\nu}^{\sigma} \eta_{\mu\rho}
\quad
{e}(\tope{U}_\mu) = - {e}(\tope{U}_\mu^\dag) = +1 ,
       \end{align}
	\end{subequations}
where $\sigma^{\mu\nu}:=\frac{\iu}{2}[\gamma^\mu,\gamma^\nu]_{\_}$ with
$\gamma^\mu$ being the Dirac $\gamma$\ndash
matrices~\cite{Bjorken&Drell-1,Bogolyubov&Shirkov}, is proved
in~\cite{bp-QFTinMP-scalars,bp-QFTinMP-spinors,bp-QFTinMP-vectors},
respectively. Besides, in \emph{loc.\ cit.}\ is proved that
equations~\eref{5.18} hold for scalar and vector fields, but not for a
spinor field.%
\footnote{~%
The problem for the validity of assertions~(i)--(iii) or
equations~\eref{5.18} in the general case of arbitrary fields (Lagrangians) is
not a subject of the present work.%
}

	Thus, we see that the Heisenberg relations~\eref{5.1}--\eref{5.3} are
stronger than the commutation relations~\eref{5.9}--\eref{5.17}, when imposed
on the Lagrangian formalism as subsidiary restrictions.


\section{Types of possible commutation relations}
\label{Sect6}

	In a broad sense, by a \emph{commutation relation} we shall
understand any \emph{algebraic} relation between the creation and
annihilation operators imposed as  subsidiary restriction on the Lagrangian
formalism. In a narrow sense, the \emph{commutation relations} are the
equations~\eref{6.12}, with $\varepsilon=-1$, written below and satisfied by
the bose creation and annihilation operators. As \emph{anticommutation
relations} are known the equations~\eref{6.12}, with $\varepsilon=+1$,
written below and satisfied by the fermi creation and annihilation operators.
The last two types of relations are often referred as the \emph{bilinear
commutation relations}~\cite{Ohnuki&Kamefuchi}. Theoretically are possible
also \emph{trilinear commutation relations}, an example being the
\emph{paracommutation relations}~\cite{Green-1953,Ohnuki&Kamefuchi}
represented below by equations~\eref{6.16} (or~\eref{6.18}).

	Generally said, the commutation relations should be postulated.
Alternatively, they could be derived from  (equivalent to them) different
assumptions added to the Lagrangian formalism. The purpose of this section is
to be explored possible classes of commutation relations, which follow from
some natural restrictions on the Lagrangian formalism that are consequences
from the considerations in the previous sections. Special attention will be
paid on some consequences of the charge symmetric Lagrangians as the free
fields possess such a
symmetry~\cite{Bjorken&Drell-2,Bogolyubov&Shirkov,Roman-QFT,Itzykson&Zuber}.

	As pointed in Sect~\ref{Sect3}, the Euler-Lagrange equations for the
Lagrangians $\tope{L}^{\prime}$, $\tope{L}^{\prime\prime}$ and
$\tope{L}^{\prime\prime\prime}$ coincide and, in quantum field theory, the
role of these equations is to be singled out the independent degrees of
freedom of the fields in the form of creation and annihilation operators
$a_s^{\pm}(\bk)$ and $a_s^{\dag\,\pm}(\bk)$ (which are identical for
$\tope{L}^{\prime}$, $\tope{L}^{\prime\prime}$ and
$\tope{L}^{\prime\prime\prime}$). Further specialization of these operators
is provided by the commutation relations (in broad sense) which play a role
of field equations in this situation (with respect to the mentioned
operators).

	Before proceeding on, we would like to simplify our notation. As a
spin variable, $s$ say, is always coupled with a 3\ndash momentum one, $\bk$
say, we shall use the letters $l$, $m$ and $n$ to denote pairs like
$l=(s,\bk)$, $m=(t,\bs p)$ and $n=(r,\bs q)$. Equipped with this convention,
we shall write, e.g., $a_l^{\pm}$ for $a_s^{\pm}(\bk)$  and $a_l^{\dag\,\pm}$
for $a_s^{\dag\,\pm}(\bk)$.
We set $\delta_{lm}:=\delta_{st}\delta^3(\bk-\bs p)$
and a summation sign like $\sum_{l}$ should be understood as
$\sum_{s}\int\Id^3 \bk$, where the range of the polarization variable $s$
will be clear from the context (see, e.g.,~\eref{3.7}--\eref{3.10}).

\subsection{Restrictions related to the momentum operator}
\label{Subsect6.1}

	First of all, let us examine the consequences of the Heisenberg
relation~\eref{5.1} involving the momentum operator. Since in terms of
creation and annihilation operators it reads~\cite{Bogolyubov&Shirkov,
bp-QFTinMP-scalars,bp-QFTinMP-spinors,bp-QFTinMP-vectors}
	\begin{align}
			\label{6.1}
[a_s^\pm(\bk),\ope{P}_\mu]_{\_} = \mp k_\mu a_s^\pm(\bk)
\quad
[a_s^{\dag\,\pm}(\bk),\ope{P}_\mu]_{\_} = \mp k_\mu a_s^{\dag\,\pm}(\bk)
\qquad
k_0 =\sqrt{m^2c^2+{\bs k}^2} ,
	\end{align}
the field equations in terms of creation and annihilation operators for the
Lagrangians~\eref{3.1}, \eref{3.3} and~\eref{3.4} respectively are
(see~\cite{bp-QFTinMP-scalars,bp-QFTinMP-spinors,bp-QFTinMP-vectors}
or~\eref{6.1} and~\eref{3.7}):
	\begin{subequations}	\label{6.2}
	\begin{align}
			\label{6.2a}
	\begin{split}
\sum_{t=1}^{2j+1-\delta_{0m}(1-\delta_{0j})} \int
q_\mu\big|_{q_0=\sqrt{m^2c^2+{\bs q}^2}}
\bigl\{
\bigl[ a_s^{\pm}(\bk)
& ,
a_t^{\dag\,+}(\bs q) \circ a_t^{-}(\bs q) + \varepsilon
a_t^{\dag\,-}(\bs q) \circ a_t^{+}(\bs q)
\bigr]_{-}
\\
& \pm (1+\tau) a_s^{\pm}(\bk) \delta_{st} \delta^3(\bk-\bs q)
\bigr\} \Id^3\bs q = 0
	\end{split}
\\			\label{6.2b}
	\begin{split}
\sum_{t=1}^{2j+1-\delta_{0m}(1-\delta_{0j})} \int
q_\mu\big|_{q_0=\sqrt{m^2c^2+{\bs q}^2}}
\bigl\{
\bigl[ a_s^{\dag\,\pm}(\bk)
& ,
a_t^{\dag\,+}(\bs q) \circ a_t^{-}(\bs q) + \varepsilon
a_t^{\dag\,-}(\bs q) \circ a_t^{+}(\bs q)
\bigr]_{-}
\\
& \pm (1+\tau) a_s^{\dag\,\pm}(\bk) \delta_{st} \delta^3(\bk-\bs q)
\bigr\} \Id^3\bs q = 0
	\end{split}
	\end{align}
	\end{subequations}	
\vspace{-3ex}
	\begin{subequations}	\label{6.3}
	\begin{align}
			\label{6.3a}
	\begin{split}
\sum_{t=1}^{2j+1-\delta_{0m}(1-\delta_{0j})} \int
q_\mu\big|_{q_0=\sqrt{m^2c^2+{\bs q}^2}}
\bigl\{
\bigl[ a_s^{\pm}(\bk)
& ,
a_t^{+}(\bs q) \circ a_t^{\dag\,-}(\bs q) + \varepsilon
a_t^{-}(\bs q) \circ a_t^{\dag\,+}(\bs q)
\bigr]_{-}
\\
& \pm (1+\tau) a_s^{\pm}(\bk) \delta_{st} \delta^3(\bk-\bs q)
\bigr\} \Id^3\bs q = 0
	\end{split}
\\			\label{6.3b}
	\begin{split}
\sum_{t=1}^{2j+1-\delta_{0m}(1-\delta_{0j})} \int
q_\mu\big|_{q_0=\sqrt{m^2c^2+{\bs q}^2}}
\bigl\{
\bigl[ a_s^{\dag\,\pm}(\bk)
& ,
a_t^{+}(\bs q) \circ a_t^{\dag\,-}(\bs q) + \varepsilon
a_t^{-}(\bs q) \circ a_t^{\dag\,+}(\bs q)
\bigr]_{-}
\\
& \pm (1+\tau) a_s^{\dag\,\pm}(\bk) \delta_{st} \delta^3(\bk-\bs q)
\bigr\} \Id^3\bs q = 0
	\end{split}
	\end{align}
	\end{subequations}      
\vspace{-3ex}
	\begin{subequations}	\label{6.4}
	\begin{align}
			\label{6.4a}
	\begin{split}
\sum_{t=1}^{2j+1-\delta_{0m}(1-\delta_{0j})} \int
q_\mu\big|_{q_0=\sqrt{m^2c^2+{\bs q}^2}}
\bigl\{
\bigl[ a_s^{\pm}(\bk)
& ,
[ a_t^{\dag\,+}(\bs q) , a_t^{-}(\bs q) ]_\varepsilon +
[ a_t^{+}(\bs q) , a_t^{\dag\,-}(\bs q) ]_\varepsilon
\bigr]_{-}
\\
& \pm (1+\tau) a_s^{\pm}(\bk) \delta_{st} \delta^3(\bk-\bs q)
\bigr\} \Id^3\bs q = 0
	\end{split}
\\			\label{6.4b}
	\begin{split}
\sum_{t=1}^{2j+1-\delta_{0m}(1-\delta_{0j})} \int
q_\mu\big|_{q_0=\sqrt{m^2c^2+{\bs q}^2}}
\bigl\{
\bigl[ a_s^{\dag\,\pm}(\bk)
& ,
[ a_t^{\dag\,+}(\bs q) , a_t^{-}(\bs q) ]_\varepsilon +
[ a_t^{+}(\bs q) , a_t^{\dag\,-}(\bs q) ]_\varepsilon
\bigr]_{-}
\\
& \pm (1+\tau) a_s^{\dag\,\pm}(\bk) \delta_{st} \delta^3(\bk-\bs q)
\bigr\} \Id^3\bs q = 0 ,
	\end{split}
	\end{align}
	\end{subequations}	
where $j$ and $\varepsilon$ are given via~\eref{3.12}, the generalized
commutation function $[\cdot,\cdot]_{\varepsilon}$ is defined by~\eref{4.14},
and the polarization indices take the values
	\begin{equation}	\label{6.6}
s,t = 1,\dots,2j+1-\delta_{0m}(1-\delta_{0j})
=
	\begin{cases}
1	&\text{for $j=0$ or for $j=\frac{1}{2}$ and $m=0$} \\
1,2	&\text{for $j=\frac{1}{2}$ and $m\not=0$ or for $j=1$ and $m=0$} \\
1,2,3	&\text{for $j=1$ and $m\not=0$}
	\end{cases}
.
	\end{equation}
The ``b'' versions of the equations~\eref{6.2}--\eref{6.4} are consequences of
the ``a'' versions and the equalities
	\begin{gather}	\label{6.7}
(a_l^{\pm})^\dag = a_l^{\dag\,\mp}\quad
(a_l^{\dag\,\pm}) = a_l^{\mp}
\\			\label{6.8}
\bigl( [A,B]_\eta \bigr)^\dag
=
\eta [A^\dag,B^\dag]_\eta
\qquad\text{for }
[A,B]_\eta = \eta [B,A]_\eta
\qquad \eta=\pm1  .
	\end{gather}

	Applying~\eref{6.2}--\eref{6.4} and the identity
	\begin{equation}	\label{6.13}
[A,B\circ C]_{\_}  = [A,B]_\eta \circ C - \eta B\circ [A,C]_\eta
\qquad\text{for } \eta=\pm1
	\end{equation}
for the choice $\eta=-1$, one can prove by a direct calculation that
	\begin{equation}	\label{6.5}
	\begin{split}
&
[ \tope{P}_\mu , \tope{P}_\nu ]_{\_} = 0		\quad
[ \tope{Q} , \tope{P}_\mu ]_{\_} = 0                    \quad
[ \tope{S}_{\mu\nu} , \tope{P}_\lambda ]_{\_} = 0       \quad
\\
&
[ \tope{L}_{\mu\nu} , \tope{P}_\lambda ]_{\_}
= -\ih \{ \eta_{\lambda\mu} \tope{P}_\nu - \eta_{\lambda\nu} \tope{P}_\mu \}
\quad
[ \tope{M}_{\mu\nu} , \tope{P}_\lambda ]_{\_}
= -\ih \{ \eta_{\lambda\mu} \tope{P}_\nu - \eta_{\lambda\nu} \tope{P}_\mu \}
,
	\end{split}
	\end{equation}
where the operators
$\tope{P}_{\mu}$, $\tope{Q}$, $\tope{S}_{\mu\nu}$, $\tope{L}_{\mu\nu}$, and
$\tope{M}_{\mu\nu}$
denote the momentum, charge, spin, orbital and total angular momentum
operators, respectively, of the system considered and are calculated from one
and the same initial Lagrangian. This result confirms the supposition, made in
Sect.~\ref{Sect5}, that the assertion~(i) before~\eref{5.18} holds for the
fields investigated here.

	Below we shall study only those solutions of~\eref{6.2}--\eref{6.4}
for which the integrands in them vanish, \ie we shall replace the systems of
\emph{integral} equations~\eref{6.2}--\eref{6.4} with the following systems of
\emph{algebraic} equations (see the above convention on the indices $l$ and
$m$ and do not sum over indices repeated on one and the same level):
	\begin{subequations}	\label{6.9}
	\begin{align}
			\label{6.9a}
	\begin{split}
\bigl[ a_l^{\pm}
& ,
a_m^{\dag\,+} \circ a_m^{-} + \varepsilon
a_m^{\dag\,-} \circ a_m^{+}
\bigr]_{-}
\pm (1+\tau) \delta_{lm} a_l^{\pm} = 0
	\end{split}
\\ 			\label{6.9b}
	\begin{split}
\bigl[ a_l^{\dag\,\pm}
& ,
a_m^{\dag\,+} \circ a_m^{-} + \varepsilon
a_m^{\dag\,-} \circ a_m^{+}
\bigr]_{-}
\pm (1+\tau) \delta_{lm} a_l^{\dag\,\pm} = 0
	\end{split}
	\end{align}
	\end{subequations}	
\vspace{-5ex}
	\begin{subequations}	\label{6.10}
	\begin{align}
			\label{6.10a}
	\begin{split}
\bigl[ a_l^{\pm}
& ,
a_m^{+} \circ a_m^{\dag\,-} + \varepsilon
a_m^{-} \circ a_m^{\dag\,+}
\bigr]_{-}
\pm (1+\tau) \delta_{lm} a_l^{\pm} = 0
	\end{split}
\\			\label{6.10b}
	\begin{split}
\bigl[ a_l^{\dag\,\pm}
& ,
a_m^{+} \circ a_m^{\dag\,-} + \varepsilon
a_m^{-} \circ a_m^{\dag\,+}
\bigr]_{-}
\pm (1+\tau) \delta_{lm} a_l^{\dag\,\pm} = 0
	\end{split}
	\end{align}
	\end{subequations}      
\vspace{-5ex}
	\begin{subequations}	\label{6.11}
	\begin{align}
			\label{6.11a}
	\begin{split}
\bigl[ a_l^{\pm}
& ,
[ a_m^{\dag\,+} , a_m^{-} ]_\varepsilon +
[ a_m^{+} , a_m^{\dag\,-} ]_\varepsilon
\bigr]_{-}
\pm 2 (1+\tau) \delta_{lm} a_l^{\pm} = 0
	\end{split}
\\			\label{6.11b}
	\begin{split}
\bigl[ a_l^{\dag\,\pm}
& ,
[ a_m^{\dag\,+} , a_m^{-} ]_\varepsilon +
[ a_m^{+} , a_m^{\dag\,-} ]_\varepsilon
\bigr]_{-}
\pm 2 (1+\tau) \delta_{lm} a_l^{\dag\,\pm} = 0 .
	\end{split}
	\end{align}
	\end{subequations}	
It seems, these are the most general and sensible \emph{trilinear commutation
relations} one may impose on the creation and annihilation operators.

	First of all, we should mentioned that the \emph{standard bilinear
commutation relations},
viz.~\cite{Bjorken&Drell-2,Bogolyubov&Shirkov,Roman-QFT,Itzykson&Zuber,
bp-QFTinMP-scalars,bp-QFTinMP-spinors,bp-QFTinMP-vectors}
	\begin{align}	\notag
&[a_{l}^{\pm}, a_{m}^{\pm} ]_{-\varepsilon}
	= 0
&&
[a_{l}^{\dag\,\pm}, a_{m}^{\dag\,\pm} ]_{-\varepsilon}
	= 0
\\	\notag
&[a_{l}^{\mp}, a_{m}^{\pm} ]_{-\varepsilon}
	= (\pm1)^{2j+1} \tau \delta_{lm} \id_\Hil
&&
[a_{l}^{\dag\,\mp}, a_{m}^{\dag\,\pm} ]_{-\varepsilon}
	= (\pm1)^{2j+1} \tau \delta_{lm} \id_\Hil
\\	\notag
&[a_{l}^{\pm}, a_{m}^{\dag\,\pm} ]_{-\varepsilon}
	= 0
&&
[a_{l}^{\dag\,\pm}, a_{m}^{\pm} ]_{-\varepsilon}
	= 0
\\	\label{6.12}
&[a_{l}^{\mp}, a_{m}^{\dag\,\pm} ]_{-\varepsilon}
	= (\pm1)^{2j+1} \delta_{lm} \id_\Hil
&&
[a_{l}^{\dag\,\mp}, a_{m}^{\pm} ]_{-\varepsilon}
	= (\pm1)^{2j+1} \delta_{lm} \id_\Hil ,
	\end{align}
provide a solution of any one of the equations~\eref{6.9}--\eref{6.11} in a
sense that, due to~\eref{3.12} and~\eref{6.13}, with $\eta=-\varepsilon$
any set of operators satisfying~\eref{6.12}
converts~\eref{6.9}--\eref{6.11} into identities.

	Besides, this conclusion remains valid also if the normal ordering is
taken into account, \ie if, in this particular case, the changes
\(
a_{m}^{\dag\,-} \circ a_{m}^{+}
\mapsto
\varepsilon a_{m}^{+} \circ a_{m}^{\dag\,-}
\)
and
\(
a_{m}^{-} \circ a_{m}^{\dag\,+}
\mapsto
\varepsilon a_{m}^{\dag\,+} \circ a_{m}^{-}
\)
are made in~\eref{6.9}--\eref{6.11}.

	Now we shall demonstrate how the trilinear relations~\eref{6.11} lead
to the paracommutation relations. Equations~\eref{6.11} can be `split' into
different kinds of trilinear commutation relations into infinitely many ways.
For example, the system of equations
	\begin{subequations}	\label{6.014}
	\begin{align}
			\label{6.014a}
	\begin{split}
\bigl[ a_l^{\pm} & , [ a_m^{+} , a_m^{\dag\,-} ]_\varepsilon \bigr]_{-}
\pm (1+\tau) \delta_{lm} a_l^{\pm} = 0
	\end{split}
\\			\label{6.014b}
	\begin{split}
\bigl[ a_l^{\pm} & , [ a_m^{\dag\,+} , a_m^{-} ]_\varepsilon \bigr]_{-}
\pm (1+\tau) \delta_{lm} a_l^{\pm} = 0
	\end{split}
\\			\label{6.014c}
	\begin{split}
\bigl[ a_l^{\dag\,\pm} & , [ a_m^{+} , a_m^{\dag\,-} ]_\varepsilon \bigr]_{-}
\pm (1+\tau) \delta_{lm} a_l^{\dag\,\pm} = 0
	\end{split}
\\			\label{6.014d}
	\begin{split}
\bigl[ a_l^{\dag\,\pm} & , [ a_m^{\dag\,+} , a_m^{-} ]_\varepsilon \bigr]_{-}
\pm (1+\tau) \delta_{lm} a_l^{\dag\,\pm} = 0
	\end{split}
	\end{align}
	\end{subequations}	
provides an evident solution of~\eref{6.11}. However, it is a simple algebra
to be seen that these relations are incompatible with the standard
(anti)commutation relations~\eref{6.12} and, in this sense, are not suitable
as subsidiary restrictions on the Lagrangian formalism. For our purpose, the
equations
	\begin{subequations}	\label{6.14}
	\begin{align}
			\label{6.14a}
	\begin{split}
\bigl[ a_l^{+} & , [ a_m^{+} , a_m^{\dag\,-} ]_\varepsilon \bigr]_{-}
+2 \delta_{lm} a_l^{+} = 0
	\end{split}
\\			\label{6.14b}
	\begin{split}
\bigl[ a_l^{+} & , [ a_m^{\dag\,+} , a_m^{-} ]_\varepsilon \bigr]_{-}
+ 2 \tau \delta_{lm} a_l^{+} = 0
	\end{split}
\\			\label{6.14c}
	\begin{split}
\bigl[ a_l^{-} & , [ a_m^{+} , a_m^{\dag\,-} ]_\varepsilon \bigr]_{-}
- 2 \tau \delta_{lm} a_l^{-} = 0
	\end{split}
\\			\label{6.14d}
	\begin{split}
\bigl[ a_l^{-} & , [ a_m^{\dag\,+} , a_m^{-} ]_\varepsilon \bigr]_{-}
-2 \delta_{lm} a_l^{-} = 0
	\end{split}
	\end{align}
	\end{subequations}	
and their Hermitian conjugate provide a solution of~\eref{6.11}, which is
compatible with~\eref{6.12}, \ie if~\eref{6.12} hold, the
equations~\eref{6.14} are converted into identities.

	The idea of the paraquantization is in the following generalization
of~\eref{6.14}
	\begin{subequations}	\label{6.140}
	\begin{align}
			\label{6.140a}
	\begin{split}
\bigl[ a_l^{+} & , [ a_m^{+} , a_n^{\dag\,-} ]_\varepsilon \bigr]_{-}
+2 \delta_{ln} a_m^{+} = 0
	\end{split}
\\			\label{6.140b}
	\begin{split}
\bigl[ a_l^{+} & , [ a_m^{\dag\,+} , a_n^{-} ]_\varepsilon \bigr]_{-}
+ 2 \tau \delta_{ln} a_m^{+} = 0
	\end{split}
\\			\label{6.140c}
	\begin{split}
\bigl[ a_l^{-} & , [ a_m^{+} , a_n^{\dag\,-} ]_\varepsilon \bigr]_{-}
- 2 \tau \delta_{lm} a_n^{-} = 0
	\end{split}
\\			\label{6.140d}
	\begin{split}
\bigl[ a_l^{-} & , [ a_m^{\dag\,+} , a_n^{-} ]_\varepsilon \bigr]_{-}
-2 \delta_{lm} a_n^{-} = 0
	\end{split}
	\end{align}
	\end{subequations}	
which reduces to~\eref{6.14} for $n=m$  and is a generalization of~\eref{6.12}
in a sense that any set of operators satisfying~\eref{6.12}
converts~\eref{6.140} into identities, the opposite being generally not
valid.%
\footnote{~%
Other generalizations of~\eref{6.14} are also possible, but they do not agree
with~\eref{6.12}. Moreover, it is easy to be proved, any other (non\ndash
trivial) arrangement of the indices in~\eref{6.140} is incompatible
with~\eref{6.12}.%
}

	Suppose that the field considered consists of a single sort of
particles, \eg electrons or photons, created by $b_l^\dag:=a_l^{\dag}$ and
annihilated by $b_l:=a_l^{\dag\,-}$. Then the equation Hermitian conjugated
to~\eref{6.14a} reads
	\begin{equation}	\label{6.15}
[b_l , [ b_m^\dag, b_m ]_\varepsilon ]_{\_} = 2 \delta_{lm} b_m .
	\end{equation}
This is the main relation from which the paper~\cite{Green-1953} starts. The
basic \emph{paracommutation relations}
are~\cite{Green-1953,Greenberg&Messiah-1965,Ohnuki&Kamefuchi,Govorkov}:
	\begin{subequations}	\label{6.16}
	\begin{align}	\label{6.16a}
[b_l , [ b_m^\dag, b_n ]_\varepsilon ]_{\_} & = 2 \delta_{lm} b_n
\\			\label{6.16b}
[b_l , [ b_m     , b_n ]_\varepsilon ]_{\_} & = 0 .
	\end{align}
	\end{subequations}
The first of them is a generalization (stronger version) of~\eref{6.15} by
replacing the second index $m$ with an arbitrary one, say $n$, and the second
one is added (by "hands") in the theory as an additional assumption.
Obviously,~\eref{6.16} are a solution of~\eref{6.14} and therefore
of~\eref{6.11} in the considered case of a field consisting of only one sort
of particles.

	The equations~\eref{6.14} contain also the relativistic version of
the paracommutation relations, when the existence of antiparticles must be
respected~\cite[sec.~18.1]{Ohnuki&Kamefuchi}. Indeed, noticing that the
field's particles (resp.\ antiparticles) are created by
$b_l^\dag:=a_l^{+}$ (resp.\ $c_l^\dag:=a_l^{\dag\,+}$) and annihilated by
$b_l:=a_l^{\dag\,-}$ (resp.\ $c_l:=a_l^{-}$),
from~\eref{6.14} and the Hermitian conjugate to them equations, we get
	\begin{subequations}	\label{6.17}
	\begin{alignat}{2}	\label{6.17a}
[b_l , [ b_m^\dag, b_m ]_\varepsilon ]_{\_} & = 2 \delta_{lm} b_m  &\quad
[c_l , [ c_m^\dag, c_m ]_\varepsilon ]_{\_} & = 2 \delta_{lm} c_m
\\			\label{6.17b}
[b_l^\dag , [ c_m^\dag, c_m ]_\varepsilon ]_{\_} &=-2\tau\delta_{lm} b_m^\dag
&\quad
[c_l^\dag , [ b_m^\dag, b_m ]_\varepsilon ]_{\_} & =-2\tau\delta_{lm} c_m^\dag
.
	\end{alignat}
	\end{subequations}
Generalizing these equations in a way similar to the transition
from~\eref{6.15} to~\eref{6.16}, we obtain the \emph{relativistic
paracommutation relations} as (cf.~\eref{6.140})
	\begin{subequations}	\label{6.18}
	\begin{alignat}{2}	\label{6.18a}
[b_l , [ b_m^\dag, b_n ]_\varepsilon ]_{\_} & = 2 \delta_{lm} b_n  &\quad
[b_l , [ b_m     , b_n ]_\varepsilon ]_{\_} & = 0
\\				\label{6.18b}
[c_l , [ c_m^\dag, c_n ]_\varepsilon ]_{\_} & = 2 \delta_{lm} c_n  &\quad
[c_l , [ c_m     , c_n ]_\varepsilon ]_{\_} & = 0
\\ 				\label{6.18c}
[b_l^\dag , [ c_m^\dag, c_n ]_\varepsilon ]_{\_} & =-2\tau\delta_{ln} b_m^\dag
&\quad
[c_l^\dag , [ b_m^\dag, b_n ]_\varepsilon ]_{\_} & =-2\tau\delta_{ln} c_m^\dag
.
	\end{alignat}
	\end{subequations}
The equations~\eref{6.18a} (resp.~\eref{6.18b}) represent the paracommutation
relations for the field's particles (resp.\ antiparticles) as independent
objects, while~\eref{6.18c} describe a pure relativistic effect of some
``interaction'' (or its absents) between field's particles and antiparticles
and fixes the paracommutation relations involving the $b_l$'s and $c_l$'s, as
pointed in~\cite[p.~207]{Ohnuki&Kamefuchi} (where $b_l$ is denoted by $a_l$
and $c_l$ by $b_l$). The relations~\eref{6.15} and~\eref{6.18} for
$\varepsilon=+1$ (resp.\ $\varepsilon=-1$) are referred as the
\emph{parabose} (resp.\ \emph{parafermi})
\emph{commutation relations}~\cite{Ohnuki&Kamefuchi}. This terminology is a
natural one also with respect to the commutation relations~\eref{6.140},
which will be referred as the \emph{paracommutation relations} too.

	As first noted in~\cite{Green-1953}, the equations~\eref{6.12}
provide a solution of~\eref{6.18} (or~\eref{6.16} in the nonrelativistic
case) but the latter equations admit also an infinite number of other
solutions. Besides, by taking Hermitian conjugations of (some of) the
equations~\eref{6.16} or~\eref{6.18} and applying generalized Jacobi
identities, like
	\begin{equation}	\label{6.19}
	\begin{split}
 &
\alpha  [ [ A,B]_{\xi} , C ]_{\eta} +
\xi\eta [ [ A,C]_{-\alpha/\xi} , B ]_{-\alpha/\eta} -
\alpha^2 [ [ B,C]_{\xi\eta/\alpha} , A ]_{1/\alpha}
= 0
\quad \alpha\xi\eta \not=0
\\ &
\beta  [A,[B,C ]_{\alpha}, ]_{-\beta\gamma} +
\gamma [B,[C,A ]_{\beta}, ]_{-\gamma\alpha} +
\alpha [C,[A,B ]_{\gamma}, ]_{-\alpha\beta}
= 0
\quad \alpha,\beta,\gamma = \pm 1
\\ &
[[A,B]_{\eta}, C]_{-} +
[[B,C]_{\eta}, A]_{-} +
[[C,A]_{\eta}, B]_{-}
= 0
\quad \eta = \pm 1
\\ &
[[A,B]_{\xi}, [C,D]_{\eta}]_{-} =
[[A,B]_{\xi}, C]_{-}, D]_{\eta} + \eta
[[A,B]_{\xi}, D]_{-}, C]_{1/\eta}
\quad \eta \not = 0 ,
	\end{split}
	\end{equation}
one can obtain a number of other (para)commutation relations for which the
reader is referred
to~\cite{Green-1953,Ohnuki&Kamefuchi,Govorkov}.

	Of course, the paracommutation relations~\eref{6.140}, in
particular~\eref{6.16} and~\eref{6.18} as their stronger versions, do not
give the general solution of the trilinear relations~\eref{6.11}. For
instance, one may replace~\eref{6.11} with the equations
	\begin{subequations}	\label{6.20}
	\begin{align}
			\label{6.20a}
	\begin{split}
\bigl[ a_l^{+}
& ,
[ a_m^{\dag\,+} , a_n^{-} ]_\varepsilon +
[ a_m^{+} , a_n^{\dag\,-} ]_\varepsilon
\bigr]_{-}
+ 2 (1+\tau) \delta_{ln} a_m^{+} = 0
	\end{split}
\\			\label{6.20b}
	\begin{split}
\bigl[ a_l^{-}
& ,
[ a_m^{\dag\,+} , a_n^{-} ]_\varepsilon +
[ a_m^{+} , a_n^{\dag\,-} ]_\varepsilon
\bigr]_{-}
- 2 (1+\tau) \delta_{lm} a_n^{-} = 0 .
	\end{split}
	\end{align}
	\end{subequations}	
and their Hermitian conjugate, which in terms of the operators $b_l$ and
$c_l$ introduced above read
	\begin{subequations}	\label{6.21}
	\begin{alignat}{2}	\label{6.21a}
[b_l , [ b_m^\dag, b_n ]_\varepsilon  +
       [ c_m^\dag, c_m ]_\varepsilon ]_{\_}
& = 2 (1+\tau) \delta_{lm} b_n
\\			\label{6.21b}
[c_l , [ b_m^\dag, b_n ]_\varepsilon  +
       [ c_m^\dag, c_m ]_\varepsilon ]_{\_}
& = 2 (1+\tau) \delta_{lm} c_n ,
	\end{alignat}
	\end{subequations}
and supplement these relations with equations like~\eref{6.16b}. Obviously,
equations~\eref{6.140} convert~\eref{6.20} into identities and, consequently,
the (standard) paracommutation relations~\eref{6.18} provide a solution
of~\eref{6.21}. On the base of~\eref{6.21} or other similar equations that
can be obtained by generalizing the ones in~\eref{6.9}--\eref{6.11}, further
research on particular classes of trilinear commutation relations can be done,
but, however, this is not a subject of the present work.

	Let us now pay attention to the fact that
equations~\eref{6.9},~\eref{6.10} and~\eref{6.11} are generally different
(regardless of existence of some connections between their solutions).
The cause for this being that the momentum operators for the Lagrangians
$\ope{L}^{\prime}$, $\ope{L}^{\prime\prime}$ and
$\ope{L}^{\prime\prime\prime}$ are generally different unless some additional
restrictions are added to the Lagrangian formalism (see Sect.~\ref{Sect4}). A
necessary and sufficient condition for~\eref{6.9}--\eref{6.11} to be
identical is
	\begin{gather}	\label{6.22}
[ a_l^\pm , [a_m^{\dag\,+} , a_m^{-} ]_{-\varepsilon}
	  - [a_m^{+} , a_m^{\dag\,-} ]_{-\varepsilon} ]_{\_}
= 0,
\intertext{which certainly is valid if the condition~\eref{4.9'}, \viz}
			\label{6.23}
  [a_m^{\dag\,+} , a_m^{-} ]_{-\varepsilon}
- [a_m^{+} , a_m^{\dag\,-} ]_{-\varepsilon}
= 0,
	\end{gather}
ensuring the uniqueness of the momentum operator are, holds. If one adopts the
standard bilinear commutation relations~\eref{6.12}, then~\eref{6.23}, and
hence~\eref{6.22}, is identically valid, but in the framework of, e.g., the
paracommutation relations~\eref{6.140} (or~\eref{6.18} in other form) the
equations~\eref{6.23} should be postulated to ensure uniqueness of the
momentum operator and therefore of the field equations.

	On the base of~\eref{6.9} or~\eref{6.10} one may invent other types
of commutation relations, which will not be investigated in this paper
because we shall be interested mainly in the case
when~\eref{6.9},~\eref{6.10} and~\eref{6.11} are identical (see~\eref{6.22})
or, more generally, when the dynamical variables are unique in the sense
pointed in Sect.~\ref{Sect4}.

\subsection{Restrictions related to the charge operator}
\label{Subsect6.2}

	The consequences of the Heisenberg relations~\eref{5.2}, involving the
charge operator for a \emph{charged} field, $q\not=0$ (and hence
$\tau=0$ -- see~\eref{3.12}), will be examined in this subsection. In
terms of creation and annihilation operators it is equivalent
to~\cite{Bogolyubov&Shirkov,
bp-QFTinMP-scalars,bp-QFTinMP-spinors,bp-QFTinMP-vectors}
	\begin{equation}	\label{6.24}
  [a_s^{\pm}(\bk) , \ope{Q}]_{\_} = q a_s^{\pm}(\bk)
\quad
  [a_s^{\dag\,\pm}(\bk) , \ope{Q}]_{\_} = - q a_s^{\dag\,\pm}(\bk) ,
	\end{equation}
the values of the polarization indices being specified by~\eref{6.6}.
Substituting here~\eref{3.8}, we see that, for a \emph{charged} field, the
field equations for the Lagrangians
$\ope{L}^{\prime}$, $\ope{L}^{\prime\prime}$ and
$\ope{L}^{\prime\prime\prime}$
(see Sect.~\ref{Sect3}) respectively are:
	\begin{subequations}	\label{6.25}
	\begin{align}	\label{6.25a}
&
\sum_{t=1}^{2j+1-\delta_{0m}(1-\delta_{0j})} \int \Id^3\bs p
\{
[ a_s^{\pm}(\bk) ,
a_t^{\dag\,+}(\bs p)\circ a_t^-(\bs p) - \varepsilon
a_t^{\dag\,-}(\bs p)\circ a_t^+(\bs p)
]_{\_}
-  a_s^{\pm}(\bk) \delta_{st} \delta^3(\bk-\bs p)
\}
= 0
\\			\label{6.25b}
&
\sum_{t=1}^{2j+1-\delta_{0m}(1-\delta_{0j})} \!\!\!\! \int \!\!\! \Id^3\bs p
\{
[ a_s^{\dag\,\pm}(\bk) ,
a_t^{\dag\,+}(\bs p)\circ a_t^-(\bs p) - \varepsilon
a_t^{\dag\,-}(\bs p)\circ a_t^+(\bs p)
]_{\_}
+  a_s^{\dag\,\pm}(\bk) \delta_{st} \delta^3(\bk-\bs p)
\}
= 0
	\end{align}
	\end{subequations}
\vspace{-3ex}
	\begin{subequations}	\label{6.26}
	\begin{align}	\label{6.26a}
&
\sum_{t=1}^{2j+1-\delta_{0m}(1-\delta_{0j})} \int \Id^3\bs p
\{
[ a_s^{\pm}(\bk) ,
a_t^{+}(\bs p)\circ a_t^{\dag\,-}(\bs p) - \varepsilon
a_t^{-}(\bs p)\circ a_t^{\dag\,+}(\bs p)
]_{\_}
+  a_s^{\pm}(\bk) \delta_{st} \delta^3(\bk-\bs p)
\}
= 0
\\			\label{6.26b}
&
\sum_{t=1}^{2j+1-\delta_{0m}(1-\delta_{0j})} \!\!\!\! \int \!\!\! \Id^3\bs p
\{
[ a_s^{\dag\,\pm}(\bk) ,
a_t^{+}(\bs p)\circ a_t^{\dag\,-}(\bs p) - \varepsilon
a_t^{-}(\bs p)\circ a_t^{\dag\,+}(\bs p)
]_{\_}
-  a_s^{\dag\,\pm}(\bk) \delta_{st} \delta^3(\bk-\bs p)
\}
= 0
	\end{align}
	\end{subequations}
\vspace{-3ex}
	\begin{subequations}	\label{6.27}
	\begin{align}	\label{6.27a}
&
\sum_{t=1}^{2j+1-\delta_{0m}(1-\delta_{0j})} \!\!\!\! \int \!\!\!\! \Id^3\bs p
\{
[ a_s^{\pm}(\bk) ,
[ a_t^{\dag\,+}(\bs p) , a_t^-(\bs p) ]_\varepsilon -
[ a_t^{+}(\bs p) , a_t^{\dag\,-}(\bs p)_\varepsilon
]_{\_}
-  2 a_s^{\pm}(\bk) \delta_{st} \delta^3(\bk-\bs p)
\}
= 0
\\			\label{6.27b}
&  \!\!\!
\sum_{t=1}^{2j+1-\delta_{0m}(1-\delta_{0j})} \!\!\!\! \int \!\!\!\! \Id^3\bs p
\{
[ a_s^{\dag\,\pm}(\bk) ,
[ a_t^{\dag\,+}(\bs p) , a_t^-(\bs p) ]_\varepsilon -
[ a_t^{+}(\bs p) , a_t^{\dag\,-}(\bs p)_\varepsilon
]_{\_}
+  2 a_s^{\dag\,\pm}(\bk) \delta_{st} \delta^3(\bk-\bs p)
\}
\!=\! 0 .
	\end{align}
	\end{subequations}

	Using~\eref{6.25}--\eref{6.27} and~\eref{6.13}, with
$\eta=\varepsilon=-1$, or simply~\eref{6.24}, one can easily verify the
validity of the equations
	\begin{equation}	\label{6.28}
	\begin{split}
[\tope{P}_\mu,\tope{Q}]_{\_} = 0 \quad &
[\tope{L}_{\mu\nu},\tope{Q}]_{\_} = 0
\\
[\tope{S}_{\mu\nu},\tope{Q}]_{\_} = 0  \quad &
[\tope{M}_{\mu\nu},\tope{Q}]_{\_} = 0  ,
	\end{split}
	\end{equation}
where the operators $\tope{P}_{\mu}$, $\tope{Q}$,  $\tope{S}_{\mu\nu}$,
$\tope{L}_{\mu\nu}$ and $\tope{M}_{\mu\nu}$ are calculated from one and the
same initial Lagrangian according to~\eref{3.7}--\eref{3.10}.
This result confirms the validity of assertion~(ii) before~\eref{5.18} for
the fields considered.

	Following the above considerations, concerning the momentum operator,
we shall now replace the systems of \emph{integral}
equations~\eref{6.25}--\eref{6.27} with respectively the following stronger
systems of \emph{algebraic} equations (by equating to zero the integrands
in~\eref{6.25}--\eref{6.27}):
	\begin{subequations}	\label{6.29}
	\begin{align}	\label{6.29a}
	\begin{split}
\bigl[ a_l^{\pm}
& ,
a_m^{\dag\,+} \circ a_m^{-} - \varepsilon
a_m^{\dag\,-} \circ a_m^{+}
\bigr]_{-}
-  \delta_{lm} a_l^{\pm} = 0
	\end{split}
\\			\label{6.29b}
	\begin{split}
\bigl[ a_l^{\dag\,\pm}
& ,
a_m^{\dag\,+} \circ a_m^{-} - \varepsilon
a_m^{\dag\,-} \circ a_m^{+}
\bigr]_{-}
+  \delta_{lm} a_l^{\dag\,\pm} = 0
	\end{split}
	\end{align}
	\end{subequations}
\vspace{-5ex}
	\begin{subequations}	\label{6.30}
	\begin{align}	\label{6.30a}
	\begin{split}
\bigl[ a_l^{\pm}
& ,
a_m^{+} \circ a_m^{\dag\,-} - \varepsilon
a_m^{-} \circ a_m^{\dag\,+}
\bigr]_{-}
+ \delta_{lm} a_l^{\pm} = 0
	\end{split}
\\			\label{6.30b}
	\begin{split}
\bigl[ a_l^{\dag\,\pm}
& ,
a_m^{+} \circ a_m^{\dag\,-} - \varepsilon
a_m^{-} \circ a_m^{\dag\,+}
\bigr]_{-}
- \delta_{lm} a_l^{\dag\,\pm} = 0
	\end{split}
	\end{align}
	\end{subequations}
\vspace{-5ex}
	\begin{subequations}	\label{6.31}
	\begin{align}	\label{6.31a}
	\begin{split}
\bigl[ a_l^{\pm}
& ,
[ a_m^{\dag\,+} , a_m^{-} ]_\varepsilon -
[ a_m^{+} , a_m^{\dag\,-} ]_\varepsilon
\bigr]_{-}
- 2 \delta_{lm} a_l^{\pm} = 0
	\end{split}
\\			\label{6.31b}
	\begin{split}
\bigl[ a_l^{\dag\,\pm}
& ,
[ a_m^{\dag\,+} , a_m^{-} ]_\varepsilon -
[ a_m^{+} , a_m^{\dag\,-} ]_\varepsilon
\bigr]_{-}
+ 2 \delta_{lm} a_l^{\dag\,\pm} = 0 .
	\end{split}
	\end{align}
	\end{subequations}
These \emph{trilinear commutation relations} are similar
to~\eref{6.9}--\eref{6.11} and, consequently, can be treated in analogous
way.

	By invoking~\eref{6.13}, it is a simple algebra to be proved that the
standard bilinear commutation relations~\eref{6.12}
convert~\eref{6.29}--\eref{6.31} into identities. Thus~\eref{6.12} are
stronger version of~\eref{6.29}--\eref{6.31} and, in this sense, any type of
commutation relations, which provide a solution of~\eref{6.29}--\eref{6.31}
and is compatible with~\eref{6.12}, is a suitable candidate for
generalizing~\eref{6.12}. To illustrate that idea, we shall proceed
with~\eref{6.31} in a way similar to the `derivation' of the paracommutation
relations from~\eref{6.11}.

	Obviously, the equations (cf.~\eref{6.014} with $\tau=0$, as now
$q\not=0$)
	\begin{subequations}	\label{6.32}
	\begin{align}	\label{6.32a}
[ a_l^{\pm} , [ a_m^{+} , a_m^{\dag\,-}]_\varepsilon ]_{\_}
+ \delta_{lm}  a_m^{\pm}
& = 0
\\			\label{6.32b}
[ a_l^{\pm} , [ a_m^{\dag\,+} , a_m^{-}]_\varepsilon ]_{\_}
- \delta_{lm}  a_m^{\pm}
& = 0
	\end{align}
	\end{subequations}
and their Hermitian conjugate provide  a solution of~\eref{6.31}, but, as a
direct calculations shows, they do not agree with the standard
(anti)commutation relations~\eref{6.12}. A solution of~\eref{6.31}
compatible with~\eref{6.12} is given by the equations~\eref{6.14}, with
$\tau=0$ as the field considered is charged one --- see~\eref{3.12}. Therefore
equations~\eref{6.140}, with $\tau=0$, also provide a compatible
with~\eref{6.12} solution of~\eref{6.31}, from where immediately follows that
the paracommutation relations~\eref{6.18}, with $\tau=0$, convert~\eref{6.31}
into identities. To conclude, we can say that the paracommutation
relations~\eref{6.18}, in particular their special case~\eref{6.12}, ensure
the simultaneous validity of the Heisenberg relations~\eref{5.1}
and~\eref{5.2} for free scalar, spinor and vector fields.

	Similarly to~\eref{6.20}, one may generalize~\eref{6.31} to
	\begin{subequations}	\label{6.33}
	\begin{align}
			\label{6.33a}
	\begin{split}
\bigl[ a_l^{+}
& ,
[ a_m^{\dag\,+} , a_n^{-} ]_\varepsilon -
[ a_m^{+} , a_n^{\dag\,-} ]_\varepsilon
\bigr]_{-}
- 2 \delta_{ln} a_m^{+} = 0
	\end{split}
\\			\label{6.33b}
	\begin{split}
\bigl[ a_l^{-}
& ,
[ a_m^{\dag\,+} , a_n^{-} ]_\varepsilon -
[ a_m^{+} , a_n^{\dag\,-} ]_\varepsilon
\bigr]_{-}
- 2 \delta_{lm} a_n^{-} = 0 .
	\end{split}
	\end{align}
	\end{subequations}	
which equations agree with~\eref{6.12}, \eref{6.14}, \eref{6.140}
and~\eref{6.18}, but generally do not agree with~\eref{6.20}, with $\tau=0$,
unless the equations~\eref{6.140}, with $\tau=0$, hold.

	More generally, we can assert that~\eref{6.31} and~\eref{6.11}, with
$\tau=0$, hold simultaneously if and only if~\eref{6.14}, with $\tau=0$, is
fulfilled. From here, again, it follows that the paracommutation relations
ensure the simultaneous validity of~\eref{5.1} and~\eref{5.2}.

	Let us say now some words on the uniqueness problem for the
Heisenberg equations involving the charge operator. The systems of
equations~\eref{6.29}--\eref{6.31} are identical iff
	\begin{equation}	\label{6.34}
\bigl[ a_l^{\pm} ,
[ a_m^{\dag\,+} , a_m^{-} ]_{-\varepsilon} +
[ a_m^{+} , a_m^{\dag\,-} ]_{-\varepsilon}
\bigr]_{-}
= 0 ,
	\end{equation}
which, in particular, is satisfied if the condition
	\begin{equation}	\label{6.35}
[ a_m^{\dag\,+} , a_m^{-} ]_{-\varepsilon} +
[ a_m^{+} , a_m^{\dag\,-} ]_{-\varepsilon}
= 0 ,
	\end{equation}
ensuring the uniqueness of the charge operator (see~\eref{4.10'}), is valid.
Evidently, equations~\eref{6.34} and~\eref{6.22} are compatible iff
	\begin{equation}	\label{6.36}
\bigl[ a_l^{+} ,
[ a_m^{\dag\,\pm} , a_m^{\mp} ]_{-\varepsilon} \bigr]_{-} = 0
\quad
\bigl[ a_l^{-} ,
[ a_m^{\dag\,\pm} , a_m^{\mp} ]_{-\varepsilon} \bigr]_{-} = 0
	\end{equation}
which is a weaker form of~\eref{4.15} ensuring simultaneous uniqueness of the
momentum and charge operator.

\subsection{Restrictions related to the angular momentum operator(s)}
\label{Subsect6.3}

	It is now turn to be investigated the restrictions on the creation
and annihilation operators that follow from the Heisenberg
relations~\eref{5.3} concerning the angular momentum operator. They can be
obtained by inserting the equations~\eref{3.9} and~\eref{3.10}
into~\eref{5.3}. As pointed in Sect.~\ref{Sect5}, the resulting equalities,
however, depend not only on the particular Lagrangian employed, but also on
the geometric nature of the field considered; the last dependence being
explicitly given via~\eref{5.19} and the polarization functions
$\sigma_{\mu\nu}^{ss'm\pm}(\bk)$ and $l_{\mu\nu}^{ss'm\pm}(\bk)$ (see
also~\eref{3.13}).

	Consider the terms containing derivatives in~\eref{5.3},
	\begin{equation}	\label{6.37}
\tope{L}_{\mu\nu}^{\mathrm{or}}
:=
\ih
\Bigl( x_\mu \frac{\pd}{\pd x^\nu}  - x_\nu \frac{\pd}{\pd x^\mu} \Bigr)
\tope{\varphi}_i(x) .
	\end{equation}
If $\underline{\tope{\varphi}}_i(k)$ denotes the Fourier image of
$\tope{\varphi}_i(x)$, \ie
	\begin{equation}	\label{6.38}
\tope{\varphi}_i(x)
=
\Lambda \int \Id^4 k
\e^{-\iih k^\mu x_\mu} \underline{\tope{\varphi}}_i(k),
	\end{equation}
with $\Lambda$ being a normalization constant, then the Fourier image
of~\eref{6.37} is
	\begin{equation}	\label{6.39}
\underline{\tope{L}}_{\mu\nu}^{\mathrm{or}}
=
\ih
\Bigl( k_\mu \frac{\pd}{\pd k^\nu}  - k_\nu \frac{\pd}{\pd k^\mu} \Bigr)
\underline{\tope{\varphi}}_i(k) .
	\end{equation}
Comparing this expression with equations~\eref{3.10}, we see that the terms
containing derivatives in~\eref{3.10} should be responsible for the
term~\eref{6.37} in~\eref{5.3}.%
\footnote{~%
The terms proportional to the momentum operator in~\eref{3.10} disappear if
the creation and annihilation operators~\eref{2.28-2} in Heisenberg picture are
employed (see
also~\cite{bp-QFTinMP-scalars,bp-QFTinMP-spinors,bp-QFTinMP-vectors}).%
}
For this reason, we shall suppose that the momentum operator
$\tope{M}_{\mu\nu}$ admits a representation
	\begin{equation}	\label{6.40}
\tope{M}_{\mu\nu}
=
\tope{M}_{\mu\nu}^{\mathrm{or}} + \tope{M}_{\mu\nu}^{\mathrm{sp}}
	\end{equation}
such that the operators $\tope{M}_{\mu\nu}^{\mathrm{or}}$ and
$\tope{M}_{\mu\nu}^{\mathrm{sp}}$ satisfy the relations~\eref{5.4}
and~\eref{5.5}, respectively. Thus we shall replace~\eref{5.3} with the
stronger system of equations~\eref{5.4}--\eref{5.5}. Besides, we shall admit
that the explicit form of the operators$\tope{M}_{\mu\nu}^{\mathrm{or}}$ and
$\tope{M}_{\mu\nu}^{\mathrm{sp}}$ are given via~\eref{5.7-5} and~\eref{5.7-4}
for the fields investigated in the present work.

	Let us consider at first the `orbital' Heisenberg
relations~\eref{5.4}, which is independent of the particular geometrical
nature of the fields studied. Substituting~\eref{5.7-5} and~\eref{6.38}
into~\eref{5.4}, using that $\underline{\tope{\varphi}}_i(\pm k)$, with
$k^2=m^2c^2$, is a linear combination of $\ta_s^\pm(\bk)$ with classical,
not operator\ndash valued, functions of $\bk$ as
coefficients~\cite{Bogolyubov&Shirkov,
bp-QFTinMP-scalars,bp-QFTinMP-spinors,bp-QFTinMP-vectors} and introducing for
brevity the operator
	\begin{equation}	\label{6.41}
\omega_{\mu\nu}(k)
:= k_\mu \frac{\pd}{\pd k^\nu} -  k_\nu \frac{\pd}{\pd k^\mu},
	\end{equation}
we arrive to the following \emph{integro\ndash differential} systems of
equations:
	\begin{subequations}	\label{6.42} 
	\begin{multline}	\label{6.42a}
\sum_{t=1}^{2j+1-\delta_{0m}(1-\delta_{0j})} \int \Id^3 \bs p
\bigl\{
 \bigl(
( -\omega_{\mu\nu}(p) + \omega_{\mu\nu}(q) )
( [\ta_s^{\pm}(\bk) ,
\ta_t^{\dag\,+}(\bs p) \circ \ta_t^{-}(\bs q)
\\
- \varepsilon
\ta_t^{\dag\,-}(\bs p) \circ \ta_t^{+}(\bs q)
]_{\_} )
\bigr) \big|_{q=p}
\bigr\} \big|_{ p_0=\sqrt{m^2c^2+{\bs p}^2} }
=
2 (1+\tau) \omega_{\mu\nu}(k) (\ta_s^{\pm}(\bk))
	\end{multline}
\vspace{-5.3ex}
	\begin{multline}	\label{6.42b}
\sum_{t=1}^{2j+1-\delta_{0m}(1-\delta_{0j})} \int \Id^3 \bs p
\bigl\{
 \bigl(
( -\omega_{\mu\nu}(p) + \omega_{\mu\nu}(q) )
( [\ta_s^{\dag\,\pm}(\bk) ,
\ta_t^{\dag\,+}(\bs p) \circ \ta_t^{-}(\bs q)
\\
- \varepsilon
\ta_t^{\dag\,-}(\bs p) \circ \ta_t^{+}(\bs q)
]_{\_} )
\bigr) \big|_{q=p}
\bigr\} \big|_{ p_0=\sqrt{m^2c^2+{\bs p}^2} }
=
2 (1+\tau) \omega_{\mu\nu}(k) (\ta_s^{\dag\,\pm}(\bk))
	\end{multline}
	\end{subequations}		
\vspace{-3.1ex}
	\begin{subequations} 	\label{6.43}	
	\begin{multline}	\label{6.43a}
\sum_{t=1}^{2j+1-\delta_{0m}(1-\delta_{0j})} \int \Id^3 \bs p
\bigl\{
 \bigl(
( -\omega_{\mu\nu}(p) + \omega_{\mu\nu}(q) )
( [\ta_s^{\pm}(\bk) ,
\ta_t^{+}(\bs p) \circ \ta_t^{\dag\,-}(\bs q)
\\
- \varepsilon
\ta_t^{-}(\bs p) \circ \ta_t^{\dag\,+}(\bs q)
]_{\_} )
\bigr) \big|_{q=p}
\bigr\} \big|_{ p_0=\sqrt{m^2c^2+{\bs p}^2} }
=
2 (1+\tau) \omega_{\mu\nu}(k) (\ta_s^{\pm}(\bk))
	\end{multline}			
\vspace{-5.3ex}
	\begin{multline}	\label{6.43b}
\sum_{t=1}^{2j+1-\delta_{0m}(1-\delta_{0j})} \int \Id^3 \bs p
\bigl\{
 \bigl(
( -\omega_{\mu\nu}(p) + \omega_{\mu\nu}(q) )
( [\ta_s^{\dag\,\pm}(\bk) ,
\ta_t^{+}(\bs p) \circ \ta_t^{\dag\,-}(\bs q)
\\
- \varepsilon
\ta_t^{-}(\bs p) \circ \ta_t^{\dag\,+}(\bs q)
]_{\_} )
\bigr) \big|_{q=p}
\bigr\} \big|_{ p_0=\sqrt{m^2c^2+{\bs p}^2} }
=
2 (1+\tau) \omega_{\mu\nu}(k) (\ta_s^{\dag\,\pm}(\bk))
	\end{multline}			
	\end{subequations}
\vspace{-3.1ex}
	\begin{subequations}	\label{6.44} 
	\begin{multline}	\label{6.44a}
\sum_{t=1}^{2j+1-\delta_{0m}(1-\delta_{0j})} \int \Id^3 \bs p
\bigl\{
 \bigl(
( -\omega_{\mu\nu}(p) + \omega_{\mu\nu}(q) )
( [\ta_s^{\pm}(\bk) ,
[ \ta_t^{\dag\,+}(\bs p) , \ta_t^{-}(\bs q) ]_{\varepsilon}
\\
+
[ \ta_t^{+}(\bs p) , \ta_t^{\dag\,-}(\bs q) ]_{\varepsilon}
]_{\_} )
\bigr) \big|_{q=p}
\bigr\} \big|_{ p_0=\sqrt{m^2c^2+{\bs p}^2} }
=
4 (1+\tau) \omega_{\mu\nu}(k) (\ta_s^{\pm}(\bk))
	\end{multline}
\vspace{-5.3ex}
	\begin{multline}	\label{6.44b}
\sum_{t=1}^{2j+1-\delta_{0m}(1-\delta_{0j})} \int \Id^3 \bs p
\bigl\{
 \bigl(
( -\omega_{\mu\nu}(p) + \omega_{\mu\nu}(q) )
( [\ta_s^{\dag\,\pm}(\bk) ,
[ \ta_t^{\dag\,+}(\bs p) , \ta_t^{-}(\bs q) ]_{\varepsilon}
\\
+
[ \ta_t^{+}(\bs p) , \ta_t^{\dag\,-}(\bs q) ]_{\varepsilon}
]_{\_} )
\bigr) \big|_{q=p}
\bigr\} \big|_{ p_0=\sqrt{m^2c^2+{\bs p}^2} }
=
4 (1+\tau) \omega_{\mu\nu}(k) (\ta_s^{\dag\,\pm}(\bk)) ,
	\end{multline}
	\end{subequations}		
where $k_0=\sqrt{m^2c^2+\bk^2}$ is set after the differentiations are
performed (see~\eref{6.41}). Following the procedure of the previous
considerations, we replace the \emph{integro\ndash differential}
equations~\eref{6.42}--\eref{6.44} with the following \emph{differential}
ones:
	\begin{subequations}	\label{6.45} 
	\begin{equation}	\label{6.45a}
 \bigl\{
( -\omega_{\mu\nu}^\circ(m) + \omega_{\mu\nu}^\circ(n) )
( [\ta_l^{\pm} ,
\ta_m^{\dag\,+} \circ \ta_n^{-}
- \varepsilon
\ta_m^{\dag\,-} \circ \ta_n^{+}
]_{\_} )
\bigr\} \big|_{n=m}
=
2 (1+\tau) \delta_{lm} \omega_{\mu\nu}^\circ(l) (\ta_l^{\pm})
	\end{equation}			
%
\vspace{-3.6ex}
%
	\begin{equation}	\label{6.45b}
 \bigl\{
( -\omega_{\mu\nu}^\circ(m) + \omega_{\mu\nu}^\circ(n) )
( [\ta_l^{\dag\,\pm} ,
\ta_m^{\dag\,+} \circ \ta_n^{-}
- \varepsilon
\ta_m^{\dag\,-} \circ \ta_n^{+}
]_{\_} )
\bigr\} \big|_{n=m}
=
2 (1+\tau) \delta_{lm} \omega_{\mu\nu}^\circ(l) (\ta_l^{\dag\,\pm})
	\end{equation}			
	\end{subequations}		
\vspace{-2.1ex}
	\begin{subequations} 	\label{6.46}	
	\begin{equation}	\label{6.46a}
 \bigl\{
( -\omega_{\mu\nu}^\circ(m) + \omega_{\mu\nu}^\circ(n) )
( [\ta_l^{\pm} ,
\ta_m^{+} \circ \ta_n^{\dag\,-}
- \varepsilon
\ta_m^{-} \circ \ta_n^{\dag\,+}
]_{\_} )
\bigr\} \big|_{n=m}
=
2 (1+\tau) \delta_{lm} \omega_{\mu\nu}^\circ(l) (\ta_l^{\pm})
	\end{equation}			
%
\vspace{-3.6ex}
%
	\begin{equation}	\label{6.46b}
 \bigl\{
( -\omega_{\mu\nu}^\circ(m) + \omega_{\mu\nu}^\circ(n) )
( [\ta_l^{\dag\,\pm} ,
\ta_m^{+} \circ \ta_n^{\dag\,-}
- \varepsilon
\ta_m^{-} \circ \ta_n^{\dag\,+}
]_{\_} )
\bigr\} \big|_{n=m}
=
2 (1+\tau) \delta_{lm} \omega_{\mu\nu}^\circ(l) (\ta_l^{\dag\,\pm})
	\end{equation}			
	\end{subequations}
\vspace{-2.1ex}
	\begin{subequations}	\label{6.47} 
	\begin{equation}	\label{6.47a}
 \bigl\{
( -\omega_{\mu\nu}^\circ(m) + \omega_{\mu\nu}^\circ(n) )
( [\ta_l^{\pm} ,
[ \ta_m^{\dag\,+} , \ta_n^{-} ]_\varepsilon
+
[ \ta_m^{+} , \ta_n^{\dag\,-} ]_\varepsilon
]_{\_} )
\bigr\} \big|_{n=m}
=
4 (1+\tau) \delta_{lm} \omega_{\mu\nu}^\circ(l) (\ta_l^{\pm})
	\end{equation}			
%
\vspace{-3.6ex}
%
	\begin{equation}	\label{6.47b}
 \bigl\{
( -\omega_{\mu\nu}^\circ(m) + \omega_{\mu\nu}^\circ(n) )
( [\ta_l^{\dag\,\pm} ,
[ \ta_m^{\dag\,+} , \ta_n^{-} ]_\varepsilon
+
[ \ta_m^{+} , \ta_n^{\dag\,-} ]_\varepsilon
]_{\_} )
\bigr\} \big|_{n=m}
=
4 (1+\tau) \delta_{lm} \omega_{\mu\nu}^\circ(l) (\ta_l^{\dag\,\pm}) ,
	\end{equation}			
	\end{subequations}		
where we have set (cf.~\eref{6.41})
	\begin{equation}	\label{6.48}
\omega_{\mu\nu}^\circ(l)
:= \omega_{\mu\nu}(k)
 = k_\mu \frac{\pd}{\pd k^\nu} -  k_\nu \frac{\pd}{\pd k^\mu}
\qquad\text{if } l=(s,\bk)
	\end{equation}
and $k_0=\sqrt{m^2c^2+\bk^2}$ is set after the differentiations are performed.

	\textbf{Remark.} Instead of~\eref{6.45}--\eref{6.47} one can write
similar equations in which the operator $-\omega_{\mu\nu}^\circ(m)$ or
$+\omega_{\mu\nu}^\circ(n)$ is deleted and the factor $+\frac{1}{2}$ or
$-\frac{1}{2}$, respectively, is added on their right hand sides. These
manipulations correspond to an integration by parts of some of the terms
in~\eref{6.42}--\eref{6.44}.

	The main difference of the obtained trilinear relations with respect
to the previous ones considered above is that they are
\emph{partial differential} equations of first order.

	The relations~\eref{6.47} agree with the equations~\eref{6.140} in a
sense that if~\eref{6.140} hold, then~\eref{6.47} become identically valid.
Indeed, since
	\begin{equation}	\label{6.49}
	\begin{split}
\bigl\{ ( -\omega_{\mu\nu}^\circ(m) + \omega_{\mu\nu}^\circ(n) )
	(\ta_m^{\pm} \delta_{ln})
\bigr\} \big|_{n=m}
& =
- 2 \delta_{lm} \omega_{\mu\nu}^\circ(m) (\ta_m^{\pm})
\\
\bigl\{ ( -\omega_{\mu\nu}^\circ(m) + \omega_{\mu\nu}^\circ(n) )
	(\ta_n^{\pm} \delta_{lm})
\bigr\} \big|_{n=m}
& =
+ 2 \delta_{lm} \omega_{\mu\nu}^\circ(m) (\ta_m^{\pm}) ,
	\end{split}
	\end{equation}
due to~\eref{6.48},~\eref{6.41} and the equality
$\frac{\od\delta(x)}{\od x} f(x) = - \delta(x) \frac{\od f(x)}{\od x}$
for a $C^1$ function $f$, the application of the operator
 $( -\omega_{\mu\nu}^\circ(m) + \omega_{\mu\nu}^\circ(n) )$ to~\eref{6.140}
and subsequent setting $n=m$ entails~\eref{6.47}. In particular, this means
that the paracommutation relations~\eref{6.18} and, moreover, the standard
(anti)commutation relations~\eref{6.12} convert~\eref{6.47} into identities.
Therefore the `orbital' Heisenberg relations~\eref{5.4} hold for scalar,
spinor and vector fields satisfying the bilinear or para commutation
relations.

	It should be noted, the paracommutation relations are
\emph{not} the only trilinear commutation relations that are solutions
of~\eref{6.47}. As an example, we shall present the trilinear relations
	\begin{subequations}	\label{6.50}
	\begin{align}
			\label{6.50a}
& \bigl[ a_l^{+} , [ a_m^{+} , a_n^{\dag\,-} ]_\varepsilon \bigr]_{-}
= \bigl[ a_l^{+} , [ a_m^{\dag\,+} , a_n^{-} ]_\varepsilon \bigr]_{-}
= - (1+\tau) \delta_{ln} a_m^{+}
\\			\label{6.50b}
& \bigl[ a_l^{-} , [ a_m^{+} , a_n^{\dag\,-} ]_\varepsilon \bigr]_{-}
= \bigl[ a_l^{-} , [ a_m^{\dag\,+} , a_n^{-} ]_\varepsilon \bigr]_{-}
= + (1+\tau) \delta_{lm} a_n^{+} ,
	\end{align}
	\end{subequations}	
which reduce to~\eref{6.014} for $n=m$, do not agree with~\eref{6.12}, but
convert~\eref{6.47} into identities (see~\eref{6.49}). Other example is
provided by the equations~\eref{6.20}, which are compatible with the
paracommutation relations and, as a result of~\eref{6.49},
convert~\eref{6.47} into identities. \emph{Prima facie} one may suppose that
any solution of~\eref{6.11} provides a solution of~\eref{6.47}, but this is
not the general case. A counterexample is provided by the commutation
relations
	\begin{align}	\label{6.51}
	\begin{split}
\bigl[ a_l^{\pm}
& ,
[ a_m^{\dag\,+} , a_n^{-} ]_\varepsilon +
[ a_m^{+} , a_n^{\dag\,-} ]_\varepsilon
\bigr]_{-}
\pm 2 (1+\tau) \delta_{ln} a_m^{\pm} = 0 ,
	\end{split}
	\end{align}
which reduce to~\eref{6.11} for $n=m$, satisfy~\eref{6.47} with
$\ta_l^{+}$ for $\ta_l^{\pm}$, and do \emph{not} satisfy~\eref{6.47} with
$\ta_l^{-}$ for $\ta_l^{\pm}$ (see~\eref{6.49} and cf.~\eref{6.20}).

	From~\eref{5.7-5} follows that the operator
$\tope{M}_{\mu\nu}^{\mathrm{or}}$ is independent of the Lagrangian
$\ope{L}^{\prime}$, $\ope{L}^{\prime\prime}$ or
$\ope{L}^{\prime\prime\prime}$ one starts off if and only if
(see~\eref{4.11})
	\begin{align}	\label{6.52}
\bigl\{
( -\omega_{\mu\nu}^\circ(m) + \omega_{\mu\nu}^\circ(n) )
\bigl(
[ \ta_m^{\dag\,+} , \ta_n^{-} ]_{-\varepsilon} -
[ \ta_m^{+} , \ta_n^{\dag\,-} ]_{-\varepsilon}
\bigr)
\bigr\} \big|_{n=m}
= 0 .
	\end{align}
This condition ensures the coincidence of the systems of
equations~\eref{6.45}, \eref{6.46} and~\eref{6.47} too. However, the
following necessary and sufficient condition for the coincidence of these
systems is expressed by the weaker equations
	\begin{align}	\label{6.53}
\bigl\{
( -\omega_{\mu\nu}^\circ(m) + \omega_{\mu\nu}^\circ(n) )
\bigl(
\bigl[ \ta_l^\pm ,
[ \ta_m^{\dag\,+} , \ta_n^{-} ]_{-\varepsilon} -
[ \ta_m^{+} , \ta_n^{\dag\,-} ]_{-\varepsilon}
\bigr]_{-}
\bigr)
\bigr\} \big|_{n=m}
= 0 .
	\end{align}

	It is now turn to be considered the `spin' Heisenberg
relations~\eref{5.5}.

	Recall, the field operators $\varphi_i$ for the fields considered
here admit a
representation~\cite{bp-QFTinMP-scalars,bp-QFTinMP-spinors,bp-QFTinMP-vectors}
	\begin{equation}	\label{6.54}
\varphi_i
=
\Lambda \sum_{t} \int\Id^3 \bs p
\bigl\{ v_i^{t,+}(\bs p) a_t^+(\bs p) +
	v_i^{t,-}(\bs p) a_t^-(\bs p) \bigr\} ,
	\end{equation}
where $\Lambda$ is a normalization constant and $v_i^{t,\pm}(\bs p)$ are
classical, not operator\ndash valued, complex or real functions which are
linearly independent. The particular definition of $v_i^{t,\pm}(\bs p)$
depends on the geometrical nature of $\varphi_i$ and can be found
in~\cite{bp-QFTinMP-scalars,bp-QFTinMP-spinors,bp-QFTinMP-vectors} (see
also~\cite{Bogolyubov&Shirkov}), where the reader can find also a number of
relations satisfied by $v_i^{t,\pm}(\bs p)$. Here we shall mention only that
$v_i^{t,\pm}(\bs p)=1$ for a scalar field and
\(
  v_i^{t,+}(\bs p)
= v_i^{t,-}(\bs p)
=: v_i^{t}(\bs p) = (v_i^{t}(\bs p))^\ast
\)
for a vector field.

	The explicit form of the polarization functions
$\sigma_{\mu\nu}^{ss',\pm}(\bk)$ and $l_{\mu\nu}^{ss',\pm}(\bk)$ (see
Sect.~\ref{Sect3}, in particular~\eref{3.13}) through $v_i^{t,\pm}(\bk)$
are~\cite{bp-QFTinMP-scalars,bp-QFTinMP-spinors,bp-QFTinMP-vectors}:
	\begin{equation}	\label{6.55}
	\begin{split}
\sigma_{\mu\nu}^{ss',\pm}(\bk)
& =
\frac{(-1)^j}{j+\delta_{j0}} \sum_{i,i'}
( v_i^{s,\pm}(\bk) )^\ast I_{i'\mu\nu}^{i} v_{i'}^{t,\pm}(\bk)
\\
l_{\mu\nu}^{ss',\pm}(\bk)
& =
\frac{(-1)^j}{2j+\delta_{j0}} \sum_{i}
( v_i^{s,\pm}(\bk) )^\ast
\Bigl( \xlrarrow{ k_\mu \frac{\pd}{\pd k^\nu} }
     - \xlrarrow{ k_\nu \frac{\pd}{\pd k^\mu} } \Bigr)
v_{i}^{t,\pm}(\bk) ,
	\end{split}
	\end{equation}
with an exception that
$\sigma_{0a}^{ss',\pm}(\bk)=\sigma_{a0}^{ss',\pm}(\bk)=0$, $a=1,2,3$, for a
spinor field, $j=\frac{1}{2}$,~\cite{bp-QFTinMP-spinors}. Evidently, the
equations~\eref{3.13} follow from the mentioned facts (see also~\eref{5.19}).

	Substituting~\eref{6.54} and~\eref{5.7-4} into~\eref{5.5}, we obtain
the following systems of \emph{integral} equations  (corresponding
respectively to the Lagrangians $\ope{L}^{\prime}$, $\ope{L}^{\prime\prime}$
and $\ope{L}^{\prime\prime\prime}$):
	\begin{multline}	\label{6.56}
\frac{(-1)^{j+1} j}{1+\tau} \sum_{s,s',t} \int \Id^3\bk \int \Id^3\bs p
v_i^{t,\pm}(\bs p)
\bigl\{
( \sigma_{\mu\nu}^{ss',-}(\bk) + l_{\mu\nu}^{ss',-}(\bk) )
[ a_t^\pm(\bs p) , a_{s}^{\dag\,+}(\bk) \circ a_{s'}^{-}(\bk) ]_{\_}
\\ +
( \sigma_{\mu\nu}^{ss',+}(\bk) + l_{\mu\nu}^{ss',+}(\bk) )
[ a_t^\pm(\bs p) , a_{s}^{\dag\,-}(\bk) \circ a_{s'}^{+}(\bk) ]_{\_}
\bigr\}
=
\sum_{i'} \sum_{t} \int \Id^3\bs p
I_{i\mu\nu}^{i'} v_{i'}^{t,\pm}(\bs p) a_t^\pm(\bs p)
	\end{multline}
\vspace{-4.2ex}
	\begin{multline}	\label{6.57}
\varepsilon
\frac{(-1)^{j+1} j}{1+\tau} \sum_{s,s',t} \int \Id^3\bk \int \Id^3\bs p
v_i^{t,\pm}(\bs p)
\bigl\{
( \sigma_{\mu\nu}^{ss',+}(\bk) + l_{\mu\nu}^{ss',+}(\bk) )
[ a_t^\pm(\bs p) , a_{s'}^{+}(\bk) \circ a_{s}^{\dag\,-}(\bk) ]_{\_}
\\ +
( \sigma_{\mu\nu}^{ss',-}(\bk) + l_{\mu\nu}^{ss',-}(\bk) )
[ a_t^\pm(\bs p) , a_{s'}^{-}(\bk) \circ a_{s}^{\dag\,+}(\bk) ]_{\_}
\bigr\}
=
\sum_{i'} \sum_{t} \int \Id^3\bs p
I_{i\mu\nu}^{i'} v_{i'}^{t,\pm}(\bs p) a_t^\pm(\bs p)
	\end{multline}
\vspace{-4.2ex}
	\begin{multline}	\label{6.58}
\frac{(-1)^{j+1} j}{2(1+\tau)} \sum_{s,s',t} \int \Id^3\bk \int \Id^3\bs p
v_i^{t,\pm}(\bs p)
\bigl\{
( \sigma_{\mu\nu}^{ss',-}(\bk) + l_{\mu\nu}^{ss',-}(\bk) )
\bigl[ a_t^\pm(\bs p) ,
[ a_s^{\dag\,+}(\bk) , a_{s'}^{-}(\bk) ]_{\varepsilon}
\bigr]_{-}
\\ +
( \sigma_{\mu\nu}^{ss',+}(\bk) + l_{\mu\nu}^{ss',+}(\bk) )
\bigl[ a_t^\pm(\bs p) ,
[ a_s^{\dag\,-}(\bk) , a_{s'}^{+}(\bk) ]_{\varepsilon}
\bigr]_{-}
\bigr\}
=
\sum_{i'} \sum_{t} \int \Id^3\bs p
I_{i\mu\nu}^{i'} v_{i'}^{t,\pm}(\bs p) a_t^\pm(\bs p) .
	\end{multline}

	For the difference of all previously considered systems of
\emph{integral} equations, like~\eref{6.2}--\eref{6.4},
\eref{6.25}--\eref{6.27} and~\eref{6.42}--\eref{6.44}, the
systems~\eref{6.56}--\eref{6.58} cannot be replaced by ones consisting of
algebraic (or differential) equations. The cause for this state of affairs is
that in~\eref{6.56}--\eref{6.58} enter polarization modes with arbitrary $s$
and $s'$ and, generally, one cannot `diagonalize' the integrand(s) with
respect to $s$ and $s'$; moreover, for a vector field, the modes with $s=s'$
are not presented at all (see~\eref{3.13}). That is why no commutation
relations can be extracted from~\eref{6.56}--\eref{6.58} unless further
assumptions are made. Without going into details, below we shall sketch the
proof of the assertion that the
\emph{commutation relations~\eref{6.140} convert~\eref{6.58} into identities
for massive spinor and vector fields}.%
\footnote{~%
The equations~\eref{6.56}--\eref{6.58} are identities for scalar fields as
for them $I_{\mu\nu}=0$ and $v_i^{t,\pm}(\bk)=1$, which reflects the absents
of spin for these fields.%
}
In particular, this entails that the paracommutation and the bilinear
commutation relations provide solutions of~\eref{6.58}.

	Let~\eref{6.140} holds. Combining it with~\eref{6.58}, we see that
the latter splits into the equations
	\begin{subequations}	\label{6.59}
	\begin{multline}	\label{6.59a}
\frac{(-1)^j j}{1+\tau} \sum_{s,t} \int \Id^3\bs p
v_i^{t,+}(\bs p)
\bigl\{
\tau
( \sigma_{\mu\nu}^{st,-}(\bs p) + l_{\mu\nu}^{st,-}(\bs p) )
+ \varepsilon
( \sigma_{\mu\nu}^{ts,+}(\bs p) + l_{\mu\nu}^{ts,+}(\bs p) )
\bigr\}
a_s^+(\bs p) ,
\\ =
\sum_{i'} I_{i\mu\nu}^{i'}
\sum_{s} \int \Id^3\bs p
v_{i'}^{s,+}(\bs p) a_s^+(\bs p)
	\end{multline}
\vspace{-4.2ex}
	\begin{multline}	\label{6.59b}
\frac{(-1)^{j+1} j}{1+\tau} \sum_{s,t} \int \Id^3\bs p
v_i^{t,-}(\bs p)
\bigl\{
( \sigma_{\mu\nu}^{ts,-}(\bs p) + l_{\mu\nu}^{ts,-}(\bs p) )
+ \varepsilon \tau
( \sigma_{\mu\nu}^{st,+}(\bs p) + l_{\mu\nu}^{st,+}(\bs p) )
\bigr\}
a_s^-(\bs p) ,
\\ =
\sum_{i'} I_{i\mu\nu}^{i'}
\sum_{s} \int \Id^3\bs p
v_{i'}^{s,-}(\bs p) a_s^-(\bs p) .
	\end{multline}
	\end{subequations}
Inserting here~\eref{6.55}, we see that one needs the explicit definition of
 $v_{i}^{s,\pm}(\bk)$ and formulae for sums like
\(
\rho_{ii'}(\bk) := \sum_{s} v_{i}^{s,\pm}(\bk) (v_{i'}^{s,\pm}(\bk))^\ast
\),
which are specific for any particular field and can be found
in~\cite{bp-QFTinMP-scalars,bp-QFTinMP-spinors,bp-QFTinMP-vectors}. In this
way, applying~\eref{5.19}, \eref{3.12} and the mentioned results
from~\cite{bp-QFTinMP-scalars,bp-QFTinMP-spinors,bp-QFTinMP-vectors}, one can
check the validity of~\eref{6.59} for massive fields in a way similar to the
proof of~\eref{5.3}
in~\cite{bp-QFTinMP-scalars,bp-QFTinMP-spinors,bp-QFTinMP-vectors} for
scalar, spinor and vector fields, respectively.

	We shall end the present subsection with the remark that the
equations~\eref{4.16} and~\eref{4.17}, which together with~\eref{4.15}
ensure the uniqueness of the spin and orbital operators, are sufficient
conditions for the coincidence of the equations~\eref{6.56}, \eref{6.57}
and~\eref{6.58}.


\section{Inferences}
\label{Sect7}

	To begin with, let us summarize the major conclusions from
Sect.~\ref{Sect6}. Each of the Heisenberg equations~\eref{5.1}--\eref{5.3},
the equations~\eref{5.3} being split into~\eref{5.4} and~\eref{5.5}, induces
in a natural way some relations that the creation and annihilation operators
should satisfy. These relations can be chosen  as algebraic trilinear ones in
a case of~\eref{5.1} and~\eref{5.2} (see~\eref{6.9}--\eref{6.11}
and~\eref{6.29}--\eref{6.31}, respectively). But for~\eref{5.4}
and~\eref{5.5} they need not to be algebraic and are differential ones in the
case of~\eref{5.4} (see~\eref{6.45}--\eref{6.47}) and integral equations in
the case of~\eref{5.5} (see~\eref{6.56}--\eref{6.58}). It was pointed that
the cited relations depend on the initial Lagrangian from which the theory is
derived, unless some explicitly written conditions hold (see~\eref{6.22},
\eref{6.35} and~\eref{6.53}); in particular, these conditions are true if the
equations~\eref{4.9}--\eref{4.13}, ensuring the uniqueness of the
corresponding dynamical operators, are valid. Since the `charge symmetric'
Lagrangians~\eref{3.4} seem to be the ones that best describe free fields,
the arising from them (commutation) relations~\eref{6.11}, \eref{6.31},
\eref{6.47} and~\eref{6.58} were studied in more details. It was proved that
the trilinear commutation relations~\eref{6.140} convert them into
identities, as a result of which the same property possess the
paracommutation relations~\eref{6.18} and, in particular, the bilinear
commutation relations~\eref{6.12}. Examples of trilinear commutation
relations, which are neither ordinary nor para ones, were presented; some of
them, like~\eref{6.014}, \eref{6.32} and~\eref{6.50}, do not agree
with~\eref{6.12} and other ones, like~\eref{6.140}, \eref{6.20}
and~\eref{6.33}, generalize~\eref{6.18} and hence are compatible
with~\eref{6.12}. At last, it was demonstrated that the commutators between
the dynamical variables (see~\eref{5.9}--\eref{5.17}) are uniquely defined if
a Heisenberg relation for one of the operators entering in it is postulated.

	The chief aim of the present section is to be explored the problem
whether all of the reasonable conditions, mentioned in the previous sections and
that can be imposed on the creation and annihilation operators, can hold or
not hold simultaneously. This problem is suggested by the strong evidences
that the relations~\eref{5.1}--\eref{5.3} and~\eref{5.9}--\eref{5.17}, with a
possible exception of~\eref{5.3} (more precisely, of~\eref{5.5}) in the
massless case, should be valid in a realistic quantum field
theory~\cite{Bjorken&Drell-2,Bogolyubov&Shirkov,Roman-QFT,Itzykson&Zuber,
Bogolyubov&et_al.-AxQFT,Bogolyubov&et_al.-QFT}. Besides, to the arguments in
\emph{loc.\ cit.}, we shall add the requirement for uniqueness of the
dynamical variables (see Sect.~\ref{Sect4}).

	As it was shown in Sect.~\ref{Sect6}, the relations~\eref{5.1},
\eref{5.2}, \eref{5.4} and~\eref{5.5} are compatible if one starts from a
charge symmetric Lagrangian (see~\eref{3.4}), which best describes a free
field theory; in particular, the commutation relations~\eref{6.140} (and
hence~\eref{6.18} and~\eref{6.12}) ensure their simultaneous validity.%
\footnote{~%
The special case(s) when~\eref{5.5} may not hold for a massless field will
not be considered below.%
}
For  that reason, we shall investigate below only commutation relations for
which~\eref{5.1}, \eref{5.2}, \eref{5.4} and~\eref{5.5} hold. It will be
assumed that they should be such that the equations~\eref{6.9}--\eref{6.11},
\eref{6.29}--\eref{6.31}, \eref{6.45}--\eref{6.47}
and~\eref{6.56}--\eref{6.58}, respectively, hold.

	Consider now the problem for the uniqueness of the dynamical
variables and its consistency with the commutation relations just mentioned
for a charged field. It will be assumed that this uniqueness is ensured via
the equations~\eref{4.9}--\eref{4.11}.

	The equation~\eref{4.15}, \viz
	\begin{equation}	\label{7.1}
[ a_m^{\dag\,\pm} , a_m^{\mp} ]_{-\varepsilon} = 0 ,
	\end{equation}
is a necessary and sufficient conditions for the uniqueness of the momentum
and charge operators (see Sect.~\ref{Sect4} and the notation introduced at
the beginning of Sect.~\ref{Sect6}). Before commenting on this relation, we
would like to derive some consequences of it. Applying
consequently~\eref{6.13} for $\eta=-\varepsilon$, \eref{7.1} and the identity
	\begin{equation}	\label{7.2}
[A,B\circ C]_{+}  = [A,B]_\eta \circ C - \eta B\circ [A,C]_{-\eta}
\qquad \eta=\pm1
	\end{equation}
for $\eta=+\varepsilon,-\varepsilon$, we, in view of~\eref{7.1}, obtain
	\begin{equation}	\label{7.3}
	\begin{split}
[ a_m^{+} , [a_m^{+} , a_m^{\dag\, -} ]_{\varepsilon} ]_{\_}
& =
[ a_m^{\dag\,-} , [a_m^{+} , a_m^{+} ]_{-\varepsilon} ]_{+}
=
(1-\varepsilon) [ a_m^{\dag\,-} , a_m^{+} ]_{\varepsilon} \circ a_m^{+}
\\
[ a_m^{-} , [a_m^{\dag\,+} , a_m^{-} ]_{\varepsilon} ]_{\_}
& = \varepsilon
[ a_m^{\dag\,+} , [a_m^{-} , a_m^{-} ]_{-\varepsilon} ]_{+}
= \varepsilon
(1-\varepsilon) [ a_m^{\dag\,+} , a_m^{-} ]_{\varepsilon} \circ a_m^{-} .
	\end{split}
	\end{equation}

	Forming the sum and difference of~\eref{6.11a}, for $\tau=0$,
and~\eref{6.31a}, we see that the system of equations they form is equivalent
to
	\begin{subequations}	\label{7.4}
	\begin{alignat}{2}	\label{7.4a}
& [ a_l^{+} , [a_m^{\dag\,+} , a_m^{-} ]_{\varepsilon} ]_{\_} = 0
& \quad
& [ a_l^{-} , [a_m^{+} , a_m^{\dag\,-} ]_{\varepsilon} ]_{\_} = 0
\\			\label{7.4b}
& [ a_l^{+} , [a_m^{+} , a_m^{\dag\,-} ]_{\varepsilon} ]_{\_}
+ 2\delta_{lm} a_l^{+}
= 0
& \quad
& [ a_l^{-} , [a_m^{\dag\,+} , a_m^{-} ]_{\varepsilon} ]_{\_}
- 2\delta_{lm} a_l^{-}
= 0 .
	\end{alignat}
	\end{subequations}
Combining~\eref{7.4b}, for $l=m$, with~\eref{7.3}, we get
	\begin{align}	\label{7.5}
(1-\varepsilon)
[a_m^{\dag\,-} , a_m^{+} ]_{\varepsilon} \circ a_m^{+} + 2 a_m^{+} = 0
\quad
\varepsilon (1-\varepsilon)
[a_m^{\dag\,+} , a_m^{-} ]_{\varepsilon} \circ a_m^{-} - 2 a_m^{-} = 0 .
	\end{align}
Obviously, these equations reduce to
	\begin{equation}	\label{7.6}
a_m^{\pm} = 0
	\end{equation}
for bose fields as for them $\varepsilon=+1$ (see~\eref{3.12}). Since the
operators~\eref{7.6} describe  a completely unobservable field, or, more
precisely, an absence of a field at all, the obtained result means that the
theory considered cannot describe any really existing physical field with
spin $j=0,1$. Such a conclusion should be regarded as a contradiction in the
theory. For fermi fields, $j=\frac{1}{2}$ and $\varepsilon=-1$, the
equations~\eref{7.5} have solutions different from~\eref{7.6} iff $a_m^\pm$
are degenerate operators, \ie with no inverse ones, in which case~\eref{7.4a}
is a consequence of~\eref{7.5} and~\eref{7.1} (see~\eref{6.13} and~\eref{7.3}
too).

	The source of the above contradiction is in the equation~\eref{7.1},
which does not agree with the bilinear commutation relations~\eref{6.12} and
contradicts to the existing correlation between creation and annihilation of
particles with identical characteristics ($m=(t,\bs p)$ in our case)
as~\eref{7.1} can be interpreted physically as mutual independence of the
acts of creation and annihilation of such
particles~\cite[\S~10.1]{Bogolyubov&Shirkov}.

	At this point, there are two ways for `repairing' of the theory. On one
hand, one can forget about the uniqueness of the dynamical variables (in a
sense of Sect.~\ref{Sect4}), after which the formalism can be developed by
choosing, e.g., the charge symmetric Lagrangians~\eref{3.4} and following the
usual Lagrangian formalism; in fact, this is the way the parafield theory is
build~\cite{Green-1953,Ohnuki&Kamefuchi}. On another hand, one may try to
change something at the ground of the theory in such a way that the uniqueness
of the dynamical variables to be ensured automatically. We shall follow the
second method. As a guiding idea, we shall have in mind that the bilinear
commutation relations~\eref{6.12} and the related to them normal ordering
procedure provide a base for the present-day quantum field theory, which
describes sufficiently well the discovered elementary particles/fields. On
this background, an extensive exploration of commutation relations which are
incompatible with~\eref{6.12} is justified only if there appear some
evidences for fields/particles that can be described via them. In that
connection it should be
recalled~\cite{Greenberg&Messiah-1965,Ohnuki&Kamefuchi}, it seems that all
known particles/fields are described via~\eref{6.12} and no one of them is a
para particle/field.

	Using the notation introduced at the beginning of Sect.~\ref{Sect4},
we shall look for a linear mapping (operator) $\ope{E}$ on the operator space
over the system's Hilbert space $\Hil$ of states such that
	\begin{equation}	\label{7.7}
\ope{E} (\ope{D}^{\prime}) = \ope{E} (\ope{D}^{\prime\prime}).
	\end{equation}
As it was shown in Sect.~\ref{Sect4}, an example of an operator $\ope{E}$ is
provided by the normal ordering operator $\ope{N}$. Therefore an operator
satisfying~\eref{7.7} always exists. To any such operator $\ope{E}$ there
corresponds a set of dynamical variables defined via
\begin{equation}	\label{7.8-1}
\ope{D} = \ope{E} (\ope{D}^{\prime}).
	\end{equation}

	Let us examine the properties of the mapping $\ope{E}$ that it should
possess due to the requirement~\eref{7.7}.

	First of all, as the operators of the dynamical variables should be
Hermitian, we shall require
	\begin{gather}	\label{7.8-2}
\bigl( \ope{E}(\ope{B}) \bigr)^\dag = \ope{E}(\ope{B}^\dag)
\\\intertext{for any operator $\ope{B}$, which entails}
			\label{7.8-3}
\ope{D}^\dag = \ope{D},
	\end{gather}
due to~\eref{3.7}--\eref{3.10} and~\eref{7.8-1}.

	As in Sect.~\ref{Sect4}, we shall replace the so\ndash arising
integral equations with corresponding algebraic ones. Thus the
equations~\eref{4.5}--\eref{4.19} remain valid if the operator $\ope{E}$ is
applied to their left hand sides.

	Consider the general case of a charged field, $q\not=0$. So, the
analogue of~\eref{4.15} reads
	\begin{equation}	\label{7.8}
\ope{E} \bigl( [ a_m^{\dag\,\pm} , a_m^{\mp} ]_{-\varepsilon} \bigr)
= 0 ,
	\end{equation}
which equation ensures the uniqueness of the momentum and charge operators.
Respectively, the condition~\eref{4.11} transforms into
	\begin{equation}	\label{7.9}
\bigl\{
( - \omega_{\mu\nu}^{\circ}(m) + \omega_{\mu\nu}^{\circ}(n) )
\bigl(
\ope{E} ( [ a_m^{\dag\,+} , a_n^{-} ]_{-\varepsilon} ) -
\ope{E} ( [ a_m^{+} , a_n^{\dag\,-} ]_{-\varepsilon} )
\bigr)
\bigr\} \big|_{n=m} = 0 ,
	\end{equation}
which, by means of~\eref{7.8} can be rewritten as (cf.~\eref{4.15new})
	\begin{equation}	\label{7.10}
\bigl\{
 \omega_{\mu\nu}^{\circ}(n)
\bigl(
\ope{E} ( [ a_m^{\dag\,+} , a_n^{-} ]_{-\varepsilon} ) -
\ope{E} ( [ a_m^{+} , a_n^{\dag\,-} ]_{-\varepsilon} )
\bigr)
\bigr\} \big|_{n=m} = 0 .
	\end{equation}
At the end, equations~\eref{4.16} and~\eref{4.17} now should be written as
	\begin{align}	\label{7.11}
\sum_{s,s'} \bigl\{
\sigma_{\mu\nu}^{ss',-}(\bk)
\ope{E}
\bigl( [ a_s^{\dag\,+}(\bk) , a_{s'}^{-}(\bk) ]_{-\varepsilon} \bigr) +
\sigma_{\mu\nu}^{ss',+}(\bk)
\ope{E}
\bigl( [ a_{s}^{\dag\,-}(\bk) , a_{s'}^{+}(\bk) ]_{-\varepsilon} \bigr)
\bigr\}
& = 0
\\			\label{7.12}
\sum_{s,s'} \bigl\{
l_{\mu\nu}^{ss',-}(\bk)
\ope{E}
\bigl( [ a_s^{\dag\,+}(\bk) , a_{s'}^{-}(\bk) ]_{-\varepsilon}  \bigr) +
l_{\mu\nu}^{ss',+}(\bk)
\ope{E}
\bigl( [ a_{s}^{\dag\,-}(\bk) , a_{s'}^{+}(\bk) ]_{-\varepsilon} \bigr)
\bigr\}
& = 0  .
	\end{align}
These equations can be satisfied if we generalize~\eref{7.8} to
(cf.~\eref{4.19})
	\begin{equation}	\label{7.13}
\ope{E}
\bigl( [ a_s^{\dag\,\pm}(\bk) , a_{s'}^{\mp}(\bk) ]_{-\varepsilon} \bigr) = 0
	\end{equation}
for any $s$ and $s'$. At last, the following stronger version of~\eref{7.13}
	\begin{equation}	\label{7.13-1}
\ope{E}
\bigl( [ a_m^{\dag\,\pm} , a_{n}^{\mp} ]_{-\varepsilon} \bigr) = 0,
	\end{equation}
for any $m=(t,\bs p)$ and $n=(r,\bs q)$, ensures the validity of~\eref{7.11}
and~\eref{7.12} and thus of the uniqueness of all dynamical variables.

	It is time now to call attention to the possible commutation
relations. The replacement
\(
\ope{D}^{\prime}, \ope{D}^{\prime\prime}, \ope{D}^{\prime\prime\prime}
\mapsto
\ope{D}
:=
  \ope{E} (\ope{D}^{\prime})
= \ope{E} (\ope{D}^{\prime\prime})
= \ope{E} (\ope{D}^{\prime\prime\prime})
\)
 results in corresponding changes in the whole of the material of
Sect.~\ref{Sect6}. In particular, the systems of commutation
relations~\eref{6.9}--\eref{6.11}, \eref{6.29}--\eref{6.31},
\eref{6.45}--\eref{6.47} and~\eref{6.56}--\eref{6.58} should be replaced
respectively with:%
\footnote{~%
To save some space, we do not write the Hermitian conjugate of the
below-written equations.%
}
	\begin{align}
			\label{7.14}
	\begin{split}
\bigl[ a_l^{\pm}
& ,
\ope{E} ( a_m^{\dag\,+} \circ a_m^{-} ) + \varepsilon
\ope{E} ( a_m^{\dag\,-} \circ a_m^{+} )
\bigr]_{-}
\pm (1+\tau) \delta_{lm} a_l^{\pm} = 0
	\end{split}
\\ 			\label{7.15}
	\begin{split}
\bigl[ a_l^{\pm}
& ,
\ope{E} ( a_m^{\dag\,+} \circ a_m^{-} ) - \varepsilon
\ope{E} ( a_m^{\dag\,-} \circ a_m^{+} )
\bigr]_{-}
-  \delta_{lm} a_l^{\pm} = 0
	\end{split}
\\				\label{7.16}
 \bigl\{
( - & \omega_{\mu\nu}^\circ(m) + \omega_{\mu\nu}^\circ(n) )
( [\ta_l^{\pm} ,
\ope{E}( \ta_m^{\dag\,+} \circ \ta_n^{-} )
- \varepsilon
\ope{E}( \ta_m^{\dag\,-} \circ \ta_n^{+} )
]_{\_} )
\bigr\} \big|_{n=m}
=
2 (1+\tau) \delta_{lm} \omega_{\mu\nu}^\circ(l) (\ta_l^{\pm})
	\end{align}
\vspace{-4.8ex}
	\begin{multline}	\label{7.17}
\frac{(-1)^{j+1} j}{1+\tau} \sum_{s,s',t} \int \Id^3\bk \int \Id^3\bs p
v_i^{t,\pm}(\bs p)
\bigl\{
( \sigma_{\mu\nu}^{ss',-}(\bk) + l_{\mu\nu}^{ss',-}(\bk) )
[ a_t^\pm(\bs p) ,
\ope{E}( a_{s}^{\dag\,+}(\bk) \circ a_{s'}^{-}(\bk) ) ]_{\_}
\\ +
( \sigma_{\mu\nu}^{ss',+}(\bk) + l_{\mu\nu}^{ss',+}(\bk) )
[ a_t^\pm(\bs p) ,
\ope{E}( a_{s}^{\dag\,-}(\bk) \circ a_{s'}^{+}(\bk) ) ]_{\_}
\bigr\}
=
\sum_{i'} \sum_{t} \int \Id^3\bs p
I_{i\mu\nu}^{i'} v_{i'}^{t,\pm}(\bs p) a_t^\pm(\bs p) .
	\end{multline}
Due to the uniqueness conditions~\eref{7.8}--\eref{7.11}, one can rewrite the
terms
$\ope{E}( a_m^{\dag\,\pm} \circ a_m^{\mp} )$
in~\eref{7.14}--\eref{7.17} in a number of equivalent ways; \eg
(see~\eref{7.8})
	\begin{equation}	\label{7.18}
\ope{E}( a_m^{\dag\,\pm} \circ a_m^{\mp} )
= \varepsilon \ope{E}( a_m^{\mp} \circ a_m^{\dag\,\pm} )
= \frac{1}{2} \ope{E}( [ a_m^{\dag\,\pm} , a_m^{\mp} ]_{\varepsilon} ) .
	\end{equation}

	Consider the general case of a charged field, $q\not=0$ (and hence
$\tau=0$). The system of equations~\eref{7.14}--\eref{7.15} is then
equivalent to
	\begin{subequations}	\label{7.19}
	\begin{align}
			\label{7.19a}
\bigl[ a_l^{\pm} & , \ope{E} ( a_m^{\dag\,\pm} \circ a_m^{\mp} ) \bigr]_{-}
= 0
\\ 			\label{7.19b}
\bigl[ a_l^{+}   & , \ope{E} ( a_m^{\dag\,-} \circ a_m^{+} ) \bigr]_{-}
+ \varepsilon \delta_{lm} a_l^{+} = 0
\\				\label{7.19c}
\bigl[ a_l^{-}   & , \ope{E} ( a_m^{\dag\,+} \circ a_m^{-} ) \bigr]_{-}
- \delta_{lm} a_l^{-} = 0 .
	\end{align}
	\end{subequations}
These (commutation) relations ensure the simultaneous fulfillment of the
Heisenberg relations~\eref{5.1} and~\eref{5.2} involving the momentum and
charge operators, respectively. To ensure also the validity of~\eref{7.16},
with $\tau=0$, and, consequently, of~\eref{5.4}, we generalize~\eref{7.19} to
	\begin{subequations}	\label{7.20}
	\begin{align}
			\label{7.20a}
\bigl[ a_l^{\pm} & , \ope{E} ( a_m^{\dag\,\pm} \circ a_n^{\mp} ) \bigr]_{-}
= 0
\\ 			\label{7.20b}
\bigl[ a_l^{+}   & , \ope{E} ( a_m^{\dag\,-} \circ a_n^{+} ) \bigr]_{-}
+ \varepsilon \delta_{lm} a_n^{+} = 0
\\				\label{7.20c}
\bigl[ a_l^{-}   & , \ope{E} ( a_m^{\dag\,+} \circ a_n^{-} ) \bigr]_{-}
- \delta_{lm} a_n^{-} = 0 ,
	\end{align}
	\end{subequations}
for any $l=(s,\bk)$, $m=(t,\bs p)$ and $n=(t,\bs q)$ (see also~\eref{6.49}).
In the way pointed in Sect.~\ref{Sect6}, one can verify that~\eref{7.20} for
any $l=(s,\bk)$, $m=(t,\bs p)$ and $n=(r,\bs p)$ entails~\eref{7.17} and
hence~\eref{5.5}. At last, to ensure the validity of all of the mentioned
conditions and a suitable transition to a case of Hermitian field, for which
$q=0$ and $\tau=1$ (see~\eref{3.12}), we generalize~\eref{7.20} to
	\begin{subequations}	\label{7.21}
	\begin{align}
			\label{7.21a}
\bigl[ a_l^{+} & , \ope{E} ( a_m^{\dag\,+} \circ a_n^{-} ) \bigr]_{-}
+ \tau \delta_{ln} a_m^{+} = 0
\\ 			\label{7.21b}
\bigl[ a_l^{-} & , \ope{E} ( a_m^{\dag\,-} \circ a_n^{+} ) \bigr]_{-}
- \varepsilon \tau \delta_{ln} a_m^{-} = 0
\\				\label{7.21c}
\bigl[ a_l^{+} & , \ope{E} ( a_m^{\dag\,-} \circ a_n^{+} ) \bigr]_{-}
+ \varepsilon \delta_{lm} a_n^{+} = 0 ,
\\				\label{7.21d}
\bigl[ a_l^{-} & , \ope{E} ( a_m^{\dag\,+} \circ a_n^{-} ) \bigr]_{-}
- \delta_{lm} a_n^{-} = 0
	\end{align}
	\end{subequations}
where $l$, $m$ and $n$ are arbitrary. As a result of~\eref{7.13-1}, which we
assume to hold, and $\tau a_l^{\dag\,\pm}=\tau a_l^{\pm}$ (see~\eref{3.12}),
the equations~\eref{7.21a} and~\eref{7.21c} (resp.~\eref{7.21b}
and~\eref{7.21d}) become identical when $\tau=1$ (and hence
$a_l^{\dag\,\pm}=a_l^{\pm}$); for $\tau=0$ the system~\eref{7.21} reduces
to~\eref{7.20}. Recalling that $\varepsilon=(-1)^{2j}$ (see~\eref{3.12}), we
can rewrite~\eref{7.21} in a more compact form as
	\begin{subequations}	\label{7.22}
	\begin{align}
			\label{7.22a}
\bigl[ a_l^{\pm} & , \ope{E} ( a_m^{\dag\,\pm} \circ a_n^{\mp} ) \bigr]_{-}
+ (\pm 1)^{2j+1} \tau \delta_{ln} a_m^{\pm} = 0
\\ 			\label{7.22b}
\bigl[ a_l^{\pm} & , \ope{E} ( a_m^{\dag\,\mp} \circ a_n^{\pm} ) \bigr]_{-}
- (\mp 1)^{2j+1} \tau \delta_{lm} a_n^{\pm} = 0 .
	\end{align}
	\end{subequations}
Since the last equation is equivalent to (see~\eref{7.13-1}) and use that
$\varepsilon=(-1)^{2j}$)
	\begin{align}
	\tag{\ref{7.22b}$^\prime$}	\label{7.22b'}
\bigl[ a_l^{\pm} & , \ope{E} ( a_m^{\pm} \circ a_n^{\dag\,\mp} ) \bigr]_{-}
+ (\pm 1)^{2j+1} \delta_{ln} a_m^{\pm} = 0 ,
	\end{align}
it is evident that the equations~\eref{7.22a} and ~\eref{7.22b} coincide for
a neutral field.

	Let us draw the main moral from the above considerations: the
equations~\eref{7.13-1} are sufficient conditions for the uniqueness of the
dynamical variables, while~\eref{7.22} are such conditions for the validity
of the Heisenberg relations~\eref{5.1}--\eref{5.5}, in which the dynamical
variables are redefined according to~\eref{7.8-1}. So, any set of operators
$a_l^\pm$ and $\ope{E}$, which are simultaneous solutions of~\eref{7.13-1}
and~\eref{7.22}, ensure uniqueness of the dynamical variables and at the same
time the validity of the Heisenberg relations.

	Consider the uniqueness problem for the solutions of the system of
equations consisting of~\eref{7.13-1}and~\eref{7.22}. Writing~\eref{7.13-1}
as
	\begin{equation}	\label{7.23}
\ope{E}( a_m^{\dag\,\pm} \circ a_n^{\mp} )
= \varepsilon \ope{E}( a_n^{\mp} \circ a_m^{\dag\,\pm} )
= \frac{1}{2} \ope{E}( [ a_m^{\dag\,\pm} , a_n^{\mp} ]_{\varepsilon} ) ,
	\end{equation}
which reduces to~\eref{7.18} for $n=m$, and using $\varepsilon=(-1)^{2j}$
(see~\eref{3.12}), one can verify that~\eref{7.22} is equivalent to
	\begin{subequations}	\label{7.24}
	\begin{align}
			\label{7.24a}
	\begin{split}
\bigl[ a_l^{+} & ,
\ope{E}( [ a_m^{+} , a_n^{\dag\,-} ]_\varepsilon ) \bigr]_{-}
+2 \delta_{ln} a_m^{+} = 0
	\end{split}
\\			\label{7.24b}
	\begin{split}
\bigl[ a_l^{+} & ,
\ope{E}( [ a_m^{\dag\,+} , a_n^{-} ]_\varepsilon ) \bigr]_{-}
+ 2 \tau \delta_{ln} a_m^{+} = 0
	\end{split}
\\			\label{7.24c}
	\begin{split}
\bigl[ a_l^{-} & ,
\ope{E}( [ a_m^{+} , a_n^{\dag\,-} ]_\varepsilon ) \bigr]_{-}
- 2 \tau \delta_{lm} a_n^{-} = 0
	\end{split}
\\			\label{7.24d}
	\begin{split}
\bigl[ a_l^{-} & ,
\ope{E}( [ a_m^{\dag\,+} , a_n^{-} ]_\varepsilon ) \bigr]_{-}
-2 \delta_{lm} a_n^{-} = 0 .
	\end{split}
	\end{align}
	\end{subequations}	
The similarity between this system of equations and~\eref{6.140} is more than
evident:~\eref{7.24} can be obtained from~\eref{6.140} by replacing
$[\cdot,\cdot]_{\varepsilon}$ with $\ope{E}([\cdot,\cdot]_{\varepsilon})$.

	As it was said earlier, the bilinear commutation
relations~\eref{6.12} and the identification of $\ope{E}$ with the normal
ordering operator $\ope{N}$,
	\begin{equation}	\label{7.25}
\ope{E} = \ope{N} ,
	\end{equation}
convert~\eref{7.23}--\eref{7.24} into identities; by invoking~\eref{6.13},
for $\eta=-\varepsilon$, the reader can check this via a direct calculation
(see also~\eref{4.21}). However, this is not the only possible solution
of~\eref{7.23}--\eref{7.24}. For example, if, in the particular case, one
defines an `anti\ndash normal' ordering operator $\ope{A}$ as a linear
mapping such that
	\begin{equation}	\label{7.26}
	\begin{split}
\ope{A}( a_m^{+}\circ a_n^{\dag\,-} )
& := \varepsilon a_n^{\dag\,-}\circ a_m^{+}  \quad
\ope{A}( a_m^{\dag\,+}\circ a_n^{-} )
:= \varepsilon a_n^{-} \circ a_m^{\dag\,+}
\\
\ope{A}( a_m^{-}\circ a_n^{\dag\,+} ) & := a_m^{-}\circ a_n^{\dag\,+}
\quad\ \ \!
\ope{A}( a_m^{\dag\,-}\circ a_n^{+} ) := a_m^{\dag\,-}\circ a_n^{+} ,
	\end{split}
	\end{equation}
then the bilinear commutation relations~\eref{6.12} and the setting
$\ope{E}=\ope{A}$ provide a solution of~\eref{7.23}--\eref{7.24}; to prove
this, apply~\eref{6.13} for $\eta=-\varepsilon$. Evidently, a linear
combination of $\ope{N}$ and $\ope{A}$, together with~\eref{6.12}, also
provides a solution of~\eref{7.23}--\eref{7.24}.%
\footnote{~%
If we admit $a_l^\pm$ to satisfy the `anomalous" bilinear commutation
relations~\eref{8.16} (see below), i.e.~\eref{6.12} with $\varepsilon$ for
$-\varepsilon$ and $(\pm1)^{2j}$ for $(\pm1)^{2j+1}$, then
$\ope{E}=\ope{N},\ope{A}$ also provides a solution
of~\eref{7.23}--\eref{7.24}. However, as it was demonstrated
in~\cite{bp-QFTinMP-scalars,bp-QFTinMP-spinors,bp-QFTinMP-vectors}, the
anomalous commutation relations are rejected if one works with the charge
symmetric Lagrangians~\eref{3.4}.%
}
Other solution of the same system of equations is given by
$\ope{E}=\id$ and operators $a_l^\pm$ satisfying~\eref{6.140}, in
particular the paracommutation relations~\eref{6.18}, and
\(
a_m^{\dag\,\pm}\circ a_n^{,\mp}
= \varepsilon a_n^{,\mp} \circ a_m^{\dag\,\pm}.
\)
The problem for the general solution of~\eref{7.23}--\eref{7.24} with respect
to $\ope{E}$ and $a_l^{\pm}$ is open at present.

	Let us introduce the \emph{particle} and \emph{antiparticle number
operators} respectively by (see~\eref{7.23}, \eref{7.8-2} and~\eref{3.15})
	\begin{equation}	\label{7.27}
	\begin{split}
\ope{N}_l
& := \frac{1}{2} \ope{E}\big( [a_l^{+},a_l^{\dag\,-}] \big)
   = \ope{E}(a_l^{+}\circ a_l^{\dag\,-})
   = (\ope{N}_l)^\dag
   =: \ope{N}_l^\dag
\\
\lindex[\mspace{-6mu}\ope{N}_l]{}{\dag}
& := \frac{1}{2} \ope{E}\big( [a_l^{\dag\,+},a_l^{-}] \big)
   = \ope{E}(a_l^{\dag\,+}\circ a_l^{-})
   = ( \lindex[\mspace{-6mu}\ope{N}_l]{}{\dag} )^\dag
  =: \lindex[\mspace{-6mu}\ope{N}_l]{}{\dag} ^\dag .
	\end{split}
	\end{equation}
As a result of the commutation relations~\eref{7.24}, with $n=m$, they
satisfy the equations%
\footnote{~%
The equations~\eref{7.28a} and~\eref{7.28b} correspond to~\eref{7.24a}
and~\eref{7.24b}, respectively, and~\eref{7.28c} and~\eref{7.28d} correspond
to the Hermitian conjugate to~\eref{7.24c} and~\eref{7.24d}, respectively.%
}
	\begin{subequations}	\label{7.28}
	\begin{align}
			\label{7.28a}
[ \ope{N}_l & , a_m^{+}]_{-} =  \delta_{lm} a_l^{+}
\\			\label{7.28b}
[ \lindex[\mspace{-6mu}\ope{N}_l]{}{\dag} & , a_m^{+}]_{-}
					= \tau \delta_{lm} a_l^{+}
\\			\label{7.28c}
[ \ope{N}_l & , a_m^{\dag\,+}]_{-} = \tau \delta_{lm} a_l^{\dag\,+}
\\			\label{7.28d}
[ \lindex[\mspace{-6mu}\ope{N}_l]{}{\dag} & , a_m^{\dag\,+}]_{-}
					= \delta_{lm} a_l^{\dag\,+}  .
	\end{align}
	\end{subequations}	
Combining~\eref{3.7}--\eref{3.10} and~\eref{5.7-3}--\eref{5.7-5}
with~\eref{7.8-1}, \eref{7.23} and~\eref{7.27}, we get the following
expressions for the operators of the (redefined) dynamical variables:
	\begin{align}	\label{7.29}	
\tope{P}_\mu
= &\; \frac{1}{1+\tau}\sum_{l} k_\mu|_{k_0=\sqrt{m^2c^2+\bk^2}}
		( \ope{N}_l + \lindex[\mspace{-6mu}\ope{N}_l]{}{\dag} )
\qquad l=(s,\bk)
\\			\label{7.30}
\tope{Q}
= &\; q \sum_{l} ( - \ope{N}_l + \lindex[\mspace{-6mu}\ope{N}_l]{}{\dag} )
\\			\label{7.31}
\tope{S}_{\mu\nu}
= &\; \frac{(-1)^{j-1/2} j \hbar}{1+\tau}\sum_{m,n}
\{ \varepsilon \sigma_{\mu\nu}^{mn,+} \ope{N}_{nm} +
	\sigma_{\mu\nu}^{mn,-} \lindex[\mspace{-6mu}\ope{N}_{mn}]{}{\dag} )
\}\big|_{ \substack{m=(s,\bk)\\ n=(s',\bk) }}
\\			 \notag
\tope{L}_{\mu\nu}
= &\; x_{0\,\mu} \tope{P}_\nu - x_{0\,\nu} \tope{P}_\mu
+ \frac{(-1)^{j-1/2} j \hbar}{1+\tau}\sum_{m,n}
  \{ \varepsilon l_{\mu\nu}^{mn,+} \ope{N}_{nm} +
	       l_{\mu\nu}^{mn,-} \lindex[\mspace{-6mu}\ope{N}_{mn}]{}{\dag} )
  \}\big|_{ \substack{m=(s,\bk)\\ n=(s',\bk) }}
\\			\label{7.32}
  &\; + \frac{\ih}{2(1+\tau)} \sum_{l}
\bigl\{
  \bigl( - \omega_{\mu\nu}^\circ(l) + \omega_{\mu\nu}^\circ(m) \bigr)
  ( \ope{N}_l + \lindex[\mspace{-6mu}\ope{N}_l]{}{\dag} )
\bigr\}\big|_{m=l=(s,\bk)}
\\			\label{7.33}
\tope{M}_{\mu\nu}^{\mathrm{sp}}
= &\; \frac{(-1)^{j-1/2} j \hbar}{1+\tau}\sum_{m,n}
\{ \varepsilon ( \sigma_{\mu\nu}^{mn,+} + l_{\mu\nu}^{mn,+} ) \ope{N}_{nm} +
	       ( \sigma_{\mu\nu}^{mn,-} + l_{\mu\nu}^{mn,-} )
				\lindex[\mspace{-6mu}\ope{N}_{mn}]{}{\dag} )
\}\big|_{ \substack{m=(s,\bk)\\ n=(s',\bk) }}
\\			\label{7.34}
\tope{M}_{\mu\nu}^{\mathrm{or}}
= &\; \frac{\ih}{2(1+\tau)} \sum_{l}
\bigl\{
  \bigl( - \omega_{\mu\nu}^\circ(l) + \omega_{\mu\nu}^\circ(m) \bigr)
  ( \ope{N}_l + \lindex[\mspace{-6mu}\ope{N}_l]{}{\dag} )
\bigr\}\big|_{m=l=(s,\bk)} .
	\end{align}	
Here $\omega_{\mu\nu}^\circ(l)$ is defined via~\eref{6.48}, we have set
	\begin{gather}	\label{7.35}
\sigma_{\mu\nu}^{mn,\pm} := \sigma_{\mu\nu}^{ss',\pm}(\bk) \quad
     l_{\mu\nu}^{mn,\pm} := 	 l_{\mu\nu}^{ss',\pm}(\bk)
\qquad\text{for } m=(s,\bk) \text{ and } n=(s',\bk) ,
\\\intertext{and (see~\eref{7.23})}
			\label{7.36}
	\begin{split}
\ope{N}_{lm}
& := \frac{1}{2} \ope{E}\big( [a_l^{+},a_m^{\dag\,-}] \big)
   = \ope{E}(a_l^{+}\circ a_m^{\dag\,-})
   = (\ope{N}_{ml})^\dag
   =: \ope{N}_{ml}^\dag
\\
\lindex[\mspace{-6mu}\ope{N}_{lm}]{}{\dag}
& := \frac{1}{2} \ope{E}\big( [a_l^{\dag\,+},a_m^{-}] \big)
   = \ope{E}(a_l^{\dag\,+}\circ a_m^{-})
   = ( \lindex[\mspace{-6mu}\ope{N}_{ml}]{}{\dag} )^\dag
  =: \lindex[\mspace{-6mu}\ope{N}_{ml}]{}{\dag} ^\dag
	\end{split}
	\end{gather}
are respectively the \emph{particle} and \emph{antiparticle transition
operators} (cf.~\cite[sec.~1]{Govorkov} in a case of parafields). Obviously,
we have
	\begin{equation}	\label{7.37}
\ope{N}_l = \ope{N}_{l l} \quad
\lindex[\mspace{-6mu}\ope{N}_l]{}{\dag}
= \lindex[\mspace{-6mu}\ope{N}_{ll}]{}{\dag}.
	\end{equation}
The choice~\eref{7.25}, evidently, reduces~\eref{7.29}--\eref{7.32}
to~\eref{4.22}, \eref{4.23}, \eref{4.29} and~\eref{4.30}, respectively.

	In terms of the operators~\eref{7.34}, the commutation
relations~\eref{7.24} can equivalently be rewritten as (see
also~\eref{7.8-2})
	\begin{subequations}	\label{7.38}
	\begin{align}
			\label{7.38a}
[ \ope{N}_{lm} & , a_n^{+}]_{-} =  \delta_{mn} a_l^{+}
\\			\label{7.38b}
[ \lindex[\mspace{-6mu}\ope{N}_{lm}]{}{\dag} & , a_n^{+}]_{-}
					= \tau \delta_{mn} a_l^{+}
\\			\label{7.38c}
[ \ope{N}_{lm} & , a_n^{\dag\,+}]_{-} = \tau \delta_{mn} a_l^{\dag\,+}
\\			\label{7.38d}
[ \lindex[\mspace{-6mu}\ope{N}_{lm}]{}{\dag} & , a_n^{\dag\,+}]_{-}
					= \delta_{mn} a_l^{\dag\,+}  .
	\end{align}
	\end{subequations}	
If $m=l$, these relations reduce to~\eref{7.28}, due to~\eref{7.35}.

	We shall end this section with the remark that the conditions for
the uniqueness of the dynamical variables and the validity of the Heisenberg
relations are quite general and are not enough for fixing some commutation
relations regardless of a number of additional assumptions made to reduce
these conditions to the system of equations~\eref{7.23}--\eref{7.24}.


\section{State vectors, vacuum and mean values}
\label{Sect8}

	Until now we have looked on the commutation relations only from pure
mathematical view-point. In this way, making a number of assumptions, we
arrived to the system~\eref{7.23}--\eref{7.24} of commutation relations.
Further specialization of this system is, however, almost impossible without
making contact with physics. For the purpose, we have to
recall~\cite{Bjorken&Drell-2,Bogolyubov&Shirkov,Roman-QFT,Itzykson&Zuber}
that the physically measurable quantities are the mean (expectation) values
of the dynamical variables (in some state) and the transition amplitudes
between different states. To make some conclusions from these basic
assumption of the quantum theory, we must rigorously said how the states are
described as vectors in system's Hilbert space $\Hil$ of states, on which all
operators considered act.

	For the purpose, we shall need the notion of the vacuum or, more
precisely, the assumption of the existence of unique vacuum state (vector)
(known also as the no\ndash particle condition). Before defining rigorously
this state, which will be denoted by $\ope{X}_0$, we shall heuristically
analyze the properties it should possess.

	First of all, the vacuum state vector $\ope{X}_0$ should represent a
state of the field without any particles. From here two conclusions may be
drawn:
	(i)~as a field is thought as a collection of particles and a
`missing' particle should have vanishing dynamical variables, those of the
vacuum should vanish too (or, more generally, to be finite constants, which
can be set equal to zero by rescaling some theory's parameters) and
	(ii)~since the operators $a_l^{-}$ and $a_l^{\dag\,-}$ are
interpreted as ones that annihilate a particle characterize by $l=(s,\bk)$
and charge $-q$ or $+q$, respectively, and one cannot destroy an `absent'
particle, these operators should transform the vacuum into the zero vector,
which may be interpreted as a complete absents of the field.  Thus, we can
expect that
	\begin{subequations}	\label{8.1}
	\begin{align}	\label{8.1a}
& \ope{D}(\ope{X}_0) = 0
\\			\label{8.1b}
& a_l^{-} (\ope{X}_0) = 0 \quad
  a_l^{\dag\,-} (\ope{X}_0) = 0 .
	\end{align}
	\end{subequations}

	Further, as the operators $a_l^{+}$ and $a_l^{\dag\,+}$ are
interpreted as ones creating a particle characterize by $l=(s,\bk)$ and charge
$-q$ or $+q$, respectively, state vectors like $a_l^{+}(\ope{X}_0)$ and
$a_l^{\dag\,+}(\ope{X}_0)$ should correspond to 1\ndash particle states. Of
course, a necessary condition for this is
	\begin{equation}	\label{8.1-1}
\ope{X}_0 \not= 0 ,
	\end{equation}
due to which the vacuum can be normalize to unit,
	\begin{equation}	\label{8.1-2}
\langle \ope{X}_0  | \ope{X}_0 \rangle = 1 ,
	\end{equation}
where $\langle \cdot|\cdot \rangle\colon\Hil\times\Hil\to\field[C]$ is the
Hermitian scalar (inner) product of $\Hil$. More generally, if
$\ope{M}(a_{l_1}^{+},a_{l_2}^{\dag\,+},\ldots)$ is a monomial only in
$i\in\field[N]$ creation operators, the vector
	\begin{equation}	\label{8.2}
\psi_{l_1 l_2 \ldots}
:=
\ope{M}(a_{l_1}^{+},a_{l_2}^{\dag\,+},\ldots) (\ope{X}_0)
	\end{equation}
may be expected to describe an $i$-particle state (with $i_1$ particles and
$i_2$ antiparticles, $i_1+i_2=i$, where $i_1$ and $i_2$ are the number of
operators $a_{l}^{+}$ and $a_{l}^{\dag\,+}$, respectively, in
$\ope{M}(a_{l_1}^{+},a_{l_2}^{\dag\,+},\ldots)$).
Moreover, as a free field is intuitively thought as a collection of particles
and antiparticles, it is natural to suppose that the vectors~\eref{8.2}
form a basis in the Hilbert space $\Hil$. But the validity of this assumption
depends on the accepted commutation relations; for its proof, when the
paracommutation relations are adopted, see the proof
of~\cite[p.~26, theorem~I-1]{Ohnuki&Kamefuchi}.

	Accepting the last assumption and recalling that the transition
amplitude between two states  is represented via the scalar product of the
corresponding to them state vectors, it is clear that for the calculation of
such an amplitude is needed an effective procedure for calculation of scalar
products of the form
	\begin{equation}	\label{8.3}
\langle \psi_{l_1 l_2 \ldots} | \varphi_{m_1 m_2 \ldots} \rangle
:=
\langle \ope{X}_0 |
( \ope{M} (a_{l_1}^{+},a_{l_2}^{\dag\,+},\ldots) )^\dag
\circ
\ope{M}' (a_{m_1}^{+},a_{m_2}^{\dag\,+},\ldots)
\ope{X}_0\rangle ,
	\end{equation}
with $\ope{M}$ and $\ope{M}'$ being monomials only in the creation operators.
Similarly, for computation of the mean value of some dynamical operator
$\ope{D}$ in a certain state, one should be equipped with a method for
calculation of scalar products like
	\begin{equation}	\label{8.4}
\langle \psi_{l_1 l_2 \ldots} | \ope{D} \varphi_{m_1 m_2 \ldots} \rangle
:=
\langle \ope{X}_0 |
( \ope{M} (a_{l_1}^{+},a_{l_2}^{\dag\,+},\ldots) )^\dag
\circ \ope{D} \circ
\ope{M}' (a_{m_1}^{+},a_{m_2}^{\dag\,+},\ldots)
\ope{X}_0\rangle .
	\end{equation}

	Supposing, for the moment, the vacuum to be defined via~\eref{8.1},
let us analyze~\eref{8.1}--\eref{8.4}. Besides, the validity
of~\eref{7.23}--\eref{7.24} will be assumed.

	From the expressions~\eref{7.8-1} and~\eref{3.7}--\eref{3.10} for the
dynamical variables, it is clear that the condition~\eref{8.1a} can be
satisfied if
	\begin{equation}	\label{8.5}
\ope{E} ( a_m^{\dag\,\pm} \circ a_n^{\mp} ) (\ope{X}_0) = 0,
	\end{equation}
which, in view of~\eref{7.23}, is equivalent to any one of the equations
    \begin{subequations}    \label{8.6}
	\begin{align}	\label{8.6a}
\ope{E} ( a_m^{\pm} \circ a_n^{\dag\,\mp} ) (\ope{X}_0) & = 0
\\		    	\label{8.6b}
\ope{E} ( [a_m^{\pm} , a_n^{\dag\,\mp}]_{\varepsilon} ) (\ope{X}_0) & = 0.
	\end{align}
    \end{subequations}
Equation~\eref{8.5} is quite natural as it expresses the vanishment of all
modes of the vacuum corresponding to different polarizations, 4\ndash
momentum and charge. It will be accepted hereafter.

	By means of~\eref{8.6} and the commutation relations~\eref{7.24} in
the form~\eref{7.38}, in particular~\eref{7.28}, one can explicitly calculate
the action of any one of the operators~\eref{7.29}--\eref{7.34} on the
vectors~\eref{8.2}: for the purpose one should simply to commute the
operators $\ope{N}_{lm}$ (or $\ope{N}_{l}=\ope{N}_{ll}$) with the creation
operators in~\eref{8.2} according to~\eref{7.38} (resp.~\eref{7.28}) until
they act on the vacuum and, hence, giving zero, as a result of~\eref{8.6}
and~\eref{7.38} (resp.~\eref{7.28}). In particular, we have the equations
   ($k_0=\sqrt{m^2c^2+\bk^2}$):
	\begin{align}	\label{8.6-1}	
	\begin{split}
&
\tope{P}_\mu \bigl( a_l^{+}(\ope{X}_0) \bigr)
   = k_\mu a_l^{+}(\ope{X}_0) \quad
\tope{P}_\mu \bigl( a_l^{\dag\,+}(\ope{X}_0) \bigr)
   = k_\mu a_l^{\dag\,+}(\ope{X}_0)
\quad l=(s,\bk)
	\end{split}
\displaybreak[2]\\		\label{8.6-2}
&
\tope{Q} \bigl( a_l^{+}(\ope{X}_0) \bigr) = -q a_l^{+}(\ope{X}_0) \quad
\tope{Q} \bigl( a_l^{\dag\,+}(\ope{X}_0) \bigr) = + q a_l^{\dag\,+}(\ope{X}_0)
\displaybreak[2]\\		\label{8.6-3}
	\begin{split}
&
\tope{S}_{\mu\nu} \bigl( a_l^{+}\big|_{ l=(s,\bk) } (\ope{X}_0) \bigr)
=
\frac{(-1)^{j-1/2} j \hbar}{1+\tau}\sum_{t}
\{ \varepsilon \sigma_{\mu\nu}^{lm,+} + \tau \sigma_{\mu\nu}^{ml,-}
\} \big|_{ m=(t,\bk) }
   a_m^{+}|_{m=(t,k)}(\ope{X}_0)
\\
&
\tope{S}_{\mu\nu} \bigl( a_l^{\dag\,+}\big|_{ l=(s,\bk) } (\ope{X}_0) \bigr)
=
\frac{(-1)^{j-1/2} j \hbar}{1+\tau}\sum_{t}
\{ \varepsilon \tau \sigma_{\mu\nu}^{lm,+} + \sigma_{\mu\nu}^{ml,-}
\} \big|_{ m=(t,\bk) }
   a_m^{\dag\,+}|_{m=(t,k)}(\ope{X}_0)
	\end{split}
\displaybreak[2]\\		\label{8.6-4}
	\begin{split}
&
\tope{L}_{\mu\nu} \bigl( a_l^{+}\big|_{ l=(s,\bk) } (\ope{X}_0) \bigr)
=
(x_{0\,\mu} k_\nu-x_{0\,\nu} k_\mu) (a_l^{+}) (\ope{X}_0)
- \ih \bigl( \omega_{\mu\nu}^\circ(l) (a_l^{+})\bigr) (\ope{X}_0)
\\ &
\hphantom{
\tope{L}_{\mu\nu} \bigl( a_l^{+}\big|_{ l=(s,\bk) } (\ope{X}_0) \bigr) =
}
+
\frac{(-1)^{j-1/2} j \hbar}{1+\tau}\sum_{t}
\{ \varepsilon l_{\mu\nu}^{lm,+} + \tau l_{\mu\nu}^{ml,-}
\} \big|_{ m=(t,\bk) }
   a_m^{+}|_{m=(t,k)}(\ope{X}_0)
\\
&
\tope{L}_{\mu\nu} \bigl( a_l^{\dag\,+}\big|_{ l=(s,\bk) } (\ope{X}_0) \bigr)
=
(x_{0\,\mu} k_\nu-x_{0\,\nu} k_\mu) (a_l^{\dag\,+})(\ope{X}_0)
- \ih \bigl( \omega_{\mu\nu}^\circ(l) (a_l^{\dag\,+})\bigr) (\ope{X}_0)
\\ &
\hphantom{
\tope{L}_{\mu\nu} \bigl( a_l^{\dag\,+}\big|_{ l=(s,\bk) } (\ope{X}_0) \bigr)
}
+
\frac{(-1)^{j-1/2} j \hbar}{1+\tau}\sum_{t}
\{ \varepsilon \tau l_{\mu\nu}^{lm,+} + l_{\mu\nu}^{ml,-}
\} \big|_{ m=(t,\bk) }
   a_m^{\dag\,+}|_{m=(t,k)}(\ope{X}_0)
	\end{split}
\displaybreak[2]\\		\label{8.6-5}
	\begin{split}
&
\tope{M}_{\mu\nu}^{\mathrm{sp}}
   		\bigl( a_l^{+}\big|_{ l=(s,\bk) }  (\ope{X}_0) \bigr)
= \frac{(-1)^{j-1/2} j \hbar}{1+\tau}\sum_{t}
\{ \varepsilon (\sigma_{\mu\nu}^{lm,+} + l_{\mu\nu}^{lm,+} )
\\ & \mspace{300mu}
    +
    	 \tau (\sigma_{\mu\nu}^{ml,-} + l_{\mu\nu}^{ml,-} )
\} \big|_{ m=(t,\bk) }
   a_m^{+}|_{m=(t,k)}(\ope{X}_0)
\\
&
\tope{M}_{\mu\nu}^{\mathrm{sp}}
   		\bigl( a_l^{\dag\,+}\big|_{ l=(s,\bk) } (\ope{X}_0) \bigr)
= \frac{(-1)^{j-1/2} j \hbar}{1+\tau}\sum_{t}
\{ \varepsilon \tau ( \sigma_{\mu\nu}^{lm,+} + l_{\mu\nu}^{lm,+} )
\\ & \mspace{300mu}
   +
   		    ( \sigma_{\mu\nu}^{ml,-} + l_{\mu\nu}^{ml,-} )
\} \big|_{ m=(t,\bk) }
   a_m^{\dag\,+}|_{m=(t,k)}(\ope{X}_0)
	\end{split}
\displaybreak[2]\\		\label{8.6-6}
	\begin{split}
   &
\tope{M}_{\mu\nu}^{\mathrm{or}} \bigl( \ta_l^{+} (\ope{X}_0) \bigr)
=
- \ih \bigl( \omega_{\mu\nu}^\circ(l) (\ta_l^{+})\bigr) (\ope{X}_0)
\quad
\tope{M}_{\mu\nu}^{\mathrm{or}} \bigl( \ta_l^{\dag\,+} (\ope{X}_0) \bigr)
=
- \ih \bigl( \omega_{\mu\nu}^\circ(l) (\ta_l^{\dag\,+})\bigr) (\ope{X}_0) .
	\end{split}
	\end{align}	
   These equations and similar, but more complicated, ones with an arbitrary
   monomial in the creation operators for $a_l^{+}$ or $a_l^{\dag\,+}$ are the
   base for the particle interpretation of the quantum theory of free fields.
   For instance, in view of~\eref{8.6-1} and~\eref{8.6-2}, the state vectors
    $a_l^{+}(\ope{X}_0)$ and $a_l^{\dag\,+}(\ope{X}_0)$ are interpreted as
   ones representing particles with 4\ndash momentum
   $(\sqrt{m^2c^2+\bk^2},\bk)$ and charges $-q$ and $+q$, respectively;
   similar multiparticle interpretation can be given to the general
   vectors~\eref{8.2} too.

	The equations~\eref{8.6-1}--\eref{8.6-4} completely agree with
similar ones obtained
in~\cite{bp-QFTinMP-scalars,bp-QFTinMP-spinors,bp-QFTinMP-vectors} on the
base of the bilinear commutation relations~\eref{6.12}.

	By means of~\eref{8.5}, the expression~\eref{8.4} can be represented
as a linear combination of terms like~\eref{8.3}. Indeed, as $\ope{D}$ is a
linear combinations of terms like $\ope{E}(a_m^{\dag\,\pm} \circ a_n^{\mp})$,
by means of the relations~\eref{7.24} we can commute each of these terms with
the creation (resp.\ annihilation) operators in the monomial
$\ope{M}' (a_{m_1}^{+},a_{m_2}^{\dag\,+},\ldots)$
(resp.\
\(
( \ope{M} (a_{l_1}^{+},a_{l_2}^{\dag\,+},\ldots) )^\dag
=
\ope{M}^{\prime\prime} (a_{l_1}^{\dag\,-},a_{l_2}^{-},\ldots)
\)%
)
and thus moving them to the right (resp.\ left) until they act on the vacuum
$\ope{X}_0$, giving the zero vector --- see~\eref{8.5}. In this way the
matrix elements of the dynamical variables, in particular their mean values,
can be expressed as linear combinations of scalar products of the
form~\eref{8.3}. Therefore the supposition~\eref{8.5} reduces the computation
of mean values of dynamical variables to the one of the vacuum mean value of
a product (composition) of creation and annihilation operators in which the
former operators stand to the right of the latter ones. (Such a product of
creation and annihilation operators can be called  their `antinormal'
product; \cf the properties~\eref{7.26} of the antinormal ordering operator
$\ope{A}$.)

	The calculation of such mean values, like~\eref{8.3} for states
$\psi,\varphi\not=\ope{X}_0$, however, cannot be done (on the base
of~\eref{7.23}--\eref{7.24}, \eref{8.5} and~\eref{8.1a}) unless additional
assumption are made. For the purpose one needs some kind of commutation
relations by means of which the creation (resp.\ annihilation) operators on
the r.h.s.\ of~\eref{8.3} to be moved to the left (resp.\ right) until they
act on the left (resp.\ right) vacuum vector $\ope{X}_0$; as a result of this
operation, the expressions between the two vacuum vectors in~\eref{8.3}
should transform into a linear combination of constant terms and such with no
contribution in~\eref{8.3}. (Examples of the last type of terms are
$\ope{E}(a_m^{\dag\,\pm}\circ a^\mp)$ and normally ordered products of
creation and annihilation operators.) An alternative procedure may consists
in defining axiomatically the values of all or some of the mean
values~\eref{8.3} or, more stronger, the explicit action of all or some of
the operators, entering in the r.h.s.\ of~\eref{8.3}, on the vacuum.%
\footnote{~%
Such an approach resembles the axiomatic description of the scattering
matrix~\cite{Bogolyubov&et_al.-QFT,Bogolyubov&et_al.-AxQFT,Bogolyubov&Shirkov}.%
}
It is clear, both proposed schemes should be consistent with the
relations~\eref{7.23}--\eref{7.24}, \eref{8.1b} and~\eref{8.5}--\eref{8.6}.

	Let us summarize the problem before us: the operator $\ope{E}$
in~\eref{7.23}--\eref{7.24} has to be fixed and a method for computation of
scalar products like~\eref{8.3} should be given provided the vacuum vector
$\ope{X}_0$ satisfies~\eref{8.1b}, \eref{8.1-1}, \eref{8.1-2} and~\eref{8.5}.
Two possible ways for exploration of this problem were indicated above.

	Consider the operator $\ope{E}$. Supposing
$\ope{E}(a_m^{\dag\,\pm}\circ a_n^\mp)$
to be a function only of $a_m^{\dag\,\pm}$ and $a_n^\mp$, we, in view
of~\eref{8.1b}, can write
\(
\ope{E}(a_m^{\dag\,\pm}\circ a_n^\mp)
=
f^\pm(a_m^{\dag\,\pm}\circ a_n^\mp) \circ b
\)
with $b=a_n^-$ (upper sign) or $b=a_m^{\dag\,-}$ (lower sign) and some
functions $f^\pm$. Applying~\eref{7.23}, we obtain (do not sum over $l$)
	\begin{alignat*}{2}
&
\ope{E}(a_m^{\dag\,+}\circ a_l^-)
= f^+(a_m^{\dag\,+} , a_l^-) \circ a_l^-
& \quad  &
\ope{E}(a_m^{+}\circ a_l^{\dag\,-})
= f^-(a_m^{+} , a_l^{\dag\,-}) \circ a_l^{\dag\,-}
\\
&
\ope{E}(a_l^{-}\circ a_m^{\dag\,+})
= \varepsilon f^+(a_m^{\dag\,+} , a_l^-) \circ a_l^-
& \quad  &
\ope{E}(a_l^{\dag\,-}\circ a_m^{+})
= \varepsilon f^-(a_m^{+} , a_l^{\dag\,-}) \circ a_l^{\dag\,-} .
	\end{alignat*}
Since $\ope{E}$ is a linear operator, the expression
$\ope{E}(a_m^{\dag\,\pm}\circ a_n^\mp)$
turns to be a linear and homogeneous function of $a_m^{\dag\,\pm}$ and
$a_n^\mp$, which immediately implies $f^\pm(A,B)=\lambda^\pm A$ for
operators $A$ and $B$ and some constants $\lambda^\pm\in\field[C]$. For
future convenience, we assume $\lambda^\pm=1$, which can be achieved via a
suitable renormalization of the creation and annihilation operators.%
\footnote{~%
Since $\lambda^+=0$ or/and  $\lambda^-=0$ implies $\ope{D}=0$, due
to~\eref{7.8-1}, these values are excluded for evident reasons.%
}
Thus, the last equations reduce to
	\begin{subequations}	\label{8.7}
	\begin{alignat}{2}	\label{8.7a}
&
\ope{E}(a_m^{\dag\,+}\circ a_l^-) = a_m^{\dag\,+} \circ a_l^-
& \quad  &
\ope{E}(a_m^{+}\circ a_l^{\dag\,-}) = a_m^{+} \circ a_l^{\dag\,-}
\\				\label{8.7b}
&
\ope{E}(a_l^{-}\circ a_m^{\dag\,+}) = \varepsilon a_m^{\dag\,+} \circ a_l^-
& \quad  &
\ope{E}(a_l^{\dag\,-}\circ a_m^{+}) = \varepsilon a_m^{+} \circ a_l^{\dag\,-} .
	\end{alignat}
	\end{subequations}
Evidently, these equations convert~\eref{7.23},~\eref{8.5} and~\eref{8.6}
into identities. Comparing~\eref{8.7} and~\eref{4.20}, we see that the
identification
	\begin{equation}	\label{8.8}
\ope{E} = \ope{N}
	\end{equation}
of the operator $\ope{E}$ with the normal ordering operator $\ope{N}$ is
quite natural. However, for our purposes, this identification is not
necessary as only the equations~\eref{8.7}, not the general definition of
$\ope{N}$, will be employed.

	As a result of~\eref{8.7}, the commutation relations~\eref{7.24} now
read:
	\begin{subequations}	\label{8.9}
	\begin{align}
			\label{8.9a}
	\begin{split}
[ a_l^{+} & , a_m^{+} \circ a_n^{\dag\,-} ]_{\_}
+  \delta_{ln} a_m^{+} = 0
	\end{split}
\\			\label{8.9b}
	\begin{split}
[ a_l^{+} & , a_m^{\dag\,+} \circ a_n^{-} ]_{\_}
+  \tau \delta_{ln} a_m^{+} = 0
	\end{split}
\\			\label{8.9c}
	\begin{split}
[ a_l^{-} & , a_m^{+} \circ a_n^{\dag\,-} ]_{\_}
-  \tau \delta_{lm} a_n^{-} = 0
	\end{split}
\\			\label{8.9d}
	\begin{split}
[ a_l^{-} & , a_m^{\dag\,+} \circ a_n^{-} ]_{\_}
- \delta_{lm} a_n^{-} = 0 .
	\end{split}
	\end{align}
	\end{subequations}	
(In a sense, these relations are `one half' of the (para)commutation
relations~\eref{6.140}: the latter are a sum of the former and the ones
obtained from~\eref{8.9} via the changes
\(
a_m^+\circ a_n^{\dag\,-} \mapsto \varepsilon a_n^{\dag\,-}\circ a_m^+
\)
and
\(
a_m^{\dag\,+}\circ a_n^{-} \mapsto \varepsilon a_n^{-}\circ a_m^{\dag\,+} ;
\)
the last relations correspond to~\eref{7.24} with $\ope{E}=\ope{A}$,
$\ope{A}$ being the antinormal ordering operator --- see~\eref{7.26}. Said
differently, up to the replacement
$a_i^\pm\mapsto \sqrt{2}a_l^\pm$ for all $l$,
the relations~\eref{8.9} are identical with~\eref{6.140} for $\varepsilon=0$;
as noted in~\cite[the remarks following theorem~2 in sec.~1]{Govorkov}, this
is a quite exceptional case from the view\ndash point of parastatistics
theory.)
By means of~\eref{6.13} for $\eta=-\varepsilon$, one can verify
that equations~\eref{8.9} agree with the bilinear commutation
relations~\eref{6.12}, i.e.~\eref{6.12} convert~\eref{8.9} into identities.

	The equations~\eref{8.7} imply the following explicit forms of the
number operators~\eref{7.27} and the transition operators~\eref{7.36}:
	\begin{gather}	\label{8.9-1}
\ope{N}_l = a_{l}^{+}\circ a_{l}^{\dag\,-} \quad
\lindex[\mspace{-6mu}\ope{N}_l]{}{\dag} = a_{l}^{\dag\,+}\circ a_{l}^{-}
\\			\label{8.9-2}
\ope{N}_{lm} = a_{l}^{+}\circ a_{m}^{\dag\,-} \quad
\lindex[\mspace{-6mu}\ope{N}_{lm}]{}{\dag} = a_{l}^{\dag\,+}\circ a_{m}^{-} .
	\end{gather}
As a result of them, the equations~\eref{7.29}--\eref{7.32} are simply a
different form of writing of~\eref{4.22}, \eref{4.23}, \eref{4.29}
and~\eref{4.30}, respectively.

	Let us return to the problem of calculation of vacuum mean values of
antinormal ordered products like~\eref{8.3}. In view of~\eref{8.1b}
and~\eref{8.1-2}, the simplest of them are
	\begin{equation}	\label{8.10}
\langle\ope{X}_0 |\lambda\id_\Hil (\ope{X}_0)\rangle = \lambda
\quad
\langle\ope{X}_0 | \ope{M}^\pm(\ope{X}_0)\rangle = 0
	\end{equation}
where $\lambda\in\field[C]$ and $\ope{M}^+$ (resp.\ $\ope{M}^-$) is any
monomial of degree not less than~1 only in the creation (resp.\ annihilation)
operators; \eg
\(
\ope{M}^\pm =
a_{l}^{\pm},
a_{l}^{\dag\,\pm},
a_{l_1}^{\pm}\circ a_{l_2}^{\pm},
a_{l_1}^{\pm}\circ a_{l_2}^{\dag\,\pm} .
\)
These equations, with $\lambda=1$, are another form of what is called the
\emph{stability of the vacuum}: if $\ope{X}_{i}$ denotes an $i$\ndash
particle state, $i\in\field[N]\cup\{0\}$, then, by virtue of~\eref{8.10} and
the particle interpretation of~\eref{8.2}, we have
	\begin{equation}	\label{8.11}
\langle\ope{X}_i | \ope{X}_0 \rangle = \delta_{i0} ,
	\end{equation}
\ie the only non-forbidden transition into (from) the vacuum is from (into)
the vacuum. More generally, if $\ope{X}_{i^{\prime},0}$ and
$\ope{X}_{0,j^{\prime\prime}}$ denote respectively $i^{\prime}$\ndash particle
and $j^{\prime\prime}$\ndash antiparticle states, with
$\ope{X}_{0,0}:=\ope{X}_0$, then
	\begin{equation}	\label{8.12}
\langle\ope{X}_{i^{\prime},0} | \ope{X}_{0,j^{\prime\prime}} \rangle
= \delta_{i^{\prime} 0} \delta_{0 j^{\prime\prime}} ,
	\end{equation}
\ie transitions between two states consisting entirely of particles and
antiparticles, respectively, are forbidden unless both states coincide with
the vacuum.
	Since we are dealing with free fields, one can expect that the
amplitude of a transitions from an
($i^{\prime}$\ndash particle~$+$~$j^{\prime}$\ndash antiparticle)
state $\ope{X}_{i^{\prime},j^{\prime}}$
into an
($i^{\prime\prime}$\ndash particle~$+$~$j^{\prime\prime}$\ndash antiparticle)
state $\ope{X}_{i^{\prime\prime},j^{\prime\prime}}$
is
	\begin{equation}	\label{8.12-1}
\langle \ope{X}_{i^{\prime},j^{\prime}} |
\ope{X}_{i^{\prime\prime},j^{\prime\prime}} \rangle
=
\delta_{i^{\prime} i^{\prime\prime}}
\delta_{j^{\prime} j^{\prime\prime}} ,
	\end{equation}
but, however, the proof of this hypothesis requires new assumptions
(\emph{vide infra}).

	Let us try to employ~\eref{8.9} for calculation of expressions
like~\eref{8.3}. Acting with~\eref{8.9} and their Hermitian conjugate on the
vacuum, in view of~\eref{8.1b}, we get
	\begin{equation}	\label{8.13}
	\begin{split}
&
a_m^{+} \circ (- a_n^{\dag\,-} \circ a_l^{+} + \delta_{ln} \id_\Hil )
(\ope{X}_0) = 0
\quad\
a_n^{\dag\,+} \circ ( a_m^{-} \circ a_l^{\dag\,+} - \delta_{lm} \id_\Hil )
(\ope{X}_0) = 0
\\
&
a_m^{\dag\,+} \circ (- a_n^{-} \circ a_l^{+} + \tau \delta_{ln} \id_\Hil )
(\ope{X}_0) = 0
\quad
a_n^{+} \circ ( a_m^{\dag\,-} \circ a_l^{\dag\,+} - \tau \delta_{lm} \id_\Hil )
(\ope{X}_0) = 0 .
	\end{split}
	\end{equation}
These equalities, as well as~\eref{8.9}, cannot help directly to compute
vacuum mean values of antinormally ordered products of creation and
annihilation operators. But the equations~\eref{8.13} suggest the
restrictions%
\footnote{~%
Since the operators $a_l^{\pm}$ and $a_l^{\dag\,\pm}$ are, generally,
degenerate (with no inverse ones), we cannot say that~\eref{8.13}
implies~\eref{8.14}.%
}
	\begin{equation}	\label{8.14}
	\begin{split}
&
a_l^{\dag\,-} \circ a_m^{+} (\ope{X}_0) = \delta_{lm} \ope{X}_0
\quad
a_l^{-} \circ a_m^{\dag\,+}(\ope{X}_0) = \delta_{lm} \ope{X}_0
\\
&
a_l^{-} \circ a_m^{+} (\ope{X}_0) = \tau \delta_{lm} \ope{X}_0
\quad
a_l^{\dag\,-} \circ a_m^{\dag\,+}  (\ope{X}_0) = \tau \delta_{lm} \ope{X}_0
	\end{split}
	\end{equation}
to be added to the definition of the vacuum. These conditions
convert~\eref{8.13} into identities and, in this sense agree with~\eref{8.9}
and, consequently, with the bilinear commutation relations~\eref{6.12}.
Recall~\cite{Green-1953,Ohnuki&Kamefuchi}, the relations~\eref{8.14} are
similar to ones accepted in the parafield theory and coincide with that for
parastatistics of order $p=1$; however, here we do not suppose the validity
of the paracommutation relations~\eref{6.18} (or~\eref{6.140}). Equipped
with~\eref{8.14}, one is able to calculate the r.h.s.\ of~\eref{8.3} for any
monomial $\ope{M}$ (resp.\ $\ope{M}'$) and monomials $\ope{M}'$ (resp.\
$\ope{M}$) of degree~1, $\deg \ope{M}'=1$ (resp.\ $\deg \ope{M}=1$).%
\footnote{~%
For $\deg \ope{M}'=0$ (resp.\ $\deg \ope{M}'=0$) --- see~\eref{8.10}.%
}
Indeed,~\eref{8.14},~\eref{8.1b} and~\eref{8.1-2} entail:
	\begin{equation}	\label{8.15}
	\begin{split}
&
  \langle \ope{X}_0 | a_l^{\dag\,-} \circ a_m^{+} (\ope{X}_0) \rangle
= \langle \ope{X}_0 | a_l^{-} \circ a_m^{\dag\,+} (\ope{X}_0) \rangle
= \delta_{lm}
\\
&
  \langle \ope{X}_0 | a_l^{-} \circ a_m^{+} (\ope{X}_0) \rangle
= \langle \ope{X}_0 | a_l^{\dag\,-} \circ a_m^{\dag\,+} (\ope{X}_0) \rangle
= \tau \delta_{lm}
\\
&
  \langle \ope{X}_0 |
	(\ope{M}(a_{l_1}^{+},a_{l_2}^{\dag\,+},\cdots))^\dag \circ
	a_m^{+} (\ope{X}_0) \rangle
= \langle \ope{X}_0 |
	(\ope{M}(a_{l_1}^{+},a_{l_2}^{\dag\,+},\cdots))^\dag \circ
	a_m^{\dag\,+} (\ope{X}_0) \rangle
= 0
\quad \deg\ope{M}\ge 2
\\
&
\langle \ope{X}_0 |
a_l^{-} \circ \ope{M}(a_{m_1}^{+},a_{m_2}^{\dag\,+},\cdots)
(\ope{X}_0) \rangle
=
\langle \ope{X}_0 |
a_l^{\dag\,-} \circ \ope{M}(a_{m_1}^{+},a_{m_2}^{\dag\,+},\cdots)
(\ope{X}_0) \rangle
= 0
\qquad \deg\ope{M}\ge 2 .
	\end{split}
	\end{equation}
Hereof the equation~\eref{8.12-1} for $i^{\prime}+j^{\prime}=1$
(resp.\ $i^{\prime\prime}+j^{\prime\prime}=1$) and arbitrary
$i^{\prime\prime}$ and $j^{\prime\prime}$
(resp.\ $i^{\prime}$ and $j^{\prime}$) follows.

	However, it is not difficult to be realized, the calculation
of~\eref{8.3} in cases more general than~\eref{8.10} and~\eref{8.15} is not
possible on the base of the assumptions made until now.%
\footnote{~%
It should be noted, the conditions~\eref{8.1b}--\eref{8.1-2} and~\eref{8.14}
are enough for calculating~\eref{8.3} if the relations~\eref{6.140}, or
their version~\eref{6.18}, are accepted (cf.~\cite{Green-1953}). The cause
for that difference is in replacements like
$[a_m^{+},a_n^{\dag-}]_{\_} \mapsto 2 a_m^{+}\circ a_n^{\dag\,-}$, when one
passes from~\eref{6.140} to~\eref{8.9}; the existence of terms like
$a_n^{\dag\,-}\circ a_m^{+}\circ a_l^{+}$ in~\eref{6.140} is responsible for
the possibility to calculate~\eref{8.3}, in case~\eref{6.140} hold.%
}
At this point, one is
free so set in an arbitrary way the r.h.s.\ of~\eref{8.3} in the mentioned
general case or to add to~\eref{8.9} (and, possibly,~\eref{8.14}) other
(commutation) relations by means of which the r.h.s.\ of~\eref{8.3} to be
calculated explicitly; other approaches, \eg some mixture of the just pointed
ones, for finding the explicit form of~\eref{8.3} are evidently also
possible.
	Since expressions like~\eref{8.3} are directly connected with
observable experimental results, the only criterion for solving the problem
for calculating the r.h.s.\ of~\eref{8.3} in the general case can be the
agreement with the existing experimental data. As it is
known~\cite{Bjorken&Drell-2,Bogolyubov&Shirkov,Roman-QFT,Itzykson&Zuber}, at
present (almost?) all of them are satisfactory described within the framework
of the bilinear commutation relations~\eref{6.12}. This means that, from
physical point of view, the theory should be considered as realistic one if
the r.h.s.\ of~\eref{8.3}  is the same as if~\eref{6.12} are valid or is
reducible to it for some particular realization of an accepted method of
calculation, \eg if one accepts some commutation relations, like the
paracommutation ones, which are a generalization of~\eref{6.12} and reduce to
them as a special case (see, e.g.,~\eref{6.18}). It should be noted, the
conditions~\eref{8.1b}--\eref{8.1-2} and~\eref{8.14} are enough for
calculating~\eref{8.3} if~\eref{6.140}, or its versions~\eref{6.15}
or~\eref{6.18}, are accepted (cf.~\cite{Green-1953}). The causes for that
difference are replacements like
$ [a_m^{+},a_n^{\dag\,-}]_{\_} \mapsto 2 a_m^{+}\circ a_n^{\dag\,-} $,
when one passes from~\eref{6.140} to~\eref{8.9}; the existence of terms like
$a_n^{\dag\,-} \circ a_m^{+} a_l^{+}$
in~\eref{6.140} are responsible for the possibility to calculate~\eref{8.3}.

	If evidences appear for events for which ~\eref{8.3} takes other
values, one should look, e.g., for other commutation relations leading to
desired mean values. As an example of the last type can be pointed the
following \emph{anomalous bilinear commutation relations} (cf.~\eref{6.12})
	\begin{align}	\notag
&[a_{l}^{\pm}, a_{m}^{\pm} ]_{\varepsilon}
	= 0
&&
[a_{l}^{\dag\,\pm}, a_{m}^{\dag\,\pm} ]_{\varepsilon}
	= 0
\\	\notag
&[a_{l}^{\mp}, a_{m}^{\pm} ]_{\varepsilon}
	= (\pm1)^{2j} \tau \delta_{lm} \id_\Hil
&&
[a_{l}^{\dag\,\mp}, a_{m}^{\dag\,\pm} ]_{\varepsilon}
	= (\pm1)^{2j} \tau \delta_{lm} \id_\Hil
\\	\notag
&[a_{l}^{\pm}, a_{m}^{\dag\,\pm} ]_{\varepsilon}
	= 0
&&
[a_{l}^{\dag\,\pm}, a_{m}^{\pm} ]_{\varepsilon}
	= 0
\\	\label{8.16}
&[a_{l}^{\mp}, a_{m}^{\dag\,\pm} ]_{\varepsilon}
	= (\pm1)^{2j} \delta_{lm} \id_\Hil
&&
[a_{l}^{\dag\,\mp}, a_{m}^{\pm} ]_{\varepsilon}
	= (\pm1)^{2j} \delta_{lm} \id_\Hil ,
	\end{align}
which should be imposed after expressions like
$\ope{E}(a_m^{\dag\,\pm}\circ a_n^{\mp})$ are explicitly calculated. These
relations convert~\eref{8.9} and~\eref{8.14} into identities and by their
means the r.h.s.\ of~\eref{8.3} can be calculated explicitly, but, as it is
well
known~\cite{Bjorken&Drell-2,Bogolyubov&Shirkov,Roman-QFT,Itzykson&Zuber,
Streater&Wightman} they lead to deep contradictions in the theory, due to
which should be rejected.%
\footnote{~%
As it was demonstrated
in~\cite{bp-QFTinMP-scalars,bp-QFTinMP-spinors,bp-QFTinMP-vectors}, a
quantization like~\eref{8.16} contradicts to (is rejected by) the charge
symmetric Lagrangians~\eref{3.4}.%
}

	At present, it seems, the bilinear commutation relations~\eref{6.12}
are the only known commutation relations which satisfy all of the mentioned
conditions and simultaneously provide an evident procedure for effective
calculation of all expressions of the form~\eref{8.3}.
(Besides, for them and for the paracommutation
relations the vectors~\eref{8.2} form a base, the Fock base, for the system's
Hilbert space of states~\cite{Ohnuki&Kamefuchi}.) In this connection, we want
to mention that the paracommutation relations~\eref{6.140} (or their
conventional version~\eref{6.18}), if imposed as additional restrictions to
the theory together with~\eref{8.9}, reduce in this particular case
to~\eref{6.12} as the conditions~\eref{8.14} show that we are dealing with a
parafield of order $p=1$, \ie with an ordinary
field~\cite{Greenberg&Messiah-1965,Ohnuki&Kamefuchi}.%
\footnote{~%
Notice, as a result of~\eref{8.9}, the relations~\eref{6.140} correspond
to~\eref{7.24} for $\ope{E}=\ope{A}$, with $\ope{A}$ being the antinormal
ordering operator (see~\eref{7.26}).%
}

	Ending this section, let us return to the definition of the vacuum
$\ope{X}_0$. It, generally, depends on the adopted commutation relations.
For instance, in a case of the bilinear commutation relations~\eref{6.12} it
consists of the equations~\eref{8.1a}--\eref{8.1-2}, while in a case of the
paracommutation relations~\eref{6.140} (or other ones
generalizing~\eref{6.12}) it includes~\eref{8.1a}--\eref{8.1-2}
and~\eref{8.14}.


\section
[Commutation relations for several coexisting different free fields]
{Commutation relations for\\ several coexisting different free fields}
\label{Sect9}

	Until now we have considered commutation relations for a single free
field, which can be scalar, or spinor or vector one. The present section is
devoted to similar treatment of a system consisting of several, not less than
two, \emph{different free} fields. In our context, the fields may differ by
their masses and/or charges and/or spins; e.g., the system may consist of
charged scalar field, neutral scalar field, massless spinor field, massive
spinor field and massless neural vector field. It is \emph{a priori} evident,
the commutation relations regarding only one field of the system should be as
discussed in the previous sections. The problem is to be derived/postulated
commutation relations concerning \emph{different} fields.  It will be shown,
the developed Lagrangian formalism provides a natural base for such an
investigation and makes superfluous some of the assumptions made, for
example, in~\cite[p.~B~1159, left column]{Greenberg&Messiah-1965} or
in~\cite[sec.~12.1]{Ohnuki&Kamefuchi}, where systems of different parafields
are explored.

	To begin with, let us introduce suitable notation. With the indices
$\alpha,\beta,\gamma=1,2,\dots,N$ will be distinguished the different fields
of the system, with $N\in\field[N]$, $N\ge2$, being their number, and the
corresponding to them quantities. Let $q^\alpha$ and $j^\alpha$ be
respectively the charge and spin of the $\alpha$\ndash th field. Similarly
to~\eref{3.12}, we define
	\begin{equation}
			\label{9.1}
	\begin{split}
& j^\alpha :=
	\begin{cases}
0  	     &\text{for scalar $\alpha$\ndash th field} \\
\frac{1}{2}  &\text{for spinor $\alpha$\ndash th field} \\
1	     &\text{for vector $\alpha$\ndash th field}
	\end{cases}
\qquad
\tau^\alpha :=
	\begin{cases}
1  &\text{for $q^\alpha=0$ (neutral (Hermitian) field)} \\
0  &\text{for $q^\alpha\not=0$ (charged (non-Hermitian) field)}
	\end{cases}
\\
& \varepsilon^\alpha
:= (-1)^{2j^\alpha}
=
	\begin{cases}
+ 1  &\text{for integer $j^\alpha$ (bose fields)} \\
- 1  &\text{for half-integer $j^\alpha$ (fermi fields)}
	\end{cases}
\ .
	\end{split}
	\end{equation} 

	Suppose $\ope{L}^\alpha$ is the Lagrangian of the $\alpha$\ndash
field. For definiteness, we assume $\ope{L}^\alpha$ for all $\alpha$ to be
given by one and the same set of equations, viz.~\eref{3.1}, or~\eref{3.3}
or~\eref{3.4}. To save some space, below the case~\eref{3.4}, corresponding
to charge symmetric Lagrangians, will be considered in more details; the
reader can explore other cases as exercises.

	Since the Lagrangian of our system of free fields is
	\begin{equation}	\label{9.2}
\ope{L} := \sum_{\alpha} \ope{L}^\alpha ,
	\end{equation}
the dynamical variables are
	\begin{equation}	\label{9.3}
\ope{D} = \sum_{\alpha} \ope{D}^\alpha
	\end{equation}
and the corresponding system of Euler-Lagrange equations consists of the
independent equations for each of the fields of the system (see~\eref{3.6}
with $\ope{L}^\alpha$ for $\ope{L}$). This allows an introduction of
independent creation and annihilation operators for each field. The ones for
the $\alpha$\ndash th field will be denoted by
 $a_{\alpha,s^\alpha}^{\pm}(\bk)$ and $a_{\alpha,s^\alpha}^{\dag\,\pm}(\bk)$;
notice, the values of the polarization variables generally depend on the
field considered and, therefore, they also are labeled with index $\alpha$ for
the $\alpha$\ndash th field. For brevity, we shall use the collective indices
$l^\alpha$, $m^\alpha$ and $n^\alpha$, with $l^\alpha:=(\alpha,s^\alpha,\bk)$
etc., in terms of which the last operators are
 $a_{l^\alpha}^{\pm}$ and $a_{l^\alpha}^{\dag\,\pm}$, respectively. The
particular expressions for the dynamical operators $\ope{D}^\alpha$ are given
via~\eref{3.7}--\eref{3.10} in which the following changes should be made:
	\begin{equation}	\label{9.3-1}
	\begin{split}
&     \tau \mapsto \tau^\alpha
\quad j \mapsto j^\alpha
\quad \varepsilon \mapsto \varepsilon^\alpha
\quad s \mapsto s^\alpha
\quad s' \mapsto s^{\prime\,\alpha}
\\
& \sigma_{\mu\nu}^{ss',\pm}(\bk) \mapsto
  \sigma_{\mu\nu}^{s^\alpha s^{\prime\,\alpha},\pm}(\bk)
\quad
l_{\mu\nu}^{ss',\pm}(\bk) \mapsto
l_{\mu\nu}^{s^\alpha s^{\prime\,\alpha},\pm}(\bk) .
	\end{split}
	\end{equation}

	The content of sections~\ref{Sect4} and~\ref{Sect5} remains valid
\emph{mutatis mutandis}, \viz provided the just pointed changes~\eref{9.3-1}
are made and the (integral) dynamical variables are understood in conformity
with~\eref{9.3}.

\subsection
[Commutation relations connected with the momentum operator.
Problems and their possible solutions]
{Commutation relations connected with the momentum operator.\\
Problems and their possible solutions}

\label{Subsect9.1}

	In sections~\ref{Sect6}--\ref{Sect8}, however, substantial changes
occur; for instance, when one passes from~\eref{6.11} or~\eref{6.14}
to~\eref{6.140}. We shall consider them briefly in a case when one starts
from the charge symmetric Lagrangians~\eref{3.4}.

	The basic relations~\eref{6.11}, which arise from the Heisenberg
relation~\eref{5.1} concerning the momentum operator, now read (here and
below, do not sum over $\alpha$, and/or $\beta$ and/or $\gamma$ if the
opposite is not indicated explicitly!)
	\begin{subequations}	\label{9.4}
	\begin{align}
			\label{9.4a}
	\begin{split}
\bigl[ a_{l^\alpha}^{\pm}
& ,
[ a_{m^\beta}^{\dag\,+} , a_{m^\beta}^{-} ]_{\varepsilon^\beta}+
[ a_{m^\beta}^{+} , a_{m^\beta}^{\dag\,-} ]_{\varepsilon^\beta}
\bigr]_{-}
\pm (1+\tau) \delta_{l^\alpha m^\beta} a_{l^\alpha}^{\pm} = 0
	\end{split}
\\			\label{9.4b}
	\begin{split}
\bigl[ a_{l^\alpha}^{\dag\,\pm}
& ,
[ a_{m^\beta}^{\dag\,+} , a_{m^\beta}^{-} ]_{\varepsilon^\beta}+
[ a_{m^\beta}^{+} , a_{m^\beta}^{\dag\,-} ]_{\varepsilon^\beta}
\bigr]_{-}
\pm (1+\tau) \delta_{l^\alpha m^\beta} a_{l^\alpha}^{\dag\,\pm} = 0 .
	\end{split}
	\end{align}
	\end{subequations}	
It is trivial to be seen, the following generalizations of
respectively~\eref{6.014} and~\eref{6.14}
	\begin{subequations}	\label{9.5}
	\begin{align}
			\label{9.5a}
	\begin{split}
\bigl[ a_{l^\alpha}^{\pm} & ,
  [ a_{m^\beta}^{+} , a_{m^\beta}^{\dag\,-} ]_{\varepsilon^\beta} \bigr]_{-}
\pm (1+{\tau^\beta}) \delta_{{l^\alpha}{m^\beta}} a_{l^\alpha}^{\pm} = 0
	\end{split}
\\			\label{9.5b}
	\begin{split}
\bigl[ a_{l^\alpha}^{\pm} & ,
  [ a_{m^\beta}^{\dag\,+} , a_{m^\beta}^{-} ]_{\varepsilon^\beta} \bigr]_{-}
\pm (1+{\tau^\beta}) \delta_{{l^\alpha}{m^\beta}} a_{l^\alpha}^{\pm} = 0
	\end{split}
\\			\label{9.5c}
	\begin{split}
\bigl[ a_{l^\alpha}^{\dag\,\pm} & ,
  [ a_{m^\beta}^{+} , a_{m^\beta}^{\dag\,-} ]_{\varepsilon^\beta} \bigr]_{-}
\pm (1+{\tau^\beta}) \delta_{{l^\alpha}{m^\beta}} a_{l^\alpha}^{\dag\,\pm} = 0
	\end{split}
\\			\label{9.5d}
	\begin{split}
\bigl[ a_{l^\alpha}^{\dag\,\pm} & ,
  [ a_{m^\beta}^{\dag\,+} , a_{m^\beta}^{-} ]_{\varepsilon^\beta} \bigr]_{-}
\pm (1+{\tau^\beta}) \delta_{{l^\alpha}{m^\beta}} a_{l^\alpha}^{\dag\,\pm} = 0
	\end{split}
	\end{align}
	\end{subequations}	
\vspace{-3.5ex}
	\begin{subequations}	\label{9.6}
	\begin{align}
			\label{9.6a}
	\begin{split}
\bigl[ a_{l^\alpha}^{+} & ,
  [ a_{m^\beta}^{+} , a_{m^\beta}^{\dag\,-} ]_{\varepsilon^\beta} \bigr]_{-}
+2 \delta_{{l^\alpha}{m^\beta}} a_{l^\alpha}^{+} = 0
	\end{split}
\\			\label{9.6b}
	\begin{split}
\bigl[ a_{l^\alpha}^{+} & ,
  [ a_{m^\beta}^{\dag\,+} , a_{m^\beta}^{-} ]_{\varepsilon^\beta} \bigr]_{-}
+ 2 {\tau^\beta} \delta_{{l^\alpha}{m^\beta}} a_{l^\alpha}^{+} = 0
	\end{split}
\\			\label{9.6c}
	\begin{split}
\bigl[ a_{l^\alpha}^{-} & ,
  [ a_{m^\beta}^{+} , a_{m^\beta}^{\dag\,-} ]_{\varepsilon^\beta} \bigr]_{-}
- 2 {\tau^\beta} \delta_{{l^\alpha}{m^\beta}} a_{l^\alpha}^{-} = 0
	\end{split}
\\			\label{9.6d}
	\begin{split}
\bigl[ a_{l^\alpha}^{-} & ,
  [ a_{m^\beta}^{\dag\,+} , a_{m^\beta}^{-} ]_{\varepsilon^\beta} \bigr]_{-}
-2 \delta_{{l^\alpha}{m^\beta}} a_{l^\alpha}^{-} = 0
	\end{split}
	\end{align}
	\end{subequations}	
provide a solution of~\eref{9.4} in a sense that they convert it into
identity. As it was said in Sect.~\ref{Sect6}, the equations~\eref{9.5}
(resp.~\eref{9.6}) for a single field, \ie for $\beta=\alpha$, agree
(resp.\ disagree) with the bilinear commutation relations~\eref{6.12}.

	The only problem arises when one tries to generalize, e.g., the
relations~\eref{9.6} in a way similar to the transition from~\eref{6.14}
to~\eref{6.140}. Its essence is in the generalization of expressions like
 $[ a_{m^\beta}^{\dag\,\pm} , a_{m^\beta}^{\mp} ]_{\varepsilon^\beta}$ and
 ${\tau^\beta} \delta_{{l^\alpha}{m^\beta}} a_{l^\alpha}^{\pm}$. When passing
from~\eref{6.14} to~\eref{6.140}, the indices $l$ and $m$ are changed so that
the obtained equations to be consistent with~\eref{6.12}; of course, the
numbers $\varepsilon$ and $\tau$ are preserved because this change does not
concern the field regarded. But the situation with~\eref{9.6} is different in
two directions:
\\
\indent
	(i) If we change the pair $(m^\beta,m^\beta)$ in
$[ a_{m^\beta}^{\dag\,\pm} , a_{m^\beta}^{\mp} ]_{\varepsilon^\beta}$
with $(m^\beta,n^\gamma)$, then with what the number $\varepsilon^\beta$
should be replace? With $\varepsilon^\beta$, or $\varepsilon^\gamma$ or with
something else? Similarly, if the mentioned changed is performed, with what
the multiplier $\tau^\beta$ in
 ${\tau^\beta} \delta_{{l^\alpha}{m^\beta}} a_{l^\alpha}^{\pm}$
should be replaced? The problem is that the numbers $\varepsilon^\beta$ and
$\tau^\beta$ are related to terms like
$a_{m^\beta}^{\dag\,\pm} \circ a_{m^\beta}^{\mp} $ and
$a_{m^\beta}^{\pm} \circ a_{m^\beta}^{\dag\,\mp} $,
in the momentum operator, as a whole and we cannot say whether the index
$\beta$ in $\varepsilon^\beta$ and $\tau^\beta$ originates from the first of
second index $m^\beta$ in these expressions.
\\
\indent
	(ii) When writing $(m^\beta,n^\gamma)$ for $(m^\beta,m^\beta)$
(see~(i) above), then shall we replace
 $\delta_{{l^\alpha}{m^\beta}} a_{l^\alpha}^{\pm}$ with
 $\delta_{{l^\alpha}{m^\beta}} a_{n^\gamma}^{\pm}$, or
 $\delta_{{l^\alpha}{n^\gamma}} a_{m^\beta}^{\pm}$, or
 $\delta_{{m^\beta}{n^\gamma}} a_{l^\alpha}^{\pm}$?
For a single field, $\gamma=\beta=\alpha$, this problem is solved by
requiring an agreement of the resulting generalization (of~\eref{6.140} in the
particular case) with the bilinear commutation relations~\eref{6.12}. So, how
shall~\eref{6.12} be generalized for several, not less than two, different
fields? Obviously, here we meet an obstacle similar to the one described
in~(i) above, with the only change that $-\varepsilon^\beta$ should stand for
$\varepsilon^\beta$.

	Let $b_{l^\alpha}$ and $c_{l^\alpha}$ denote some creation or
annihilation operator of the $\alpha$\ndash field. Consider the problem for
generalizing the (anti)commutator
$[b_{l^\alpha},c_{l^\alpha}]_{\pm\varepsilon^\alpha}$. This means that we are
looking for a replacement
	\begin{gather}	\label{9.7}
[b_{l^\alpha},c_{l^\alpha}]_{\pm\varepsilon^\alpha}
\mapsto
f^\pm (b_{l^\alpha},c_{m^\beta}; \alpha,\beta),
\intertext{where the functions $f^\pm$ are such that}
			\label{9.8}
f^\pm (b_{l^\alpha},c_{m^\beta}; \alpha,\beta) \big|_{\beta=\alpha}
=
[b_{l^\alpha},c_{l^\alpha}]_{\pm\varepsilon^\alpha} .
	\end{gather}
Unfortunately, the condition~\eref{9.8} is the only restriction on $f^\pm$
that the theory of free fields can provide. Thus the functions $f^\pm$,
subjected to equation~\eref{9.8}, become new free parameters of the quantum
theory of different free fields and it is a matter of convention how to
choose/fix them.

	It is generally accepted~\cite[appendix~F]{Ohnuki&Kamefuchi}, the
functions $f^\pm$ to have forms `maximum' similar to the (anti)commutators
they generalize. More precisely, the functions
	\begin{gather}	\label{9.8-1}
f^\pm (b_{l^\alpha},c_{m^\beta}; \alpha,\beta)
=
[b_{l^\alpha},c_{m^\beta}]_{\pm\varepsilon^{\alpha\beta}}
\intertext{where $\varepsilon^{\alpha\beta}\in\field[C]$ are such that}
			\label{9.9}
\varepsilon^{\alpha\alpha} = \varepsilon^{\alpha} ,
	\end{gather}
are usually considered as the only candidates for $f^\pm$. Notice,
in~\eref{9.8-1}, $\varepsilon^{\alpha\beta}$ are functions in $\alpha$ and
$\beta$, not in $l^\alpha$ and/or $m^\beta$. Besides, if we assume
$\varepsilon^{\alpha\beta}$ to be function only in $\varepsilon^\alpha$ and
$\varepsilon^\beta$, then the general form of $\varepsilon^{\alpha\beta}$ is
	\begin{equation}	\label{9.10}
\varepsilon^{\alpha\beta}
=
  u^{\alpha\beta} \varepsilon^\alpha
+ (1 - u^{\alpha\beta}) \varepsilon^\beta
+ v^{\alpha\beta} (1 - \varepsilon^\alpha \varepsilon^\beta)
\qquad
u^{\alpha\beta}, v^{\alpha\beta} \in\field[C] ,
	\end{equation}
due to~\eref{9.1} and~\eref{9.9}. (In view of~\eref{6.12}, the value
$\varepsilon^{\alpha\beta}=+1$ (resp.\ $\varepsilon^{\alpha\beta}=-1$)
corresponds to quantization via commutators (resp.\ anticommutators) of the
corresponding fields.)

	Call attention now on the numbers $\tau^\alpha$ which originate and
are associated with each term
$[b_{l^\alpha},c_{m^\alpha}]_{\pm\varepsilon^{\alpha}}$.
With every change~\eref{9.7} one can associate a replacement
	\begin{gather}	\label{9.11}
\tau^\alpha
\mapsto
g (b_{l^\alpha},c_{m^\beta}; \alpha,\beta),
\intertext{where the function $g$ is such that}
			\label{9.12}
g (b_{l^\alpha},c_{m^\beta}; \alpha,\beta) \big|_{\beta=\alpha}
= \tau^\alpha  .
	\end{gather}
Of course, the last condition does not define $g$ uniquely and, consequently,
the function $g$, satisfying~\eref{9.12}, enters in the theory as a new free
parameter. Suppose, as a working hypothesis similar
to~\eref{9.8-1}--\eref{9.9}, that  $g$ is of the form
	\begin{gather}	\label{9.13}
g (b_{l^\alpha},c_{m^\beta}; \alpha,\beta) = \tau^{\alpha\beta},
\intertext{where $\tau^{\alpha\beta}$ are complex numbers that may depend
only on $\alpha$ and $\beta$ and are such that}
			\label{9.14}
\tau^{\alpha\alpha} = \tau^\alpha  .
	\end{gather}
Besides, if we suppose  $\tau^{\alpha\beta}$ to be functions only in
$\tau^\alpha$ and $\tau^\beta$, then
	\begin{equation}	\label{9.15}
\tau^{\alpha\beta}
=
  x^{\alpha\beta} \tau^\alpha
+ y^{\alpha\beta} \tau^\beta
+ (1 - x^{\alpha\beta} - y^{\alpha\beta}) \tau^\alpha \tau^\beta
\qquad x^{\alpha\beta}, y^{\alpha\beta} \in\field[C] ,
	\end{equation}
as a result of~\eref{9.1} and~\eref{9.14}.

	Let us summarize the above discussion. If we suppose a preservation
of the algebraic structure of the bilinear commutation relations~\eref{6.12}
for a system of different free fields, then the replacements
	\begin{subequations}	\label{9.16}
	\begin{align}	\label{9.16a}
&
[b_{l^\alpha},c_{l^\alpha}]_{\pm\varepsilon^\alpha}
\mapsto
[b_{l^\alpha},c_{m^\beta}]_{\pm\varepsilon^{\alpha\beta}}
\qquad
\varepsilon^{\alpha\alpha} = \varepsilon^{\alpha}
\\			\label{9.16b}
& \tau^{\alpha} \mapsto \tau^{\alpha\alpha}
\qquad
\tau^{\alpha\alpha} = \tau^{\alpha}
	\end{align}
	\end{subequations}
should be made; accordingly, the relations~\eref{6.12} transform into:
	\begin{align}	\notag
&[a_{{l^\alpha}}^{\pm}, a_{{m^\beta}}^{\pm} ]_{-{\varepsilon^{\alpha\beta}}}
	= 0
&&
[a_{{l^\alpha}}^{\dag\,\pm},
		a_{{m^\beta}}^{\dag\,\pm} ]_{-{\varepsilon^{\alpha\beta}}}
	= 0
\\	\notag
&[a_{{l^\alpha}}^{\mp}, a_{{m^\beta}}^{\pm} ]_{-{\varepsilon^{\alpha\beta}}}
	= {\tau^{\alpha\beta}} \delta_{{l^\alpha}{m^\beta}} \id_\Hil
\times\big\{
	\begin{smallmatrix}
1\\
-\varepsilon^{\alpha\beta}
	\end{smallmatrix}
&&
[a_{{l^\alpha}}^{\dag\,\mp},
		a_{{m^\beta}}^{\dag\,\pm} ]_{-{\varepsilon^{\alpha\beta}}}
	= {\tau^{\alpha\beta}} \delta_{{l^\alpha}{m^\beta}} \id_\Hil
\times\big\{
	\begin{smallmatrix}
1\\
-\varepsilon^{\alpha\beta}
	\end{smallmatrix}
\\	\notag
&[a_{{l^\alpha}}^{\pm},
		a_{{m^\beta}}^{\dag\,\pm} ]_{-{\varepsilon^{\alpha\beta}}}
	= 0
&&
[a_{{l^\alpha}}^{\dag\,\pm},
		a_{{m^\beta}}^{\pm} ]_{-{\varepsilon^{\alpha\beta}}}
	= 0
\\	\label{9.17}
&[a_{{l^\alpha}}^{\mp},
		a_{{m^\beta}}^{\dag\,\pm} ]_{-{\varepsilon^{\alpha\beta}}}
	= \delta_{{l^\alpha}{m^\beta}} \id_\Hil
\times\big\{
	\begin{smallmatrix}
1\\
-\varepsilon^{\alpha\beta}
	\end{smallmatrix}
&&
[a_{{l^\alpha}}^{\dag\,\mp},
		a_{{m^\beta}}^{\pm} ]_{-{\varepsilon^{\alpha\beta}}}
	= \delta_{{l^\alpha}{m^\beta}} \id_\Hil
\times\big\{
	\begin{smallmatrix}
1\\
-\varepsilon^{\alpha\beta}
	\end{smallmatrix}
,
	\end{align}
where 1 (resp.\ $-\varepsilon^{\alpha\beta}$) in
\(
\big\{
	\begin{smallmatrix}
1\\
-\varepsilon^{\alpha\beta}
	\end{smallmatrix}
\)
corresponds to the choice of the upper (resp.\ lower) signs.
	If we suppose additionally  $\varepsilon^{\alpha\beta}$ (resp.\
$\tau^{\alpha\beta}$) to be a function only in $\varepsilon^\alpha$ and
$\varepsilon^\beta$ (resp.\ in $\tau^\alpha$ and $\tau^\beta$), then these
numbers are defined up to two sets of complex parameters:
	\begin{subequations}	\label{9.18}
	\begin{align}	\label{9.18a}
& \varepsilon^{\alpha\beta}
=
  u^{\alpha\beta} \varepsilon^\alpha
+ (1 - u^{\alpha\beta}) \varepsilon^\beta
+ v^{\alpha\beta} (1 - \varepsilon^\alpha \varepsilon^\beta)
&&\quad
u^{\alpha\beta}, v^{\alpha\beta} \in\field[C]
\\			\label{9.18b}
& \tau^{\alpha\beta}
=
  x^{\alpha\beta} \tau^\alpha
+ y^{\alpha\beta} \tau^\beta
+ (1 - x^{\alpha\beta} - y^{\alpha\beta}) \tau^\alpha \tau^\beta
&&\quad
x^{\alpha\beta}, y^{\alpha\beta} \in\field[C] .
	\end{align}
	\end{subequations}

	A reasonable further specialization of $\varepsilon^{\alpha\beta}$
and $\tau^{\alpha\beta}$ may be the assumption their ranges to coincide with
those of $\varepsilon^\alpha$ and $\tau^\alpha$, respectively. As a result
of~\eref{9.1}, this supposition is equivalent to
	\begin{subequations}	\label{9.19}
	\begin{align}	\label{9.19a}
&
v^{\alpha\beta}
=
-u^{\alpha\beta}, -u^{\alpha\beta}+1, u^{\alpha\beta}-1, u^{\alpha\beta}
\quad
u^{\alpha\beta}\in\field[C]
\\			\label{9.19b}
&
(x^{\alpha\beta},y^{\alpha\beta})
=
(0,0), (0,1), (1,0), (1,1).
	\end{align}
	\end{subequations}
Other admissible restriction on~\eref{9.18} may be the requirement
$\varepsilon^{\alpha\beta}$ and $\tau^{\alpha\beta}$ to be symmetric, \viz
	\begin{subequations}	\label{9.20}
	\begin{align}	\label{9.20a}
&
  \varepsilon^{\alpha\beta} (\varepsilon^\alpha,\varepsilon^\beta)
= \varepsilon^{\beta\alpha} (\varepsilon^\alpha,\varepsilon^\beta)
= \varepsilon^{\alpha\beta} (\varepsilon^\beta,\varepsilon^\alpha)
\\			\label{9.20b}
&
  \tau^{\alpha\beta} (\tau^\alpha,\tau^\beta)
= \tau^{\beta\alpha} (\tau^\alpha,\tau^\beta)
= \tau^{\alpha\beta} (\tau^\beta,\tau^\alpha) ,
	\end{align}
	\end{subequations}
which means that the  $\alpha$-th and $\beta$-th fields are treated on equal
footing and there is no \emph{a priori} way to number some of them as the
`first' or `second' one.%
\footnote{~%
However, nothing can prevent us to make other choices, compatible
with~\eref{9.16}, in the theory of free fields; for instance, one may set
\(
\varepsilon^{\alpha\beta}
= \varepsilon^\alpha \varepsilon^\beta \varepsilon^{\beta\alpha}
\)
and
\(
\tau^{\alpha\beta}
= \frac{1}{2}(\tau^\alpha + \tau^\beta) \tau^{\beta\alpha}.
\)%
}
In view of~\eref{9.18}, the conditions~\eref{9.20} are equivalent to
	\begin{subequations}	\label{9.21}
	\begin{align}	\label{9.21a}
&
u^{\alpha\beta} = \frac{1}{2}
\quad v^{\alpha\beta}\in\field[C]
\\			\label{9.21b}
&
y^{\alpha\beta} = x^{\alpha\beta}  .
	\end{align}
	\end{subequations}
If both of the restrictions~\eref{9.19} and~\eref{9.21} are imposed
on~\eref{9.18}, then the arbitrariness of the parameters in~\eref{9.18} is
reduced to:
	\begin{subequations}	\label{9.22}
	\begin{align}	\label{9.22a}
&
(u^{\alpha\beta},u^{\alpha\beta})
= \Bigl(\frac{1}{2},-\frac{1}{2}\Bigr), \Bigl(\frac{1}{2},\frac{1}{2}\Bigr)
\\			\label{9.22b}
&
(x^{\alpha\beta},y^{\alpha\beta})
=(0,0), (1,1)
	\end{align}
	\end{subequations}
and, for any \emph{fixed} pair $(\alpha,\beta)$, we are left with the
following candidates for respectively $\varepsilon^{\alpha\beta}$ and
$\tau^{\alpha\beta}$:
	\begin{subequations}	\label{9.23}
	\begin{align}	\label{9.23a}
&
\varepsilon_+^{\alpha\beta}
:= \frac{1}{2}
(+1 + \varepsilon^\alpha + \varepsilon^\beta -
      \varepsilon^\alpha \varepsilon^\beta )
\\			\label{9.23b}
&
\varepsilon_-^{\alpha\beta}
:= \frac{1}{2}
(-1 + \varepsilon^\alpha + \varepsilon^\beta +
      \varepsilon^\alpha \varepsilon^\beta )
\\			\label{9.23c}
&
\tau_0^{\alpha\beta}
:= \tau^\alpha + \tau^\beta
\\			\label{9.23d}
&
\tau_1^{\alpha\beta}
:= \tau^\alpha + \tau^\beta - \tau^\alpha \tau^\beta  .
	\end{align}
	\end{subequations}
When free fields are considered, as in our case, no further arguments from
mathematical or physical nature can help for choosing a particular
combination $(\varepsilon^{\alpha\beta},\tau^{\alpha\beta})$ from the four
possible ones according to~\eref{9.23} for a fixed pair $(\alpha,\beta)$. To
end the above considerations of $\varepsilon^{\alpha\beta}$ and
$\tau^{\alpha\beta}$, we have to say that the choice
	\begin{equation}	\label{9.24}
(\varepsilon^{\alpha\beta},\tau^{\alpha\beta})
=
(\varepsilon_+^{\alpha\beta},\tau_0^{\alpha\beta})
=
\Bigl( \frac{1}{2}
(+1 + \varepsilon^\alpha + \varepsilon^\beta -
      \varepsilon^\alpha \varepsilon^\beta )
,
\tau^\alpha + \tau^\beta
\Bigr)
	\end{equation}
is known as the \emph{normal case}~\cite[appendix~F]{Ohnuki&Kamefuchi}; in it
the relative behavior of bose (resp.\ fermi) fields is as in the case of a
single field, \ie they are quantized via commutators (resp.\ anticommutators)
as
$(\varepsilon^{\alpha\beta},\tau^{\alpha\beta})=(+1,0)$
(resp.\
$(\varepsilon^{\alpha\beta},\tau^{\alpha\beta})=(-1,0)$),
and the one of bose and fermi field is as in the case of a single fermi
field, \viz the quantization is via commutators as
$(\varepsilon^{\alpha\beta},\tau^{\alpha\beta})=(+1,0)$.
All combinations between $\varepsilon_\pm^{\alpha\beta}$ and
$\tau_{0,1}^{\alpha\beta}$ different from~\eref{9.24} are referred as
\emph{anomalous cases}.
	Above we supposed the pair $(\alpha,\beta)$ to be fixed. If $\alpha$
and $\beta$ are \emph{arbitrary}, the only essential change this implies is
in~\eref{9.23}, where the choice of the subscripts $+$, $-$, $0$ and $1$ may
depend on $\alpha$ and $\beta$. In this general situation, the \emph{normal
case} is defined as the one when~\eref{9.24} holds for all $\alpha$ and
$\beta$. All other combinations are referred as \emph{anomalous cases}; such
are, for instance, the ones when some fermi and bose operators satisfy
anticommutation relations, e.g.~\eref{9.17} with
$\varepsilon^{\alpha\beta}=-1$ for $\varepsilon^\alpha+\varepsilon^\beta=0$,
or some fermi fields are subjected to commutation relations, like~\eref{9.17}
with $\varepsilon^{\alpha\beta}=+1$ for
$\varepsilon^\alpha=\varepsilon^\beta=-1$. For some details on this topic,
see, for
instance,~\cite[appendix~F]{Ohnuki&Kamefuchi},
\cite[chapter~20]{Bogolyubov&et_al.-AxQFT}
and~\cite[sect~4-4]{Streater&Wightman}.
	Fields/operators for which $\varepsilon^{\alpha\beta}=+1$ (resp.\
$\varepsilon^{\alpha\beta}=-1$), with $\beta\not=\alpha$, are referred as
\emph{relative parabose} (resp.\ \emph{parafermi}) in the parafield
theory~\cite{Greenberg&Messiah-1965,Ohnuki&Kamefuchi}. One can transfer this
terminology in the general case and call the fields/operators for which
$\varepsilon^{\alpha\beta}=+1$ (resp.\ $\varepsilon^{\alpha\beta}=-1$), with
$\beta\not=\alpha$, \emph{relative bose} (resp.\ \emph{fermi})
fields/operators.

	Further the relations~\eref{9.17} will be referred as the
\emph{multifield bilinear commutation relations} and it will be assumed that
they represent the generalization of the bilinear commutation
relations~\eref{6.12} when we are dealing with several, not less than two,
different quantum fields. The particular values of
$\varepsilon^{\alpha\beta}$ and $\tau^{\alpha\beta}$ in them are insignificant
in the following; if one likes, one can fix them as in the normal
case~\eref{9.24}. Moreover, even the definition~\eref{9.17} of
$\tau^{\alpha\beta}$ is completely inessential at all, as
$\tau^{\alpha\beta}$ always appears in combinations like
$\tau^{\alpha\beta}\delta_{l^\alpha m^\beta}$ (see~\eref{9.17} or similar
relations, like~\eref{9.25}, below), which are non\ndash vanishing if
$\beta=\alpha$, but then $\tau^{\alpha\alpha}=\tau^\alpha$; so one can freely
write $\tau^\alpha$ for $\tau^{\alpha\beta}$ in all such cases.

	Equipped with~\eref{9.17} and~\eref{9.16}, we can
generalize~\eref{9.6} in different ways. For example, the straightforward
generalization of~\eref{6.140} is:
	\begin{subequations}	\label{9.25}
	\begin{align}
			\label{9.25a}
	\begin{split}
\bigl[ a_{l^\alpha}^{+} & ,
  [ a_{m^\beta}^{+} ,
  a_{n^\gamma}^{\dag\,-} ]_{\varepsilon^{\beta\gamma}} \bigr]_{-}
+2 \delta_{{l^\alpha}{n^\gamma}} a_{m^\beta}^{+} = 0
	\end{split}
\\			\label{9.25b}
	\begin{split}
\bigl[ a_{l^\alpha}^{+} & ,
  [ a_{m^\beta}^{\dag\,+} ,
  a_{n^\gamma}^{-} ]_{\varepsilon^{\beta\gamma}} \bigr]_{-}
+ 2 {\tau^{\alpha\gamma}} \delta_{{l^\alpha}{n^\gamma}} a_{m^\beta}^{+} = 0
	\end{split}
\\			\label{9.25c}
	\begin{split}
\bigl[ a_{l^\alpha}^{-} & ,
  [ a_{m^\beta}^{+} ,
  a_{n^\gamma}^{\dag\,-} ]_{\varepsilon^{\beta\gamma}} \bigr]_{-}
- 2 {\tau^{\alpha\beta}} \delta_{{l^\alpha}{m^\beta}} a_{n^\gamma}^{-} = 0
	\end{split}
\\			\label{9.25d}
	\begin{split}
\bigl[ a_{l^\alpha}^{-} & ,
  [ a_{m^\beta}^{\dag\,+} ,
  a_{n^\gamma}^{-} ]_{\varepsilon^{\beta\gamma}} \bigr]_{-}
-2 \delta_{{l^\alpha}{m^\beta}} a_{n^\gamma}^{-} = 0 .
	\end{split}
	\end{align}
	\end{subequations}	
However, generally, the relations~\eref{9.17} do \emph{not}
convert~\eref{9.25} into identities. The reason is that an equality/identity
like (cf.~\eref{6.13})
	\begin{gather}	\label{9.25-1}
[b_{l^\alpha},c_{m^\beta} \circ d_{n^\gamma}]_{\_}
=
[b_{l^\alpha},c_{m^\beta}]_{-\varepsilon^{\alpha\beta}} \circ d_{n^\gamma}
+ \lambda^{\alpha\beta\gamma}
c_{m^\beta} \circ [b_{l^\alpha},d_{n^\gamma}]_{-\varepsilon^{\alpha\gamma}} ,
\intertext{where $b_{l^\alpha}$, $c_{m^\beta}$ and $d_{n^\gamma}$ are some
creation/annihilation operators and
$\lambda^{\alpha\beta\gamma}\in\field[C]$, can be valid only for}
			\label{9.25-2}
\lambda^{\alpha\beta\gamma} = \varepsilon^{\alpha\beta}
\quad
\varepsilon^{\alpha\gamma} = 1/\varepsilon^{\alpha\beta}
\qquad
(\varepsilon^{\alpha\beta} \not= 0),
	\end{gather}
which, in particular, is fulfilled if $\gamma=\beta$ and
$\varepsilon^{\alpha\beta}=\pm1$.
	So, the agreement between~\eref{9.17} and~\eref{9.25} depends on the
concrete choice of the numbers $\varepsilon^{\alpha\beta}$. There exist cases
when even the normal case~\eref{9.24} cannot ensure~\eref{9.17} to
convert~\eref{9.25} into identities; \eg when the $\alpha$\ndash th field and
$\beta$\ndash th fields are fermion ones and the $\gamma$\ndash th field is a
boson one. Moreover, it can be proved that~\eref{9.17} and~\eref{9.25} are
compatible in the general case if unacceptable equalities like
$a_l^\pm\circ a_m^\pm =0$ hold.

	One may call~\eref{9.25} the \emph{multifield paracommutation
relations} as from them a corresponding generalization of~\eref{6.16}
and/or~\eref{6.18} can be derived. For completeness, we shall record the
multifield version of~\eref{6.18}:
	\begin{subequations}	\label{9.26}
	\begin{alignat}{2}	\label{9.26a}
[b_{l^\alpha} , [ b_{m^\beta}^\dag, b_{n^\gamma} ]_{\varepsilon^{\beta\gamma}}
 ]_{\_} & = 2 \delta_{{l^\alpha}{m^\beta}} b_{n^\gamma}  &\quad
[b_{l^\alpha} , [ b_{m^\beta}     , b_{n^\gamma} ]_{\varepsilon^{\beta\gamma}}
 ]_{\_} & = 0
\\				\label{9.26b}
[c_{l^\alpha} , [ c_{m^\beta}^\dag, c_{n^\gamma} ]_{\varepsilon^{\beta\gamma}}
 ]_{\_} & = 2 \delta_{{l^\alpha}{m^\beta}} c_{n^\gamma}  &\quad
[c_{l^\alpha} , [ c_{m^\beta}     , c_{n^\gamma} ]_{\varepsilon^{\beta\gamma}}
 ]_{\_} & = 0
\\ 				\label{9.26c}
[b_{l^\alpha}^\dag , [ c_{m^\beta}^\dag, c_{n^\gamma}
 ]_{\varepsilon^{\beta\gamma}} ]_{\_}
& =-2{\tau^{\alpha\gamma}}\delta_{{l^\alpha}{n^\gamma}} b_{m^\beta}^\dag
&\quad
[c_{l^\alpha}^\dag , [ b_{m^\beta}^\dag, b_{n^\gamma}
 ]_{\varepsilon^{\beta\gamma}} ]_{\_}
& =-2{\tau^{\alpha\gamma}}\delta_{{l^\alpha}{n^\gamma}} c_{m^\beta}^\dag
.
	\end{alignat}
	\end{subequations}
For details regarding these multifield paracommutation relations, the reader
is referred to~\cite{Greenberg&Messiah-1965,Ohnuki&Kamefuchi}, where the case
$\tau^\alpha=\tau^\beta=\tau^{\alpha\beta}=0$ is considered.

	We leave to the reader as exercise to write down the multifield
versions of the commutation relations~\eref{6.20} or~\eref{6.21}, which
provide examples of generalizations of~\eref{9.6} and hence of~\eref{9.17}
and~\eref{9.25}.

\subsection
[Commutation relations connected with the charge and
						angular momentum operators]
{Commutation relations connected with the charge and
					     \\ angular momentum operators}

\label{Subsect9.2}

	In a case of several, not less than two, different fields, the basic
trilinear commutation relations~\eref{6.31}, which ensure the validity of the
Heisenberg relation~\eref{5.2} concerning the charge operator, read:
	\begin{subequations}	\label{9.27}
	\begin{align}	\label{9.27a}
	\begin{split}
\bigl[ a_{l^\alpha}^{\pm}
& ,
[ a_{m^\beta}^{\dag\,+} , a_{m^\beta}^{-} ]_{\varepsilon^\beta} -
[ a_{m^\beta}^{+} , a_{m^\beta}^{\dag\,-} ]_{\varepsilon^\beta}
\bigr]_{-}
- 2 \delta_{l^{\alpha}{m^\beta}} a_{l^\alpha}^{\pm} = 0
	\end{split}
\\			\label{9.27b}
	\begin{split}
\bigl[ a_{l^\alpha}^{\dag\,\pm}
& ,
[ a_{m^\beta}^{\dag\,+} , a_{m^\beta}^{-} ]_{\varepsilon^\beta} -
[ a_{m^\beta}^{+} , a_{m^\beta}^{\dag\,-} ]_{\varepsilon^\beta}
\bigr]_{-}
+ 2 \delta_{l^{\alpha}{m^\beta}} a_{l^\alpha}^{\dag\,\pm} = 0 .
	\end{split}
	\end{align}
	\end{subequations}
Of course, these relations hold only for those fields which have non\ndash
vanishing charges, \ie in~\eref{9.27} is supposed (see~\eref{9.1})
	\begin{equation}	\label{9.28}
\tau^\alpha = 0 \quad
\tau^\beta = 0
\qquad ( \iff q^\alpha q^\beta\not=0 ) .
	\end{equation}

	The problem for generalizing~\eref{9.27} for these fields is similar
to the one for~\eref{9.6} in the case of non\ndash vanishing charges,
$\tau^\beta=0$. Without repeating the discussion of
Subsect.~\ref{Subsect9.1}, we shall adopt the rule~\eref{9.16} for
generalizing (anti)commutation relations between creation/annihilation
operators of a single field. By its means one can obtain different
generalizations of~\eref{9.27}. For instance, the commutation relations.
	\begin{subequations}	\label{9.29}
	\begin{align}
			\label{9.29a}
	\begin{split}
\bigl[ a_{l^\alpha}^{+}
& ,
[ a_{m^\beta}^{\dag\,+} , a_{n^\gamma}^{-} ]_{\varepsilon^{\beta\gamma}} -
[ a_{m^\beta}^{+} , a_{n^\gamma}^{\dag\,-} ]_{\varepsilon^{\beta\gamma}}
\bigr]_{-}
- 2 \delta_{{l^\alpha}{n^\gamma}} a_{m^\beta}^{+} = 0
	\end{split}
\\			\label{9.29b}
	\begin{split}
\bigl[ a_{l^\alpha}^{-}
& ,
[ a_{m^\beta}^{\dag\,+} , a_{n^\gamma}^{-} ]_{\varepsilon^{\beta\gamma}} -
[ a_{m^\beta}^{+} , a_{n^\gamma}^{\dag\,-} ]_{\varepsilon^{\beta\gamma}}
\bigr]_{-}
- 2 \delta_{{l^\alpha}{m^\beta}} a_{n^\gamma}^{-} = 0
	\end{split}
	\end{align}
	\end{subequations}	
and their Hermitian conjugate contain~\eref{9.27} and~\eref{6.33}  as evident
special cases and agree with~\eref{9.17} if $\gamma=\beta$ and
$\varepsilon^{\alpha\beta} \varepsilon^{\beta\gamma}=+1$. Besides, the
multifield paracommutation relations~\eref{9.25} for charged fields,
$\tau^\alpha=\tau^\beta=\tau^\gamma=0$, convert~\eref{9.29} into identities
and, in this sense,~\eref{9.29} agree with (contain as special
case)~\eref{9.25} for charged fields. As an example of commutation relations
that do not agree with~\eref{9.25} for charged fields and, consequently,
with~\eref{9.29}, we shall point the following ones:
	\begin{subequations}	\label{9.30}
	\begin{align}
			\label{9.30a}
	\begin{split}
\bigl[ a_{l^\alpha}^{\pm}
& ,
[ a_{m^\beta}^{+} , a_{n^\gamma}^{\dag\,-} ]_{\varepsilon^{\beta\gamma}}
\bigr]_{-}
+  \delta_{{l^\alpha}{n^\gamma}} a_{m^\beta}^{\pm} = 0
	\end{split}
\\			\label{9.30b}
	\begin{split}
\bigl[ a_{l^\alpha}^{\pm}
& ,
[ a_{m^\beta}^{\dag\,+} , a_{n^\gamma}^{-} ]_{\varepsilon^{\beta\gamma}} -
\bigr]_{-}
-  \delta_{{l^\alpha}{n^\gamma}} a_{m^\beta}^{\pm} = 0 ,
	\end{split}
	\end{align}
	\end{subequations}	
which are a multifield generalization of~\eref{6.32}.
\vspace{2ex}

	The consideration of commutation relations originating from the
`orbital' Heisenberg equation~\eref{5.4} is analogous to the one of the same
relations regarding the charge operator. The multifield version of~\eref{6.47}
is:
	\begin{subequations}	\label{9.31} 
	\begin{multline}	\label{9.31a}
 \bigl\{
( -\omega_{\mu\nu}^\circ({m^\beta}) + \omega_{\mu\nu}^\circ({n^\gamma}) )
( [\ta_{l^\alpha}^{\pm} ,
[ \ta_{m^\beta}^{\dag\,+} , \ta_{n^\gamma}^{-} ]_{\varepsilon^{\beta\gamma}}
\\
+
[ \ta_{m^\beta}^{+} , \ta_{n^\gamma}^{\dag\,-} ]_{\varepsilon^{\beta\gamma}}
]_{\_} )
\bigr\} \big|_{{n^\gamma}={m^\beta}}
=
4 (1+\tau^{\alpha\beta}) \delta_{{l^\alpha}{m^\beta}}
		\omega_{\mu\nu}^\circ({l^\alpha}) (\ta_{l^\alpha}^{\pm})
	\end{multline}			
\vspace{-5.3ex}
	\begin{multline}	\label{9.31b}
 \bigl\{
( -\omega_{\mu\nu}^\circ({m^\beta}) + \omega_{\mu\nu}^\circ({n^\gamma}) )
( [\ta_{l^\alpha}^{\dag\,\pm} ,
[ \ta_{m^\beta}^{\dag\,+} , \ta_{n^\gamma}^{-} ]_{\varepsilon^{\beta\gamma}}
\\
+
[ \ta_{m^\beta}^{+} , \ta_{n^\gamma}^{\dag\,-} ]_{\varepsilon^{\beta\gamma}}
]_{\_} )
\bigr\} \big|_{{n^\gamma}={m^\beta}}
=
4 (1+\tau^{\alpha\beta}) \delta_{{l^\alpha}{m^\beta}}
		\omega_{\mu\nu}^\circ({l^\alpha}) (\ta_{l^\alpha}^{\dag\,\pm})
	\end{multline}			
	\end{subequations}		
where
	\begin{equation}	\label{9.32}
\omega_{\mu\nu}^\circ(l^\alpha)
:= \omega_{\mu\nu}(k)
 = k_\mu \frac{\pd}{\pd k^\nu} -  k_\nu \frac{\pd}{\pd k^\mu}
\qquad\text{if } l^\alpha=(\alpha,s^\alpha,\bk) .
	\end{equation}

	Applying~\eref{6.49}, with $m^\beta$ for $m$ and $n^\gamma$ for $n$,
one can check that the multifield paracommutation relations~\eref{9.25}
convert~\eref{9.31} into identities and hence provide a solution
of~\eref{9.31} and ensure the validity of~\eref{5.4}, when system of different
free fields is considered. An example of a solution of~\eref{9.31} which does
not agree with~\eref{9.25} is provided by the following multifield
generalization of~\eref{6.50}:
	\begin{subequations}	\label{9.33}
	\begin{align}
			\label{9.33a}
& \bigl[ a_{l^\alpha}^{+} , [ a_{m^\beta}^{+} , a_{n^\gamma}^{\dag\,-}
			    ]_{\varepsilon^{\beta\gamma}} \bigr]_{-}
= \bigl[ a_{l^\alpha}^{+} , [ a_{m^\beta}^{\dag\,+} , a_{n^\gamma}^{-}
			    ]_{\varepsilon^{\beta\gamma}} \bigr]_{-}
= - (1+\tau^{\alpha\gamma}) \delta_{{l^\alpha}{n^\gamma}} a_{m^\beta}^{+}
\\			\label{9.33b}
& \bigl[ a_{l^\alpha}^{-} , [ a_{m^\beta}^{+} , a_{n^\gamma}^{\dag\,-}
			    ]_{\varepsilon^{\beta\gamma}} \bigr]_{-}
= \bigl[ a_{l^\alpha}^{-} , [ a_{m^\beta}^{\dag\,+} , a_{n^\gamma}^{-}
			    ]_{\varepsilon^{\beta\gamma}} \bigr]_{-}
= + (1+\tau^{\alpha\beta}) \delta_{{l^\alpha}{m^\beta}} a_{n^\gamma}^{+} ,
	\end{align}
	\end{subequations}	
which provides a solution of~\eref{9.4}. Notice, the evident multifield
version of~\eref{6.51} agrees with~\eref{9.4}, but disagrees with~\eref{9.31}
when the lower signs are used.

	At last, the multifield exploration of the `spin' Heisenberg
relations~\eref{5.5} is a \emph{mutatis mutandis} (see~\eref{9.31}) version
of the corresponding considerations in the second part of
Subsect.~\ref{Subsect6.3}. The main result here is that the multifield
bilinear commutation relations~\eref{9.17}, as well as their para
counterparts~\eref{9.25}, ensure the validity of~\eref{5.5}.

\subsection
{Commutation relations between the dynamical variables}

\label{Subsect9.3}

	The aim of this subsection is to be discussed/proved the commutation
relations~\eref{5.9}--\eref{5.18} for a system of at least two different
quantum fields from the view\ndash point of the commutation relations
considered in subsections~\ref{Subsect9.1} and~\ref{Subsect9.2}.

	To begin with, we rewrite the Heisenberg relations~\eref{5.1},
\eref{5.2} and~\eref{5.4} in terms of creation and annihilation operators for
a multifield system~\cite{Bogolyubov&Shirkov,Bjorken&Drell-2}:
	\begin{align}	\label{9.34}
&  [ a_{l^\alpha}^{\pm} , \ope{P}_\mu ]_{\_}
= \mp k_\mu a_{l^\alpha}^{\pm}
&& [ a_{l^\alpha}^{\dag\,\pm} , \ope{P}_\mu ]_{\_}
= \mp k_\mu  a_{l^\alpha}^{\dag\,\pm}
\\			\label{9.35}
&  [ a_{l^\alpha}^{\pm} , \ope{Q} ]_{\_} =  q a_{l^\alpha}^{\pm}
&& [ a_{l^\alpha}^{\dag\,\pm} , \ope{Q} ]_{\_} = - q a_{l^\alpha}^{\dag\,\pm}
\\			\label{9.36}
&  [ \ta_{l^\alpha}^{\pm} , \ope{M}_{\mu\nu}^{\mathrm{or}} ]_{\_}
= \ih \omega_{\mu\nu}^\circ(l^\alpha) \bigl( \ta_{l^\alpha}^{\pm} \bigr)
&& [ \ta_{l^\alpha}^{\dag\,\pm} , \ope{M}_{\mu\nu}^{\mathrm{or}} ]_{\_}
= \ih \omega_{\mu\nu}^\circ(l^\alpha)
      \bigl( \ta_{l^\alpha}^{\dag\,\pm} \bigr) ,
	\end{align}
where $l^\alpha=(\alpha,s^\alpha,\bk)$, $\omega^\circ(l^\alpha)$ is defined
by~\eref{9.32} and $k_0=\sqrt{m^2c^2+\bk^2}$ is set in~\eref{9.34}
and~\eref{9.36} (after the differentiations are performed in the last case).
The corresponding version of~\eref{5.5} is more complicated and depends on
the particular field considered (do not sum over $s^\alpha$!):
	\begin{equation}	\label{9.37}
	\begin{split}
f^{s^\alpha} [ a_{\alpha,s^\alpha}^{\pm} (\bk) ,
	   \ope{M}_{\mu\nu}^{\mathrm{sp}}]_{\_}
& =
\ih g_\alpha \sum_{t^\alpha}
\bigl\{
\lindex[\sigma]{}{\pm}_{\mu\nu}^{s^\alpha t^\alpha,+} (\bk)
			      a_{\alpha,t^\alpha}^{+} (\bk)
+
\lindex[\sigma]{}{\pm}_{\mu\nu}^{s^\alpha t^\alpha,-} (\bk)
			      a_{\alpha,t^\alpha}^{-} (\bk)
\bigr\}
\\
f^{s^\alpha} [ a_{\alpha,s^\alpha}^{\dag\,\pm} (\bk) ,
	   \ope{M}_{\mu\nu}^{\mathrm{sp}}]_{\_}
& =
\ih h_\alpha \sum_{t^\alpha}
\bigl\{
\lindex[\sigma]{}{\pm}_{\mu\nu}^{s^\alpha t^\alpha,-} (\bk)
			      a_{\alpha,t^\alpha}^{\dag\,+} (\bk)
+
\lindex[\sigma]{}{\pm}_{\mu\nu}^{s^\alpha t^\alpha,+} (\bk)
			      a_{\alpha,t^\alpha}^{\dag\,-} (\bk)
\bigr\} ,
	\end{split}
	\end{equation}
where $f_{s^\alpha}=-1,0,+1$ (depending on the particular field),
\(
g_\alpha
:=-h_\alpha
:=\frac{1}{j^\alpha+\delta_{j^\alpha 0}}(-1)^{j^\alpha + 1}
\)
and
$\lindex[\sigma]{}{\pm}_{\mu\nu}^{s^\alpha t^\alpha,+} (\bk)$ and
$\lindex[\sigma]{}{\pm}_{\mu\nu}^{s^\alpha t^\alpha,-} (\bk)$
are some functions which strongly depend on the particular field considered,
with
$\lindex[\sigma]{}{\pm}_{\mu\nu}^{s^\alpha t^\alpha,\pm} (\bk)$
 being related to the spin (polarization) functions
$\sigma_{\mu\nu}^{s^\alpha t^\alpha,\pm} (\bk)$
(see~\eref{3.13} and~\eref{3.9}).%
\footnote{~
If $\tope{\phi}_{i}^{\alpha}(\bk)$ are the Fourier images of the $\alpha$-th
field and
	\begin{gather}	\label{9.38}
\tope{\phi}_{i}^{\alpha}(\bk)
=
\sum_{s^\alpha}
\bigl\{
v_{i}^{s^\alpha,+}(\bk) \ta_{\alpha,s^\alpha}^+(\bk) +
v_{i}^{s^\alpha,-}(\bk) \ta_{\alpha,s^\alpha}^-(\bk)
\bigr\} ,
\intertext{where $v_{i}^{s^\alpha,\pm}(\bk)$ are linearly independent
functions normalize via the condition}
			\label{9.39}
\sum_{i} \bigl( v_{i}^{s^\alpha,\pm}(\bk) \bigr)^\ast
		v_{i}^{t^\alpha,\pm}(\bk)
=
\delta^{s^\alpha t^\alpha} f^{s^\alpha},
	\end{gather}
with
$f^{s^\alpha}=1$ for $j^\alpha=0,\frac{1}{2}$ and
$f^{s^\alpha}=0,-1$ for
$(j^\alpha,s^\alpha)=(1,3)$ or
$(j^\alpha,s^\alpha)=(1,1),(1,2)$, respectively,
then
	\begin{equation}	\label{9.40}
	\begin{split}
\lindex[\sigma]{}{+}_{\mu\nu}^{s^\alpha t^\alpha,\pm} (\bk)
:=
\frac{1}{g_\alpha} \sum_{i, i'}^{}
\bigl( v_{i}^{s^\alpha,+}(\bk) \bigr)^\ast I_{i\mu\nu}^{i'}
v_{i'}^{t^\alpha,\pm}(\bk)
\\
\lindex[\sigma]{}{-}_{\mu\nu}^{s^\alpha t^\alpha,\pm} (\bk)
:=
\frac{1}{g_\alpha} \sum_{i, i'}^{}
\bigl( v_{i}^{s^\alpha,-}(\bk) \bigr)^\ast I_{i\mu\nu}^{i'}
v_{i'}^{t^\alpha,\pm}(\bk) ,
	\end{split}
	\end{equation}
with $I_{i\mu\nu}^{i'}$ given via~\eref{5.19}. Besides,
\(
\sigma_{\mu\nu}^{s^\alpha t^\alpha,\pm} (\bk)
= \lindex[\sigma]{}{\pm}_{\mu\nu}^{s^\alpha t^\alpha,\pm} (\bk)
\)
with an exception that
$\sigma_{\mu\nu}^{s^\alpha t^\alpha,\pm} (\bk) = 0$
for  $j^\alpha=\frac{1}{2}$ and $(\mu,\nu)=(a,0),(0,a)$ with $a=1,2,3$.%
} 
As a result of~\eref{5.5-1},~\eref{9.36} and~\eref{9.37}, one can easily
write the Heisenberg relations~\eref{5.3} in a form similar
to~\eref{9.34}--\eref{9.37}.

	The commutation relations involving the momentum operator are:
	\begin{equation}	\label{9.41}
	\begin{split}
& [ \ope{P}_\mu , \ope{P}_\nu ]_{\_} = 0		\quad
  [ \ope{Q} , \ope{P}_\mu ]_{\_} = 0
\\
& [ \ope{S}_{\mu\nu} , \ope{P}_\lambda ]_{\_}
= [ \ope{M}_{\mu\nu}^{\mathrm{sp}} , \ope{P}_\lambda ]_{\_} = 0
\\
& [ \ope{L}_{\mu\nu} , \ope{P}_\lambda ]_{\_}
= [ \ope{M}_{\mu\nu}^{\mathrm{or}}  , \ope{P}_\lambda ]_{\_}
= [ \ope{M}_{\mu\nu} , \ope{P}_\lambda ]_{\_}
= -\ih \{ \eta_{\lambda\mu} \ope{P}_\nu - \eta_{\lambda\nu} \ope{P}_\mu \}
 .
	\end{split}
	\end{equation}
We claim that these equations are consequences from~\eref{9.34} and the
explicit expressions~\eref{3.7}--\eref{3.10} and~\eref{5.7-3}--\eref{5.7-5}
for the operators of the dynamical variables of the free fields considered in
the present work. In fact, since~\eref{9.34} implies
	\begin{subequations}	\label{9.42}
	\begin{align}	\label{9.42a}
& [ b_{l^\alpha}^\pm \circ c_{m^\beta}^\mp , \ope{P}_\mu ]_{\_} = 0
\qquad
l^\alpha=(\alpha,s^\alpha,\bk),\ m^\beta=(\beta,s^\beta,\bk)
\\			\label{9.42b}
& [ b_{l^\alpha}^\pm \xlrarrow{\omega_{\mu\nu}^\circ} (l^\alpha)
    \circ c_{m^\beta}^\mp , \ope{P}_\mu ]_{\_}
=
\pm 2 ( k_\mu\eta_{\nu\lambda} - k_\nu \eta_{\mu\lambda} )
b_{l^\alpha}^\pm \circ c_{m^\beta}^\mp ,
	\end{align}
	\end{subequations}
where
\(
b_{l^\alpha}^\pm , c_{l^\alpha}^\pm
= a_{l^\alpha}^{\pm}, a_{l^\alpha}^{\dag\,\pm}
\)
 and
 $\xlrarrow{\omega_{\mu\nu}^\circ}(l^\alpha)$ is defined via~\eref{9.32}
and~\eref{3.11}, the verification of~\eref{9.41} reduces to almost trivial
algebraic calculations. Further, we assert that any system of commutation
relations considered in Subsect.~\ref{Subsect9.1} entails~\eref{9.41}: as
these relations always imply~\eref{9.4} (or similar multifield versions
of~\eref{6.9} and~\eref{6.10} in the case of the Lagrangians~\eref{3.1}
or~\eref{3.3}, respectively) and, on its turn,~\eref{9.4} implies~\eref{5.1},
the required result follows from the last assertion and the remark
that~\eref{5.1} and~\eref{9.34} are equivalent. As an additional verification
of the validity of~\eref{9.41}, the reader can prove them by invoking the
identity~\eref{6.13} and any system of commutation relations mentioned in
Subsect.~\ref{Subsect9.1}, in particular~\eref{9.17} and~\eref{9.25}.

	The commutation relations concerning the charge operator read:
	\begin{equation}	\label{9.43}
	\begin{split}
& [\ope{P}_\mu,\ope{Q}]_{\_} = 0 \quad
  [\ope{Q},\ope{Q}]_{\_} = 0
\\ &
  [\ope{L}_{\mu\nu},\ope{Q}]_{\_}
= [\ope{S}_{\mu\nu},\ope{Q}]_{\_} = 0
\\ &
  [\ope{M}_{\mu\nu}^{\mathrm{or}},\ope{Q}]_{\_}
= [\ope{M}_{\mu\nu}^{\mathrm{sp}},\ope{Q}]_{\_}
= [\ope{M}_{\mu\nu},\ope{Q}]_{\_} = 0  .
	\end{split}
	\end{equation}
These equations are trivial corollaries from~\eref{3.7}--\eref{3.10}
and~\eref{5.7-3}--\eref{5.7-5} and the observation that~\eref{9.35} implies
	\begin{equation}	\label{9.44}
  [ a_{l^\alpha}^{\dag\,\pm} \circ a_{m^\beta}^{\mp} , \ope{Q} ]_{\_}
= [ a_{l^\alpha}^{\pm} \circ a_{m^\beta}^{\dag\,\mp} , \ope{Q} ]_{\_}
= 0
\qquad\text{if } q^\alpha=q^\beta ,
	\end{equation}
due to~\eref{6.13} for $\eta=-1$. Since any one of the systems of commutation
relations mentioned in Subsect.~\ref{Subsect9.2} entails~\eref{9.27} (or
systems of similar multifield versions of~\eref{6.29} and~\eref{6.30}, if the
Lagrangians~\eref{3.1} or~\eref{3.3} are employed), which is equivalent
to~\eref{9.35},  the equations~\eref{9.43} hold if some of these systems is
valid. Alternatively, one can prove via a direct calculation that the
commutation relations arising from the charge operator entail the validity
of~\eref{9.43}; for the purpose the identity~\eref{6.13} and the explicit
expressions for the dynamical variables via the creation and annihilation
operators should be applied.

	At last, consider the commutation relations involving the different
angular momentum operators:
	\begin{equation}
			\label{9.45}
	\begin{split}
& [ \ope{P}_\lambda , \ope{S}_{\mu\nu}]_{\_}
= [ \ope{P}_\lambda , \ope{M}_{\mu\nu}^{\mathrm{sp}} ]_{\_} = 0
\\
& [ \ope{P}_\lambda , \ope{L}_{\mu\nu} ]_{\_}
= [ \ope{P}_\lambda , \ope{M}_{\mu\nu}^{\mathrm{or}} ]_{\_}
= [ \ope{P}_\lambda , \ope{M}_{\mu\nu} ]_{\_}
= + \ih \{ \eta_{\lambda\mu} \ope{P}_\nu - \eta_{\lambda\nu} \ope{P}_\mu \}
\\
& [\ope{Q},\ope{L}_{\mu\nu}]_{\_}
= [\ope{Q},\ope{S}_{\mu\nu}]_{\_}
= [\ope{Q},\ope{M}_{\mu\nu}^{\mathrm{or}}]_{\_}
= [\ope{Q},\ope{M}_{\mu\nu}^{\mathrm{sp}}]_{\_}
= [\ope{Q},\ope{M}_{\mu\nu}]_{\_} = 0
\\
&
[\ope{S}_{\varkappa\lambda} , \ope{M}_{\mu\nu} ]_{\_}
=
- \ih \bigl\{
\eta_{\varkappa\mu}	\ope{S}_{\lambda\nu} -
\eta_{\lambda\mu}	\ope{S}_{\varkappa\nu} -
\eta_{\varkappa\nu}	\ope{S}_{\lambda\mu} +
\eta_{\lambda\nu}	\ope{S}_{\varkappa\mu}
\bigr\}
\\
& [\ope{L}_{\varkappa\lambda} , \ope{M}_{\mu\nu} ]_{\_}
 =
- \ih \bigl\{
\eta_{\varkappa\mu}	\ope{L}_{\lambda\nu} -
\eta_{\lambda\mu}	\ope{L}_{\varkappa\nu} -
\eta_{\varkappa\nu}	\ope{L}_{\lambda\mu} +
\eta_{\lambda\nu}	\ope{L}_{\varkappa\mu}
\bigr\}
\\
& [ \ope{M}_{\varkappa\lambda} , \ope{M}_{\mu\nu} ]_{\_}
=
- \ih \bigl\{
\eta_{\varkappa\mu}	\ope{M}_{\lambda\nu} -
\eta_{\lambda\mu}	\ope{M}_{\varkappa\nu} -
\eta_{\varkappa\nu}	\ope{M}_{\lambda\mu} +
\eta_{\lambda\nu}	\ope{M}_{\varkappa\mu}
\bigr\} .
	\end{split}
	\end{equation}
(The other commutators, that can be form from the different angular momentum
operators, are complicated and cannot be expressed in a `closed' form.)
	The proof of these relations is based on equations like
(see~\eref{9.36} and~\eref{6.13})
	\begin{equation}	\label{9.45-1}
[ b_{l^\alpha} \circ c_{m^\beta} , \ope{M}_{\mu\nu}^{\mathrm{or}} ]_{\_}
=
\ih \omega_{\mu\nu}^{\circ}(l^\alpha)
\bigl( b_{l^\alpha} \circ c_{m^\beta} \bigr)
\qquad
l^\alpha=(\alpha,s^\alpha,\bk),\ m^\beta=(\beta,s^\beta,\bk) ,
	\end{equation}
with
\(
b_{l^\alpha}, c_{l^\alpha}
= a_{l^\alpha}^{+}, a_{l^\alpha}^{-},
  a_{l^\alpha}^{\dag\,+}, a_{l^\alpha}^{\dag\,-} ,
\)
and similar, but more complicated, ones involving the other angular
momentum operators. It, generally, depends on the particular field considered
and will be omitted.

	As it was said in Subsect.~\ref{Subsect6.3}, the Heisenberg
relations concerning the angular momentum operator(s) do not give rise to
some (algebraic) commutation relations for the creation and annihilation
operators. For this reason, the only problem is which of the commutation
relations discussed in subsections~\ref{Subsect9.1} and~\ref{Subsect9.2}
imply the validity of the equations~\eref{9.45} (or part of them). The
general answer of this problem is not known but, however, a direct
calculation by means of~\eref{9.6}, if it holds, and~\eref{6.13} shows the
validity of~\eref{9.45}. Since~\eref{9.17} and~\eref{9.25} imply~\eref{9.6},
this means that the multifield bilinear and para commutation relations are
sufficient for the fulfillment of~\eref{9.45}.

	To conclude, let us draw the major moral of the above material: the
multifield bilinear commutation relations~\eref{9.17} and the multifield
paracommutation relations~\eref{9.25} ensure the validity of all `standard'
commutation relations~\eref{9.41}, \eref{9.43} and~\eref{9.45} between the
operators of the dynamical variables characterizing free scalar, spinor and
vector fields.

\subsection
{Commutation relations under the uniqueness conditions}

\label{Subsect9.4}

	As it was said at the end of the introduction to this section, the
replacements~\eref{9.3-1} ensure the validity of the material of
Sect.~\ref{Sect4} in the multifield case. Correspondingly, the considerations
in Sect.~\ref{Sect7} remain valid in this case provided the changes
	\begin{equation}	\label{9.46}
	\begin{split}
&
l\mapsto l^\alpha \quad
m\mapsto m^\beta \quad
n\mapsto n^\gamma 
\\ &
\tau \delta_{lm} \mapsto \tau^{\alpha\beta} \delta_{l^\alpha m^\beta}
= \tau^{\alpha} \delta_{l^\alpha m^\beta}
\\ &
[b_{m},b_{m}]_{\varepsilon} \mapsto
      		[b_{m^\beta},b_{m^\beta}]_{\varepsilon^\beta} \quad
[b_{m},b_{n}]_{\varepsilon} \mapsto
      		[b_{m^\beta},b_{n^\gamma}]_{\varepsilon^{\beta\gamma}} ,
	\end{split}
	\end{equation}
with $b_m$ (or $b_{m^\beta}$) being any creation/annihilation operator, and,
in some cases,~\eref{9.3-1} are made.%
\footnote{~%
As a result of~\eref{7.8},~\eref{7.13} and~\eref{7.13-1}, in expressions
like~\eref{7.14}--\eref{7.22} the number $\varepsilon$ should be replace by
$\varepsilon^{\alpha\beta}$, where $\alpha$ and $\beta$ are the corresponding
field indices of the creation/annihilation operators on which the operator
$\ope{E}$ acts, \ie
\(
\varepsilon\ope{E}(b_m\circ b_n) \mapsto
	\varepsilon^{\beta\gamma}\ope{E}(b_{m^\beta}\circ b_{n^\gamma}).
\)%
}
Without going into details, we shall write the final results.

	The multifield version of~\eref{7.23}--\eref{7.24} is:
	\begin{equation}	\label{9.47}
\ope{E}( a_{m^\beta}^{\dag\,\pm} \circ a_{n^\gamma}^{\mp} )
= {\varepsilon^{\beta\gamma}} \ope{E}( a_{n^\gamma}^{\mp} \circ a_{m^\beta}^{\dag\,\pm} )
= \frac{1}{2} \ope{E}( [ a_{m^\beta}^{\dag\,\pm} , a_{n^\gamma}^{\mp} ]_{{\varepsilon^{\beta\gamma}}} )
	\end{equation}
\vspace{-4.2ex}
	\begin{subequations}	\label{9.48}
	\begin{align}
			\label{9.48a}
	\begin{split}
\bigl[ a_{l^\alpha}^{+} & ,
\ope{E}( [ a_{m^\beta}^{+} , a_{n^\gamma}^{\dag\,-}
	 ]_{\varepsilon^{\beta\gamma}} ) \bigr]_{-}
+2 \delta_{{l^\alpha}{n^\gamma}} a_{m^\beta}^{+} = 0
	\end{split}
\\			\label{9.48b}
	\begin{split}
\bigl[ a_{l^\alpha}^{+} & ,
\ope{E}( [ a_{m^\beta}^{\dag\,+} , a_{n^\gamma}^{-}
	 ]_{\varepsilon^{\beta\gamma}} ) \bigr]_{-}
+ 2 \tau^{\alpha\gamma} \delta_{{l^\alpha}{n^\gamma}} a_{m^\beta}^{+} = 0
	\end{split}
\\			\label{9.48c}
	\begin{split}
\bigl[ a_{l^\alpha}^{-} & ,
\ope{E}( [ a_{m^\beta}^{+} , a_{n^\gamma}^{\dag\,-}
	 ]_{\varepsilon^{\beta\gamma}} ) \bigr]_{-}
- 2 \tau^{\alpha\beta} \delta_{{l^\alpha}{m^\beta}} a_{n^\gamma}^{-} = 0
	\end{split}
\\			\label{9.48d}
	\begin{split}
\bigl[ a_{l^\alpha}^{-} & ,
\ope{E}( [ a_{m^\beta}^{\dag\,+} , a_{n^\gamma}^{-}
	 ]_{\varepsilon^{\beta\gamma}} ) \bigr]_{-}
-2 \delta_{{l^\alpha}{m^\beta}} a_{n^\gamma}^{-} = 0
	\end{split}
\\			\label{9.48e}
\gamma = & \beta .
	\end{align}
	\end{subequations}	
As one can expect, the relations~\eref{9.48a}--\eref{9.48d} can be obtained
from the multifield paracommutation relations~\eref{9.25} via the replacement
\(
[\cdot,\cdot]_{\varepsilon} \mapsto
\ope{E}( [\cdot,\cdot]_{\varepsilon^{\beta\gamma}} ).
\)
It should be paid special attention on the equation~\eref{9.48e}. It is due
to the fact that in the expressions for the dynamical variables do not enter
`cross-field-products', like
$a_{l^\alpha}^{\dag\,+}\circ a_{m^\beta}^{-}$ for $\beta\not=\alpha$,
and it corresponds to the condition~(ii)
in~\cite[p.~B~1159]{Greenberg&Messiah-1965}. The equality~\eref{9.48e} is
quite important as it selects only that part of the `$\ope{E}$\ndash
transformed' multifield paracommutation relations~\eref{9.25} which is
compatible with the bilinear commutation relations~\eref{9.17}
(see~\eref{9.25-1} and~\eref{9.25-2}). Besides,~\eref{9.48e}
makes~\eref{9.48a}--\eref{9.48d} independent of the particular definition
of $\varepsilon^{\alpha\beta}$ (see~\eref{9.9}).

	The equations~\eref{9.47} are the only restrictions on the
operator $\ope{E}$; examples of this operator are provided by the normal
(resp.\ antinormal) ordering operator $\ope{N}$ (resp.\ $\ope{A}$), which has
the properties (cf.~\eref{4.20} (resp.~\eref{7.26})
	\begin{align}	\label{9.49}
	\begin{split}
\ope{N}\bigl( a_{m^\beta}^{+} \circ a_{n^\gamma}^{\dag\,-} \bigr)
& :=
a_{m^\beta}^{+} \circ a_{n^\gamma}^{\dag\,-}
\quad \quad\
\ope{N}\bigl( a_{m^\beta}^{\dag\,+} \circ a_{n^\gamma}^{-} \bigr)
:=
a_{m^\beta}^{\dag\,+} \circ a_{n^\gamma}^{-}
\\
\ope{N}\bigl( a_{m^\beta}^{-} \circ a_{n^\gamma}^{\dag\,+} \bigr)
& :=
\varepsilon^{\beta\gamma} a_{n^\gamma}^{\dag\,+} \circ a_{m^\beta}^{-}
\quad
\ope{N}\bigl( a_{m^\beta}^{\dag\,-} \circ a_{n^\gamma}^{+} \bigr)
:=
\varepsilon^{\beta\gamma} a_{n^\gamma}^{+} \circ a_{m^\beta}^{\dag\,-}
	\end{split}
\\[1ex]			\label{9.50}
	\begin{split}
\ope{A}\bigl( a_{m^\beta}^{+} \circ a_{n^\gamma}^{\dag\,-} \bigr)
& :=
\varepsilon^{\beta\gamma} a_{n^\gamma}^{\dag\,-} \circ a_{m^\beta}^{+}
\quad
\ope{A}\bigl( a_{m^\beta}^{\dag\,+} \circ a_{n^\gamma}^{-} \bigr)
:=
\varepsilon^{\beta\gamma} a_{n^\gamma}^{-} \circ a_{m^\beta}^{\dag\,+}
\\
\ope{A}\bigl( a_{m^\beta}^{-} \circ a_{n^\gamma}^{\dag\,+} \bigr)
& :=
 a_{m^\beta}^{-}\circ a_{n^\gamma}^{\dag\,+}
\quad \quad\
\ope{A}\bigl( a_{m^\beta}^{\dag\,-} \circ a_{n^\gamma}^{+} \bigr)
:=
 a_{m^\beta}^{\dag\,-} \circ a_{n^\gamma}^{+} .
	\end{split}
	\end{align}

	The material of Sect.~\ref{Sect8} has also a multifield variant that
can be obtained via the replacements~\eref{9.46} and~\eref{9.3-1}. Here is a
brief summary of the main results found in that way.

	The operator $\ope{E}$  should possess the properties~\eref{9.49}
and, in this sense, can be identified with the normal ordering operator,
	\begin{equation}	\label{9.51}
\ope{E} = \ope{N} .
	\end{equation}
As a result of this fact and $\varepsilon^{\beta\beta}=\varepsilon^\beta$
(see~\eref{9.9}), the commutation relations~\eref{9.48} take the final form:
	\begin{subequations}	\label{9.53}
	\begin{align}
			\label{9.53a}
	\begin{split}
\bigl[ a_{l^\alpha}^{+} & ,
        a_{m^\beta}^{+} \circ a_{n^\beta}^{\dag\,-} \bigr]_{-}
+ \delta_{{l^\alpha}{n^\beta}} a_{m^\beta}^{+} = 0
	\end{split}
\\			\label{9.53b}
	\begin{split}
\bigl[ a_{l^\alpha}^{+} & ,
        a_{m^\beta}^{\dag\,+} \circ a_{n^\beta}^{-} \bigr]_{-}
+  \tau^{\alpha\beta} \delta_{{l^\alpha}{n^\beta}} a_{m^\beta}^{+} = 0
	\end{split}
\\			\label{9.53c}
	\begin{split}
\bigl[ a_{l^\alpha}^{-} & ,
       a_{m^\beta}^{+} \circ a_{n^\beta}^{\dag\,-} \bigr]_{-}
-  \tau^{\alpha\beta} \delta_{{l^\alpha}{m^\beta}} a_{n^\beta}^{-} = 0
	\end{split}
\\			\label{9.53d}
	\begin{split}
\bigl[ a_{l^\alpha}^{-} & ,
       a_{m^\beta}^{\dag\,+} \circ a_{n^\beta}^{-} \bigr]_{-}
- \delta_{{l^\alpha}{m^\beta}} a_{n^\beta}^{-} = 0
	\end{split}
	\end{align}
	\end{subequations}	
which is the multifield version of~\eref{8.9} and corresponds, up to the
replacement $a_{l^\alpha}^{\pm}\mapsto \sqrt{2}a_{l^\alpha}^{\pm}$,
to~\eref{9.25} with $\varepsilon^{\beta\gamma}=0$.

	The vacuum state vector $\ope{X}_0$ is supposed to be uniquely
defined by the following equations (cf.~\eref{8.1b}--\eref{8.1-2}):
	\begin{subequations}	\label{9.54}
	\begin{align}
			\label{9.54a}
& a_{l^\alpha}^{-} \ope{X}_0 = 0 \quad
  a_{l^\alpha}^{\dag\,-} \ope{X}_0 = 0
\\			\label{9.54b}
& \ope{X}_0 \not= 0
\\			\label{9.54c}
& \langle\ope{X}_0 | \ope{X}_0\rangle = 1
\\			\label{9.54d}
	\begin{split}
& a_{l^\alpha}^{\dag\,-} \circ a_{m^\beta}^{+} (\ope{X}_0)
=  \delta_{l^\alpha m^\beta} \ope{X}_0
\quad\quad\
  a_{l^\alpha}^{-} \circ a_{m^\beta}^{\dag\,+} (\ope{X}_0)
=  \delta_{l^\alpha m^\beta} \ope{X}_0
\\
& a_{l^\alpha}^{-} \circ a_{m^\beta}^{+} (\ope{X}_0)
= \tau^{\alpha\beta} \delta_{l^\alpha m^\beta} \ope{X}_0
\quad
  a_{l^\alpha}^{-} \circ a_{m^\beta}^{+} (\ope{X}_0)
= \tau^{\alpha\beta} \delta_{l^\alpha m^\beta} \ope{X}_0 .
	\end{split}
	\end{align}
	\end{subequations}	

	The Hilbert space $\Hil$ of state vectors is a direct sum of the
Hilbert spaces $\Hil^\alpha$ of the different fields and it is supposed to be
spanned by the vectors
	\begin{equation}	\label{9.55}
\psi_{l_{1}^{\alpha_1} l_{2}^{\alpha_2}\dots }
=
\ope{M}( a_{l_{1}^{\alpha_1}}^{+}, a_{l_{2}^{\alpha_2}}^{+},\dots )
(\ope{X}_0)
	\end{equation}
with $\ope{M}( a_{l_{1}^{\alpha_1}}^{+}, a_{l_{2}^{\alpha_2}}^{+},\dots ) $
being arbitrary monomial only in the creation operators.

	Since~\eref{9.54a},~\eref{9.51} and~\eref{9.49} imply the multifield
version of~\eref{8.5}, the computation of the mean values of~\eref{8.4}, with
$l_1\mapsto l_{1}^{\alpha_1}$ etc., of the dynamical variables is reduced to
the one of scalar products like (cf.~\eref{8.3})
	\begin{equation}	\label{9.56}
\langle
\psi_{l_{1}^{\alpha_1} l_{2}^{\alpha_2}\dots }
|
\phi_{m_{1}^{\beta_1} m_{2}^{\beta_2}\dots }
\rangle
=
\langle \ope{X}_0
|
\bigl( \ope{M}( a_{l_{1}^{\alpha_1}}^{+}, a_{l_{2}^{\alpha_2}}^{+},\dots )
\bigr)^\dag
\circ
\ope{M}'( a_{m_{1}^{\beta_1}}^{+}, a_{m_{2}^{\beta_2}}^{+},\dots )
(\ope{X}_0) \rangle
	\end{equation}
of basic vectors of the form~\eref{9.55}. By means of the basic
properties~\eref{9.54} of the vacuum, one is able to calculate the simplest
forms of the vacuum mean values~\eref{9.56}, \viz the multifield versions
(see~\eref{9.46}) of~\eref{8.10} and~\eref{8.15}. But more general such
expression cannot be calculated by means of~\eref{9.53}--\eref{9.54}.
\emph{Prima facie} one can suppose that the multifield commutation
relations~\eref{9.17}, which ensure the vectors~\eref{9.55} to form a base of
the system's Hilbert space of states, can help for the calculation
of~\eref{9.56} in more complicated cases. In fact, this is the case which
works perfectly well and covers the available experimental data.  In this
connection, we must mention that the applicability of~\eref{9.17} for
calculation of~\eref{9.56} is ensured by the \emph{compatibility/agreement}
between~\eref{9.17} and~\eref{9.53}: by means of~\eref{6.13} for
$\eta=-\varepsilon^{\alpha\beta}$, one can check that~\eref{9.17}
converts~\eref{9.53} into identities.%
\footnote{~%
Recall, equations~\eref{9.17} and~\eref{9.25}, or~\eref{9.48a}--\eref{9.48d},
for $\gamma\not=\beta$ are generally incompatible. For instance, excluding some
special cases, like systems consisting of only fermi (bose) fields or one
fermi (bose) field and arbitrary number of bose (fermi) fields, the only
operators satisfying~\eref{9.17} and~\eref{9.25} for $\gamma\not=\beta$
and having normal spin-statistics connection are such that $b_{m^\beta}\circ
b_{n^\gamma}=0$, with $\gamma\not=\beta$ and $b_{m^\beta}$ and $c_{n^\gamma}$
being any creation/annihilation operators, which, in particular, means that
no states with two particles from different fields can exist.%
}

	The commutation relations~\eref{9.53} admit as a solution also the
multifield version of the anomalous bilinear commutation relations~\eref{8.16}
but it, as we said earlier, leads to contradictions and must be rejected. The
existence of solutions of~\eref{9.53} different from it and~\eref{9.17}
seems not to be investigated. If there appear date which do not fit into the
description by means of~\eref{9.17}, one should look for other, if any,
solutions of~\eref{9.53} or compatible with~\eref{9.53} effective procedures
for calculating vacuum mean values like~\eref{9.56}.


\section {Conclusion}
\label{Conclusion}

	In this paper we have investigated two sources of (algebraic)
commutation relations in the Lagrangian quantum theory of free scalar, spinor
and vector fields: the uniqueness of the dynamical variables (momentum,
charge and angular momentum) and the Heisenberg relations/equations for them.
If one ignores the former origin, which is the ordinary case, the
paracommutation relations or some their generalizations seems to be the most
suitable candidates for the most general commutation relations that ensure
the validity of all Heisenberg equations. The simultaneous consideration of
the both sources mentioned reveals, however, their incompatibility in the
general case. The outlet of this situation is in the redefinition of the
operators of the dynamical variables, similar to the normal ordering
procedure and containing it as a special case. That operation ensures the
uniqueness of the new (redefined) dynamical variables and changes the
possible types of commutation relations. Again, the commutation relations,
connected with the Heisenberg relations concerning the (redefined) momentum
operator, entail the validity of all Heisenberg equations.

	Further constraints on the possible commutation relations follow from
the defini\-tion/in\-tro\-du\-ction of the concept of the vacuum (vacuum state
vector). They practically reduce the redefined dynamical variables to the
ones obtained via normal ordering procedure, which results in the explicit
form~\eref{8.9} of the admissible commutation relations. In a sense, they
happen to be `one half' of the paracommutation ones. As a last argument in
the way for finding the `unique true' commutation relations, we require the
existence of procedure for calculation of vacuum mean values of anti\ndash
normally ordered products of creation and annihilation operators, to which
the mean values of the dynamical variables and the transition amplitudes
between different states are reduced. We have pointed that the standard
bilinear commutation relations are, at present, the only known ones that
satisfy all of the conditions imposed and do not contradict to the existing
experimental data.

	The consideration of a system of at least two different quantum
free fields meets a new problem: the general relations between
creation/annihilation operators belonging to different fields turn to be
undefined. The cause for this is that the commutation relations for any fixed
field are well defined only on the corresponding to it Hilbert subspace of
the system's Hilbert space of states and their extension on the whole space,
as well as the inclusion in them of creation/annihilation operators of other
fields, is a matter of convention (when free fields are concerned); formally
this is reflected in the structure of the dynamical variables which are sums
of those of the individual fields included in the system under consideration.
We have, however, presented argument by means of which the \emph{a priori}
existing arbitrariness in the commutation relations involving different field
operators can be reduced to the `standard' one: these relations should
contain either commutators or anticommutators of the creation/annihilation
operators belonging to different fields. A free field theory cannot make
difference between these two possibilities. Accepting these possibilities, the
admissible commutation relations~\eref{9.53} for system of several different
fields are considered. They turn to be corresponding multifield versions of
the ones regarding a single field. Similarly to the single field case, the
standard multifield bilinear commutation relations seem to be the only known
ones that satisfy all of the imposed restrictions and are in agreement with
the existing data.


\section*{Acknowledgments}

	This research was partially supported by the National Science Fund of
Bulgaria under Grant No.~F~1515/2005.


\addcontentsline{toc}{section}{References}
\bibliography{bozhopub,bozhoref}
\bibliographystyle{unsrt}
\addcontentsline{toc}{subsubsection}{This article ends at page}

\end{document}